\documentclass[
  aps,prd,preprint,longbibliography,
  showpacs,showkeys,lengthcheck,
  nofootinbib,tightenlines,onecolumn,notitlepage,
  preprintnumbers,superscriptaddress
]{revtex4-2}

\usepackage[utf8]{inputenc}
\usepackage{newtxtext,newtxmath}
\usepackage[mathcal]{euscript}

\usepackage{graphicx}
\usepackage[dvipsnames]{xcolor}
\graphicspath{{figures/}}
\usepackage{tikz-feynman}
\usetikzlibrary{decorations.text}

\usepackage{hyperref}
\hypersetup{colorlinks=true,citecolor=blue,linkcolor=blue,urlcolor=blue}
\usepackage{orcidlink}

\usepackage{diagbox}
\usepackage{booktabs}
\usepackage{siunitx}
\AtBeginDocument{%
  \heavyrulewidth=.08em
  \lightrulewidth=.05em
  \cmidrulewidth=.03em
  \belowrulesep=.65ex
  \belowbottomsep=0pt
  \aboverulesep=.4ex
  \abovetopsep=0pt
  \cmidrulesep=\doublerulesep
  \cmidrulekern=.5em
  \defaultaddspace=.5em
}

\usepackage{amsmath}
\usepackage{slashed}
\usepackage{bm}
\usepackage{bbm}
\usepackage{mathtools}
\usepackage{tensor}

\usepackage[math]{cellspace}
\usepackage[draft,commentmarkup=uwave]{changes} 
\definechangesauthor[color=purple!80!black]{Wim}
\definechangesauthor[color=green!50!black]{Adam}
\definechangesauthor[color=blue!70!black]{Alan}

\renewcommand*{\d}{\mathop{}\!\mathrm{d}}
\newcommand*{\e}{\mathop{}\!\mathrm{e}}

\newcommand{\bd}{\bm{\varDelta}}
\newcommand{\dl}{\varDelta}
\newcommand{\lrn}{\overleftrightarrow{\nabla}}
\newcommand{\blrn}{\overleftrightarrow{\bm{\nabla}}}

\newcommand{\epi}[1][]{\varepsilon^{#1}}
\newcommand{\epf}[1][]{\varepsilon'^{*#1}}

\begin{document}

\title{
  Quantum stress and torsion distributions in the deuteron
}

\author{Wim Cosyn \orcidlink{0000-0002-9312-8569}}
\email{wcosyn@fiu.edu}
\affiliation{Department of Physics, Florida International University, Miami, Florida 33199, USA}

\author{Adam Freese \orcidlink{0000-0002-0688-4121}}
\email{afreese@jlab.org}
\affiliation{Center for Nuclear Femtography, Southeastern Universities Research Association, Newport News, Virginia 23606, USA}
\affiliation{Theory Center, Jefferson Lab, Newport News, Virginia 23606, USA}

\author{Alan Sosa
\orcidlink{0000-0002-6987-6161}}
\email{asosa090@fiu.edu}
\affiliation{Department of Physics, Florida International University, Miami, Florida 33199, USA}

\begin{abstract}
  Stress distributions in the deuteron are related to form factors of the
  asymmetric energy-momentum tensor through three-dimensional Fourier transforms.
  There are eleven such form factors, which we calculate in an impulse approximation.
  We compare the obtained form factors to prior results
  for the six form factors that have been previously calculated.
  We then elaborate on the formalism for relating the form factors to internal
  distributions of mass, mass flux, momentum, stresses, and forces,
  and obtain results for all of these distributions.
  We obtain the principal stresses for the symmetric part of the stress tensor,
  and show that the antisymmetric part describes reorientation of fermion spin
  by torsion stress when the nucleon moves between the S- and D-waves.
  Force distributions in the nucleons depend on the
  so-called non-conserved form factors through the Cauchy momentum equation,
  and are non-radial owing to the presence of
  tensor forces and spin-orbit coupling.
\end{abstract}

\preprint{JLAB-THY-26-4579}

\maketitle
\tableofcontents
\newpage

\section{Introduction}

A proliferation of studies have emerged during the past few years
on mechanical properties of hadrons~\cite{Polyakov:2018zvc,Burkert:2018bqq,Shanahan:2018nnv,Lorce:2018egm,Kumericki:2019ddg,Freese:2021czn,Burkert:2021ith,Panteleeva:2021iip,Pefkou:2021fni,Lorce:2021xku,Duran:2022xag,Burkert:2023wzr,Li:2024vgv,Hu:2024edc,Lorce:2025oot,Broniowski:2025ctl}.
Form factors of the energy-momentum tensor (EMT)---or
EMT form factors (EMT-FFs)\footnote{
  These form factors are often called
  gravitational form factors (GFFs) in the literature
  because the EMT is the source of gravitation in general relativity.
  We avoid this nomenclature here to avoid any potential confusion.
}---describe spatial distributions of mechanical
quantities such as mass, energy, momentum and forces.
Following the seminal work of Maxim Polyakov~\cite{Polyakov:2002yz},
the spatial components of the EMT in particular---usually called
the stress tensor---have commonly been interpreted
as providing distributions of mechanical stresses acting on
quarks and gluons\footnote{
  Skepticism of this interpretation has been expressed~\cite{Ji:2021mfb,Ji:2025gsq,Ji:2025qax},
  although this skepticism leans into an understanding of pressure and stress
  as only constituting surface forces induced by contact interactions.
  This understanding is at odds with how these concepts
  are used throughout the physics and engineering literature,
  where they are typically considered applicable to the interior of materials
  and to include both random kinetic motion
  (which is the microscopic origin of fluid pressure)
  and long-range forces such as the electrostatic force.
  See Ref.~\cite{maranganti2007quantum} for an excellent review
  of quantum notions of stress that touches on these matters.
};
see Refs.~\cite{Polyakov:2018zvc,Burkert:2023wzr,Lorce:2025oot}
for comprehensive reviews.
Moreover, taking this interpretation at face value
allows one to obtain the average force felt by a particular parton flavor
through the Cauchy momentum equation\footnote{
  The Cauchy momentum equation is also commonly known as
  Cauchy's first law of motion~\cite{chatterjee1999mathematical,chung2007general,irgens2008continuum}
  or Cauchy's equation of motion~\cite{chadwick1999continuum,chen2006meshless,irgens2008continuum,nayak2022continuum}.
}---that is,
by taking the divergence of that parton flavor's contribution
to the stress tensor~\cite{Polyakov:2018exb,Won:2023cyd,Freese:2024rkr,Kim:2025iis}.
Studying the hadronic stress tensor
and EMT form factors
thus has promise for providing
a detailed spatial map of strong nuclear forces in hadrons.

The deuteron is an especially appealing case study for inquiries
into mechanical properties of hadrons.
As a nucleus, it is extremely well-understood,
with its properties and wave function described to remarkable precision
through ordinary quantum mechanics with a variety of
potentials~\cite{Reid:1968sq,Lacombe:1980dr,Stoks:1994wp,Wiringa:1994wb,Machleidt:2000ge}.
Since the deuteron is to reasonable accuracy a non-relativistic system,
we can also bypass controversies about the correct definition
of relativistic densities.
Moreover, the nucleons inside it appear to experience little in the way of nuclear
modification---as evidenced by the BONuS measurement of the EMC effect
in the deuteron~\cite{Griffioen:2015hxa}---allowing
increasingly realistic and trustworthy
results for the nucleon EMT-FFs from
lattice quantum chromodynamics~\cite{Hackett:2023rif,Pefkou:2023okb}
to be used in calculating the deuteron's mechanical structure.

There has also been increased attention paid to the mechanical properties
of spin-one systems as a class.
It is well-understood how to break the EMT for spin-one systems down into
form factors~\cite{Holstein:2006ud,Abidin:2008ku,Taneja:2011sy,Cosyn:2019aio,Polyakov:2019lbq},
and how these form factors are related to Mellin moments of
generalized parton distributions~\cite{Cosyn:2018thq,Cosyn:2019aio,Polyakov:2019lbq}.
The differences between the symmetric EMT (which has nine EMT-FFs)
and the asymmetric EMT (which has eleven EMT-FFs)
are also well-understood~\cite{Cosyn:2019aio}.
Additionally, recent work by Kim and Kim~\cite{Kim:2025iis}
shows how the Cauchy momentum equation generalizes to spin-one systems,
clearing the field for studies into force distributions in the deuteron.
The formal machinery needed to study mechanical properties of the deuteron is already in place.

Despite this, there are only a handful of calculations of deuteron EMT-FFs.
Moreover, none of these calculations use ordinary, non-relativistic quantum mechanics
to obtain their results,
nor do they provide an exhaustive list of all eleven EMT-FFs.
Three calculations obtain six of the eleven EMT form factors.
The first, by Freese and Cosyn~\cite{Freese:2022yur},
used a light front convolution model.
He and Zahed~\cite{He:2023ogg,He:2024vzz} later calculated the EMT form factors
through a non-relativistic reduction of relativistic Feynman diagrams.
Most recently,
Panteleeva, Epelbaum, Gasparyan and Gegelia~\cite{Panteleeva:2024abz}
calculated the deuteron's EMT form factors
in a chiral effective field theory framework.
A few other works investigated a smaller subset of these EMT-FFs.
Two studies in holographic QCD---the first by
Mondal, Chakrabarti and Zhao~\cite{Mondal:2017lph},
and the second by Mamedov, Allahverdiyeva and
Akbarova~\cite{Mamedov:2024tth}---obtained two of the EMT-FFs by examining
the light front longitudinal momentum density.
Additionally, a Skyrme model calculation of nuclear pressures by
García Martín-Caro, Huidobro and Hatta~\cite{GarciaMartin-Caro:2023klo,GarciaMartin-Caro:2023toa}
examined a variety of nuclei, including the deuteron,
but only obtained one form factor.

The purpose of this work is to address the paucity of deuteron EMT calculations.
In particular, we obtain results for
all eleven form factors of the asymmetric EMT,
and we do so in ordinary non-relativistic quantum mechanics using
the precision AV18 deuteron wave function~\cite{Wiringa:1994wb}.
The most obvious benefit to this approach is that ordinary quantum mechanics
is a tried-and-true framework that is known to describe the
properties and behavior of the deuteron to remarkable accuracy.
Additionally, working within a non-relativistic framework from the outset
allows us to avoid ambiguities that may occur with non-relativistic reduction;
by dropping quantities of order $1/c^2$ in a non-relativistic
reduction, one may mistakenly discard terms that have nothing to do with
relativistic kinematics,
and were only present because $c$ appears naturally in the potential
(as it does in the electromagnetic and one-pion exchange potentials)
or as a unit conversion factor\footnote{
  In this work, we use standard natural units with $\hbar = c = 1$.
  In a non-relativistic context, $c$ does not have a dynamical meaning,
  and only serves as a unit conversion factor.
}.

One limitation to the scope of this work is that we will consider only
one-body contributions to the deuteron's EMT form factors.
In other words, our calculation will be in the impulse approximation.
We will thus find that the energy-momentum tensor is not locally conserved.
However, this is expected,
since local energy-momentum conservation should only be obeyed
when all subcomponents of a closed system are accounted for, including
force carriers;
see Refs.~\cite{Freese:2019bhb,Freese:2024rkr}
for examples of this.
It is with this limited scope that several so-called ``non-conserved''
form factors---of which the deuteron has four---appear,
and our work presents the first calculation of these for the deuteron.
These form factors, rather than a nuisance,
are actually quite physically meaningful,
as they are precisely what quantify the forces felt by subcomponents---in
this case, the nucleons---through the Cauchy momentum
equation~\cite{Polyakov:2018exb,Won:2023cyd,Freese:2024rkr,Kim:2025iis}.

This work is organized as follows.
In Sec.~\ref{sec:mff},
we obtain the EMT form factors,
presenting both analytic and numerical results
and comparing to prior calculations.
In Sec.~\ref{sec:density},
we obtain various spatial distributions
entailed by the EMT form factors,
including the mass, mass flux and momentum densities,
and various kinds of internal stress and force distributions.
The stresses include normal and symmetric shear stresses
that can be diagonalized through a local change of frame,
as well as an irreducible antisymmetric torsion stress.
We additionally obtain force distributions in the deuteron
through the Cauchy momentum equation.
We lastly provide an outlook in Sec.~\ref{sec:end}.
Outside the main body of the text,
Appendix~\ref{sec:conventions} compares our EMT-FF breakdown to prior conventions,
and Appendix~\ref{sec:wf} reviews the deuteron wave function.


\section{Energy-momentum tensor form factors}
\label{sec:mff}

Energy-momentum tensor form factors are defined by breaking the matrix element
$\langle p'_d, s_d' | \hat{T}^{\mu\nu}(0) | p_d, s_d \rangle$
of the energy-momentum tensor
down into the most general possible set of structures compatible with
symmetries of the operator
$\hat{T}^{\mu \nu}(0)$
and the theory in which it is defined.
For relativistic calculations, this includes Lorentz covariance,
whereas Galilei covariance is the appropriate symmetry for
non-relativistic calculations.
As shown in a recent work on non-relativistic fermions~\cite{Freese:2025tqd},
this means the EMT of a non-relativistic system generally has
more form factors than it would for a relativistic system.
However, these new form factors appear only in the energy and energy flux densities.
By restricting focus on the stress tensor,
the same number of form factors are seen for relativistic and non-relativistic systems.
Thus, by the counting of Ref.~\cite{Cosyn:2019aio},
there should be nine EMT-FFs for the symmetric stress tensor,
and eleven EMT-FFs for the asymmetric stress tensor.

Several conventions already exist for breaking the spin-one stress tensor
down into EMT-FFs~\cite{Holstein:2006ud,Abidin:2008ku,Taneja:2011sy,Cosyn:2019aio,Polyakov:2019lbq}.
Rather than using an existing convention,
we introduce a new one that is informed by
the multipole structure of the stress tensor,
and by simplifications that occur for
unpolarized or tensor-polarized ensembles\footnote{
  These simplifications are strictly non-relativistic,
  as they rely on the invariance of the
  deuteron polarization vectors under Galilei boosts.
}:
\begin{multline}
  \label{eqn:mff}
  \langle p_d', s_d' | \hat{T}^{ij}(0) | p_d, s_d \rangle
  =
  M_d
  \epi_a
  \epf_b
  \Bigg\{
    \frac{P_d^i P_d^j}{M_d^2}
    \left[
      \delta^{ab}
      A_U(\bd^2)
      +
      Y_2^{ab}(\hat{\dl})
      \frac{\bd^2}{2M_d^2}
      A_T(\bd^2)
      \right]
    +
    \frac{1}{2M_d^2}
    \left(
    \delta^{a\{i}
    \dl^b
    -
    \delta^{b\{i}
    \dl^a
    \right)
    P_d^{j\}}
    J(\bd^2)
    \\
    +
    \frac{\dl^i \dl^j - \delta^{ij} \bd^2}{4M_d^2}
    \left[
      \delta^{ab}
      D_U(\bd^2)
      +
      Y_2^{ab}(\hat{\dl})
      \frac{\bd^2}{2M_d^2}
      D_{T1}(\bd^2)
      \right]
    -
    \delta^{ij}
    \left[
      \delta^{ab}
      \bar{c}_U(\bd^2)
      +
      Y_2^{ab}(\hat{\dl})
      \frac{\bd^2}{2M_d^2}
      \bar{c}_{T1}(\bd^2)
      \right]
    \\
    +
    \frac{\bd^2}{2M_d^2}
    \left[
      Q^{jlab} Y_2^{il}(\hat{\dl})
      +
      Q^{liab} Y_2^{lj}(\hat{\dl})
      -
      Q^{klab}
      Y_2^{kl}(\hat{\dl})
      \delta^{ij}
      -
      \frac{1}{3}
      Q^{ijab}
      \right]
    D_{T2}(\bd^2)
    -
    Q^{ijab}
    \bar{c}_{T2}(\bd^2)
    \\
    +
    \frac{1}{2M_d^2}
    \left(
    \delta^{a[i}
    \dl^b
    -
    \delta^{b[i}
    \dl^a
    \right)
    P_d^{j]}
    S(\bd^2)
    +
    \frac{\bd^2}{4M_d^2}
    \Big(
    \delta^{a[i}_{\phantom{2}} Y_2^{j]b}(\hat{\dl})
    +
    \delta^{b[i}_{\phantom{2}} Y_2^{j]a}(\hat{\dl})
    \Big)
    \bar{s}(\bd^2)
  \Bigg\}
  \,.
\end{multline}
Here, $\bm{P}_d = \frac{1}{2}\big( \bm{p}_d + \bm{p}'_d \big)$ is the
average of the initial and final momenta and
$\bd = \bm{p}'_d - \bm{p}_d$ is the difference.
The indices $a$, $b$, $i$ and $j$ all range over $\{1,2,3\}$
and the Einstein summation convention is assumed.
The brackets signify symmetrization and antisymmetrization without a factor $\frac{1}{2}$,
i.e.,
$a^{\{i}b^{j\}} = a^i b^j + a^j b^i$ and
$a^{[i}b^{j]} = a^i b^j - a^j b^i$.
Since this calculation is non-relativistic, a Euclidean norm
should be assumed whenever contracting, raising or lowering indices.
A dictionary for translating these form factors into prior conventions
is given in Table~\ref{tab:mff}.
By including the so-called non-conserved form factors---in this nomenclature,
$\bar{c}_U$, $\bar{c}_{T1}$, $\bar{c}_{T2}$ and $\bar{s}$---the expression
also holds when applied to contributions from individual parton flavors.
We implicitly assume that all parton flavors have been summed
over throughout the work.
Note that due to focusing on impulse approximation contributions in this work,
these non-conserved form factors still can have non-zero values
after summing over all parton flavors.

The breakdown (\ref{eqn:mff})
borrows several tensors considered by Refs.~\cite{Polyakov:2018rew,Polyakov:2019lbq},
namely the rank-$n$ irreducible harmonic tensors:
\begin{align}
  \label{eqn:harmonic}
  \begin{split}
    Y_n^{i_1\ldots i_n}(\hat{r})
    &=
    \frac{(-1)^n}{(2n-1)!!}
    r^{n+1}
    \partial_{i_1} \ldots \partial_{i_n}
    \frac{1}{r}
    \\
    Y_0(\hat{r})
    &=
    1
    \,,
    \qquad
    Y_1^i(\hat{r})
    =
    \hat{r}
    \,,
    \qquad
    Y_2^{ij}(\hat{r})
    =
    \hat{r}^i \hat{r}^j
    -
    \frac{1}{3} \delta^{ij}
    \,,
    \\
    Y_3^{ij}(\hat{r})
    &=
    \hat{r}^i \hat{r}^j \hat{r}^k
    -
    \frac{1}{5}
    \Big(
    \delta^{ij} \hat{r}^k
    +
    \delta^{ik} \hat{r}^j
    +
    \delta^{kj} \hat{r}^i
    \Big)
    \,,
    \qquad
    \mathrm{etc.}
  \end{split}
\end{align}
and quadrupole tensor:
\begin{align}
  \label{eqn:quadrupole}
  Q^{ijab}
  &=
  \frac{1}{3}
  \delta^{ij} \delta^{ab}
  -
  \frac{1}{2}
  \delta^{a\{i}
  \delta^{j\}b}
  \,.
\end{align}
Technically, Refs.~\cite{Polyakov:2018rew,Polyakov:2019lbq}
use a symmetric traceless quadrupole operator
$\hat{Q}^{ij}=\frac{1}{2}\big(\hat J^i \hat J^j + \hat J^j \hat J^i - \tfrac{2}{3}\hat J^2 \big)$
for which
$\langle s_d' | \hat{Q}^{ij} | s_d \rangle = \epi_a \epf_b Q^{ijab}$,
with $Q^{ijab}$ as in Eq.~(\ref{eqn:quadrupole}).
The presence of $Y_2^{ab}(\hat{\dl})$ and $Q^{ijab}$
in the breakdown (\ref{eqn:mff})
clearly indicate tensor-polarized and quadrupole structures.
Helpfully, since $Y_2^{ab}(\hat{\dl})$ and $Q^{ijab}$ are both
traceless---in that they both contract with $\delta_{ab}$ to zero---it
is clear at a glance that terms with these structures will vanish
for unpolarized ensembles of deuterons.

Only a subset of the EMT-FFs survive when taking an unpolarized,
vector-polarized or tensor-polarized ensemble.
These ensemble averages are calculated by setting $s_d' = s_d$
in Eq.~(\ref{eqn:mff}) and performing a weighted average over $s_d$.
This is equivalent to taking the trace of $\rho(p_d,p'_d) \hat{T}^{ij}(0)$
where the spin-density matrix $\rho(p_d,p'_d)$ is off-diagonal in momentum,
but diagonal in spin.
Using the $z$ axis as the spin quantization axis,
the unpolarized, vector-polarized and tensor polarized
spin density matrices are:
\begin{align}
  \begin{split}
    \rho_U(p_d,p'_d)
    &=
    \frac{1}{3}
    \sum_{s_d}
    | p_d, s_d \rangle \langle p'_d, s_d |
    \\
    \rho_V(p,p')
    &=
    \frac{1}{2}
    \Big(
    | p_d, +1 \rangle \langle p'_d, +1 |
    -
    | p_d, -1 \rangle \langle p'_d, -1 |
    \Big)
    \\
    \rho_T(p,p')
    &=
    | p_d, +1 \rangle \langle p'_d, +1 |
    +
    | p_d, -1 \rangle \langle p'_d, -1 |
    -
    2
    | p_d, 0 \rangle \langle p'_d, 0 |
    \,.
  \end{split}
\end{align}
When taking an unpolarized ensemble,
only the form factors with a subscript $U$ survive:
\begin{align}
  \label{eqn:mff:U}
  \mathrm{Tr}\Big[
    \rho_U(p_d,p'_d)
    \hat{T}^{ij}(0)
    \Big]
  =
  M_d
  \Bigg\{
    \frac{P_d^i P_d^j}{M_d^2}
    A_U(\bd^2)
    +
    \frac{\dl^i \dl^j - \delta^{ij} \bd^2}{4M_d^2}
    D_U(\bd^2)
    -
    \delta^{ij} \bar{c}_U(\bd^2)
    \Bigg\}
  \,.
\end{align}
Taking a vector-polarized ensemble leaves only the form factors
$J(\bd^2)$ and $S(\bd^2)$:
\begin{align}
  \label{eqn:mff:V}
  \mathrm{Tr}\Big[
    \rho_V(p_d,p'_d)
    \hat{T}^{ij}(0)
    \Big]
  =
  -iM_d
  \Bigg\{
   \frac{
     P_d^i(\bd \times \hat{z})^j(J(\bd^2)-S(\bd^2))
     +
     P_d^j(\bd \times \hat{z})^i(J(\bd^2)+S(\bd^2))
    }{2M_d^2}
  \Bigg\}
\,.
\end{align}
Lastly, taking a tensor-polarized ensemble leaves only
the form factors with a subscript $T$ and $\bar{s}$:
\begin{multline}
  \label{eqn:mff:T}
  \mathrm{Tr}\Big[
    \rho_T(p_d,p'_d)
    \hat{T}^{ij}(0)
    \Big]
  =
  M_d
  \Bigg\{
    -
    \frac{P_d^i P_d^j}{M_d^2}
    \frac{\bd^2}{M_d^2}
    \left( \frac{3}{2} \cos^2\theta_\dl - \frac{1}{2} \right)
    A_T(\bd^2)
    -
    \frac{\dl^i \dl^j - \delta^{ij} \bd^2}{4M_d^2}
    \frac{\bd^2}{M_d^2}
    \left( \frac{3}{2} \cos^2\theta_\dl - \frac{1}{2} \right)
    D_{T1}(\bd^2)
    \\
    -
    \frac{3 \bd^2}{2M_d^2}
    \Big(
      \delta^{ij}
      \cos^2\theta_\dl
      +
      \hat{z}^i \hat{z}^j
      -
      \hat{\dl}^{\{i} \hat{z}^{j\}}
      \cos\theta_\dl
      -
      2 \delta^{ij}
      +
      2 \hat{\dl}^i \hat{\dl}^j
      \Big)
    D_{T2}(\bd^2)
    \\
    +
    \delta^{ij}
    \frac{\bd^2}{M_d^2}
    \left( \frac{3}{2} \cos^2\theta_\dl - \frac{1}{2} \right)
    \bar{c}_{T1}(\bd^2)
    +
    \left( \delta^{ij} - 3 \hat{z}^i \hat{z}^j \right)
    \bar{c}_{T2}(\bd^2)
    +
    \left(
    \frac{
      3\bd^2\hat{\dl}^{[i}\hat{z}^{j]}\cos(\theta_\Delta)
    }{
      2M_d^2
    }\right)
    \bar{s}(\bd^2) \Bigg\}
  \,.
\end{multline}
This provides convenience to both the calculation of the form factors
(allowing the use of spin density matrices to project out specific EMT-FFs)
and to their interpretation.
The $T1$ structures appearing here are identical to the
corresponding unpolarized structures appearing in Eq.~(\ref{eqn:mff:U}),
but weighted by a quadrupole structure in terms of the momentum transfer
$\bd$, namely
$\frac{3}{2} \cos^2\theta_\dl - \frac{1}{2}=P_2(\cos\theta_\Delta)$,
where $P_2$ is the second-order Legendre polynomial.
An additional factor $-\frac{\bd^2}{M_d^2}$ normalizes
$D_{T1}(0)$ and $\bar{c}_{T1}(0)$
in an analogous manner to $A_T(0)$---with the latter
being normalized to the deuteron's quadrupole moment.
The $T2$ form factors are effectively leftover structures,
and are defined to coincide
with the $D_2(\bd^2)$ and $\bar{f}(\bd^2)$
form factors of prior works~\cite{Polyakov:2019lbq,He:2023ogg}.
Details on how to project out the $T1$ and $T2$ form factors
are given in Sec.~\ref{sec:mff:results}.

Nine of the EMT-FFs in Eq.~(\ref{eqn:mff}) multiply tensors that are symmetric
under $i\leftrightarrow j$, while two---$S(\bd^2)$ and $\bar{s}(\bd^2)$---multiply
antisymmetric tensors.
The symmetric EMT can accordingly be obtained simply by setting
$S(\bd^2) = \bar{s}(\bd^2) = 0$.
Most authors restrict their attention to the symmetric EMT
(and, accordingly, the symmetric stress tensor),
but the asymmetric EMT is a more general object.
Moreover, the asymmetric EMT contains information about the distribution
of angular momentum carried by intrinsic fermion spin via $S(\bd^2)$---information
that is absent in the symmetric EMT~\cite{Leader:2013jra,Lorce:2025pxt}---and
thus provides a richer description of hadron structure.
It is also worth noting that a gauge-invariant, asymmetric EMT
naturally falls out of local translation invariance of a theory
through Noether's second theorem~\cite{Freese:2025glz}.

Several of the form factors obey sum rules,
which are called as such because they hold when
the contributions of all constituents have been summed.
These are derived in Ref.~\cite{Cosyn:2019aio}.
Translating the sum rules into our notation,
and introducing a superscript $(c)$ to signify the contribution
of the $c$th constituent,
there is a momentum sum rule:
\begin{align}
  \label{eqn:sum:mom}
  \sum_c
  A_{U}^{(c)}(0)
  =
  1
  \,,
\end{align}
a spin sum rule:
\begin{align}
  \label{eqn:sum:spin}
  \sum_c
  J^{(c)}(0)
  =
  1
  \,,
\end{align}
which is a spin-one version of the Ji sum rule~\cite{Ji:1996ek};
and several rules that follow from local momentum conservation:
\begin{align}
  \label{eqn:sum:cbar}
  \sum_c
  \bar{c}_U^{(c)}(\bd^2)
  =
  \sum_c
  \bar{c}_{T1}^{(c)}(\bd^2)
  =
  \sum_c
  \bar{c}_{T2}^{(c)}(\bd^2)
  =
  \sum_c
  \bar{s}^{(c)}(\bd^2)
  =
  0
  \,.
\end{align}
It should be stressed that the sum over constituents includes a sum over
force carriers.
For instance, in an atom, the constituents include not just the nucleus
and electrons, but also the electrostatic field.
Similarly, for the deuteron, the constituents must include not just the
proton and neutron (and the quarks and gluons within them), but also the
carrier of the inter-nucleon force---be it exchanges of pions, other mesons,
or even quarks and gluons.
If the stresses in force carriers are not accounted for,
interactions appear to occur by non-local action at a distance---which
can lead to violations of \emph{local} momentum conservation,
even if momentum is globally conserved.
For example, a recent study of the hydrogen atom's EMT-FFs~\cite{Freese:2024rkr}
found the sum rule for $\bar{c}(\bd^2)$ to be violated unless
the contribution of the electrostatic field is accounted for.
Since the present work looks only at one-body contributions to the deuteron EMT-FFs,
we should not be surprised if the sum rule~(\ref{eqn:sum:cbar}) is violated;
its restoration will only occur when exchange contributions are also included.
On the other hand, in the non-relativistic regime---where
interactions are described using static potentials
rather than dynamical field degrees of freedom%
---force carriers do not carry momentum or angular momentum,
so the one body contributions should already saturate the
momentum (\ref{eqn:sum:mom}) and spin (\ref{eqn:sum:spin}) sum rules.


\subsection{One-body contributions to the stress tensor}

The stress tensor should in general break down into one-body
and two-body contributions, as follows:
\begin{align}
  \hat{T}^{ij}
  =
  \hat{T}^{ij}_p
  +
  \hat{T}^{ij}_n
  +
  \hat{T}^{ij}_{\mathrm{int}}
  \,,
\end{align}
i.e., into pieces that act only on the proton and neutron,
and a piece corresponding to interactions.
All information about stresses and momentum fluxes carried by the
inter-nucleon force itself is contained in
$\hat{T}^{ij}_{\mathrm{int}}$,
which can reasonably be called a two-body current.
In this work, we concern ourselves with only the one-body contributions
$\hat{T}^{ij}_p$ and $\hat{T}^{ij}_n$.
Note that additionally the separation between one- and multi-body currents is in principle
scheme-dependent~\cite{Bogner:2006pc,More:2015tpa}.
Here, we adopt a high-resolution perspective, using phenomenological deuteron potentials.

Our goal now is to find expressions for the matrix elements
$\langle \bm{p}'_d, s_d' | \hat{T}^{ij}_p | \bm{p}_d, s_d \rangle$
of the proton EMT between deuteron kets
in terms of matrix elements
$\langle \bm{p}_p', s_p' | \hat{T}^{ij}_p | \bm{p}_p, s_p \rangle$
between proton kets.
This will allow the proton contribution to the deuteron EMT-FFs
to be written in terms of proton EMT-FFs,
through the non-relativistic breakdown~\cite{Freese:2025tqd}:
\begin{multline}
  \label{eqn:mff:half}
  \langle \bm{p}'_N, s'_N | \hat{T}_N^{ij}(0) | \bm{p}_N, s_N \rangle
  =
  \frac{ P_N^i P_N^j }{m_N}
  A_N(\bd^2)
  \delta_{s's}
  +
  \frac{\dl^i\dl^j - \delta^{ij}\bd^2}{4m_N}
  D_N(\bd^2)
  \delta_{s's}
  -
  m_N
  \delta^{ij}
  \bar{c}_N(\bd^2)
  \delta_{s's}
  \\
  -
  \frac{
    i
    (\bd\times\bm{\sigma}_{s's})^{\{i}
    P_N^{j\}}
  }{2m_N}
  J_N(\bd^2)
  -
  \frac{
    i
    (\bd\times\bm{\sigma}_{s's})^{[i}
    P_N^{j]}
  }{2m_N}
  S_N(\bd^2)
  \,.
\end{multline}
It's worth remarking that
our nucleon matrix element differs from Eq.~(11) of He and Zahed~\cite{He:2023ogg};
while their formula is obtained from a non-relativistic reduction of
the relativistic EMT-FF breakdown,
our formula follows from treating the non-relativistic theory
as a self-contained theory constrained by Galilei symmetry.
We additionally include the antisymmetric form factor $S_N(\bd^2)$.

To start with, the deuteron is assumed to be spanned by two-nucleon states,
so effectively\footnote{
  Note that we are using non-relativistic normalization for momentum kets, so
  $\langle \bm{p}', s' | \bm{p}, s \rangle = (2\pi)^3 \delta^{(3)}(\bm{p}-\bm{p}') \delta_{ss'}$.
}:
\begin{align}
  \label{eqn:complete}
  \sum_{s_p, s_n}
  \int \frac{\d^3 p_p}{(2\pi)^3}
  \int \frac{\d^3 p_n}{(2\pi)^3}
  | \bm{p}_p, s_p; \bm{p}_n, s_n \rangle
  \langle \bm{p}_p, s_p; \bm{p}_n, s_n |
  =
  \mathbbm{1}
  \,.
\end{align}
This means that the desired matrix element can be written:
\begin{multline}
  \label{eqn:big}
  \langle \bm{p}', s_d' | \hat{T}^{ij}_p | \bm{p}, s_d \rangle
  =
  \sum_{\substack{s_p, s_n \\ s'_p, s'_n}}
  \int \frac{\d^3 p_p}{(2\pi)^3}
  \int \frac{\d^3 p_n}{(2\pi)^3}
  \int \frac{\d^3 p'_p}{(2\pi)^3}
  \int \frac{\d^3 p'_n}{(2\pi)^3}
  \bigg\{
    \\
    \langle \bm{p}', s'_d | \bm{p}'_p, s'_p; \bm{p}'_n, s'_n \rangle
    \langle \bm{p}'_p, s'_p; \bm{p}'_n, s'_n |
    \hat{T}^{ij}_p
    | \bm{p}_p, s_p; \bm{p}_n, s_n \rangle
    \langle \bm{p}_p, s_p; \bm{p}_n, s_n | \bm{p}, s_d \rangle
    \bigg\}
  \,.
\end{multline}
Since the proton part of the EMT acts only on the proton,
we can write:
\begin{align}
  \langle \bm{p}'_p, s'_p; \bm{p}'_n, s'_n |
  T^{ij}_p
  | \bm{p}_p, s_p; \bm{p}_n, s_n \rangle
  =
  (2\pi)^3 \delta^{(3)}(\bm{p}_n - \bm{p}'_n)
  \delta_{s_n s'_n}
  \langle \bm{p}'_p, s'_p |
  T^{ij}_p
  | \bm{p}_p, s_p \rangle
  \,.
\end{align}
Additionally, the inner products of deuteron and two-nucleon states
can be written in terms of the momentum-space deuteron wave function:
\begin{align}
  \label{eqn:ket2wf}
  \langle \bm{p}_p, s_p; \bm{p}_n, s_n | \bm{p}, s_d \rangle
  =
  (2\pi)^3 \delta^{(3)}(\bm{p} - \bm{p}_p - \bm{p}_n) \;
  \tilde{\psi}_d^{(s_d;s_p,s_n)}\left(\frac{\bm{p}_p-\bm{p}_n}{2}\right)
  \,.
\end{align}
See Appendix~\ref{sec:wf} for a review of the deuteron wave function
and its properties.

The presence of three matrix elements or inner products
introducing delta functions allows three of the momentum integrals
to be eliminated from Eq.~(\ref{eqn:big}).
By defining:
\begin{align}
  \bm{k}
  \equiv
  \frac{1}{4}
  \big(
  \bm{p}_p + \bm{p}'_p - \bm{p}_n - \bm{p}'_n
  \big)
  \qquad
  \qquad
  \bd
  \equiv
  \bm{p}'_d - \bm{p}_d
  =
  \bm{p}_p' - \bm{p}_p
  \,,
\end{align}
the desired matrix element can be rewritten:
\begin{align}
  \label{eqn:1body:momentum}
  \langle \bm{p}'_d, s_d' | \hat{T}^{ij}_p | \bm{p}_d, s_d \rangle
  =
  \sum_{s_p, s_n, s'_p}
  \int \frac{\d^3 k}{(2\pi)^3}
  \tilde{\psi}_d^{*(s'_d;s'_p,s_n)}\left(\bm{k} + \tfrac{\bd}{4}\right)
  \tilde{\psi}_d^{(s_d;s_p,s_n)}\left(\bm{k} - \tfrac{\bd}{4}\right)
  \langle \bm{p}'_p, s'_p | \hat{T}^{ij}_p | \bm{p}_p, s_p \rangle
  \bigg|_{
    \bm{P}_p
    =
    \tfrac{1}{2}
    \bm{P}_d
    + \,
    \bm{k}
  }
  \,,
\end{align}
where $\bm P_d (\bm P_p)$ is the average deuteron (proton) momentum.

The integral in Eq.~(\ref{eqn:1body:momentum}) can be evaluated more easily
when converted to coordinate space
through a Fourier transform of the wave function:
\begin{align}
  \tilde{\psi}^{(s_d;s_p,s_n)}_d(\bm{k})
  =
  \int \d^3 r \,
  \psi^{(s_d;s_p,s_n)}_d(\bm{r})
  \e^{-i\bm{k}\cdot\bm{r}}
  \,.
\end{align}
Any appearances of $\bm{k}$ in Eq.~(\ref{eqn:1body:momentum}) are turned by
this substitution into two-sided derivatives, giving:
\begin{align}
  \label{eqn:1body:coordinate}
  \langle \bm{p}'_d, s_d' | \hat{T}^{ij}_p | \bm{p}_d, s_d \rangle
  =
  \sum_{s_p, s_n, s'_p}
  \int \d^3 r \,
  \e^{i\frac{\bd\cdot\bm{r}}{2}}
  \psi_d^{*(s'_d;s'_p,s_n)}(\bm{r})
  \langle \bm{p}'_p, s'_p | \hat{T}^{ij}_p | \bm{p}_p, s_p \rangle
  \psi_d^{(s_d;s_p,s_n)}(\bm{r})
  \bigg|_{
    \bm{P}_p
    =
    \tfrac{1}{2} (
    \bm{P} - i\blrn
    )
  }
  \,,
\end{align}
where $f\blrn g =f(\bm{\nabla} g)-(\bm{\nabla} f)g$.


\subsection{Analytic results for the EMT form factors}
\label{sec:mff:results}

In this section, we give analytic formulas for the one-body contributions
to the EMT form factors.
Results are written in terms of the radial S- and D-wave functions;
see Appendix~\ref{sec:wf} for a review of their definitions and properties.
The form factors are ultimately obtained through Eq.~(\ref{eqn:1body:coordinate})---with
the left-hand side equated with the deuteron EMT-FF breakdown (\ref{eqn:mff})
and the nucleon breakdown (\ref{eqn:mff:half}) inserted into
the right-hand side---which does not leave
any wiggle room for ad hoc decisions that could introduce ambiguities.
In obtaining these formulas, we use $M_d = 2m_N$,
since mass is additive (and distinct from energy) in non-relativistic mechanics.
Specific deuteron EMT-FFs
can be isolated through a variety of projection procedures,
such as considering particular polarization states,
looking at specific $T^{ij}$ components,
contracting with $\dl_i$ to isolate non-conserved form factors,
or integrating out projections onto specific spherical harmonics.
We will not include line-by-line derivations here,
but will briefly describe how we isolated each form factor
along with presenting the results.

The $A$-like form factors $A_U(\bd^2)$ and $A_T(\bd^2)$ can be isolated by
looking at unpolarized and tensor-polarized ensembles
respectively---via Eqs.~(\ref{eqn:mff:U}) and (\ref{eqn:mff:T})---and
reading off terms that are proportional to $P^i P^j$.
The results we find are
\begin{align}
  \label{eqn:AU}
  A_U(\bd^2)
  =
  \sum_{N=p,n}
  \frac{ A_N(\bd^2) }{2}
  \int_0^\infty \d r \,
  j_0\left(\frac{\dl r}{2}\right)
  \Big[ u^2(r) + w^2(r) \Big]
\end{align}
and
\begin{align}
  \label{eqn:AT}
  A_T(\bd^2)
  =
  \sum_{N=p,n}
  \frac{
    6 m_N^2
    A_N(\bd^2)
  }{
    \dl^2
  }
  \int_0^\infty \d r \,
  j_2\left(\frac{\dl r}{2}\right)
  \Big[
    2\sqrt{2} u(r) w(r) - w^2(r)
    \Big]
  \,,
\end{align}
The forward limits of these form factors are:
\begin{align}
  \label{eqn:forward:A}
  \begin{split}
    A_U(0)
    &=
    \sum_{N=p,n}
    \frac{A_N(0)}{2}
    \int_0^\infty \d r \,
    \Big[ u^2(r) + w^2(r) \Big]
    =
    1
    \\
    A_T(0)
    &=
    \sum_{N=p,n} \frac{
      m_N^2
      A_N(0)
    }{
      10
    }
    \int_0^\infty \d r \,
    r^2
    \Big[
      2\sqrt{2} u(r) w(r) - w^2(r)
      \Big]
    =
    (2m_N)^2
    Q_d
    \,,
  \end{split}
\end{align}
by virtue of the nucleon sum rule $A_N(0)=1$---recall
that we are implicitly summing over all parton flavors---the
deuteron wave function normalization (\ref{eqn:wf:norm}),
and the definition of the quadrupole moment (\ref{eqn:wf:quad}).
The first of these relationships shows that the momentum sum rule (\ref{eqn:sum:mom})
is already saturated by the one-body contributions.
We want to remind the reader that this is a feature of the non-relativistic approach taken here,
with the force carrier degrees of freedom integrated out into the static potential.

The form factors $J(\bd^2)$ and $S(\bd^2)$ are related to
the angular momentum content of the deuteron,
and can be found by projecting onto vector-polarized deuteron ensembles
via Eq.~(\ref{eqn:mff:V}).
One can then find e.g.\ the sums and differences
between $T^{13}$ and $T^{31}$ components of the matrix elements
to separate these form factors.
This way, we obtain the following results:
\begin{multline}
  \label{eqn:J}
  J(\bd^2)
  =
  \sum_{N=p,n}
  \int_0^\infty \d r \,
  \Bigg\{
    \frac{
      9
      A_N(\bd^2)
    }{
      2 \dl
    }
    j_1\left(\frac{\dl r}{2}\right)
    \frac{ w^2(r) }{r}
    \\
    +
    J_N(\bd^2)
    j_0\left(\frac{\dl r}{2}\right)
    \left[
      u^2(r) - \frac{1}{2} w^2(r)
      \right]
    +
    \frac{ J_N(\bd^2) }{2}
    j_2\left(\frac{\dl r}{2}\right)
    \Big[
      w^2(r) + \sqrt{2} u(r) w(r)
      \Big]
    \Bigg\}
\end{multline}
and
\begin{align}
  \label{eqn:S}
  S(\bm \Delta^2)
  =
  \sum_{N=p,n}
  S_N(\bd^2)
  \int_0^\infty \d r \,
  \Bigg\{
    j_0\left(\frac{\dl r}{2}\right)
    \left[
      u^2(r) - \frac{1}{2} w^2(r)
      \right]
    +
    \frac{1}{2}
    j_2\left(\frac{\dl r}{2}\right)
    \Big[
      w^2(r) + \sqrt{2} u(r) w(r)
      \Big]
    \Bigg\}
  \,.
\end{align}
In the forward limit,
these form factors become:
\begin{align}
  \label{eqn:JS:0}
  \begin{split}
    J(0)
    &=
    \sum_{N=p,n}
    \left\{
      \tfrac{3}{4}
      \mathcal{P}_D
      A_N(0)
      +
      \Big( 1 - \tfrac{3}{2} \mathcal{P}_D \Big)
      J_N(0)
      \right\}
    =
    1
    \\
    S(0)
    &=
    \sum_{N=p,n}
    \Big( 1 - \tfrac{3}{2} \mathcal{P}_D \Big)
    S_N(0)
    \,,
  \end{split}
\end{align}
where $S_N(0)$ signifies the quark spin contribution to the nucleon.
Both equations feature the well-known deuteron depolarization factor
$(1-\tfrac{3}{2}\mathcal{P}_D)$,
which is of the order of 90-95\% for most realistic deuteron wave functions.
The first of these shows that the spin sum rule (\ref{eqn:sum:spin})
is already saturated by the one-body contributions.
To achieve this saturation, the depolarization of the nucleon spin is
compensated by an orbital angular momentum contribution from the deuteron D-wave.
The second relation gives the quark spin contribution to the deuteron's total spin,
since gluons do not contribute to the antisymmetric part of the EMT (and thus do not contribute to $S(\bd^2)$).

The form factor $D_U(\bd^2)$ can be found by taking an unpolarized ensemble
via Eq.~(\ref{eqn:mff:U}),
dropping any terms proportional to $\bm{P}$ from the result
to eliminate $A_U(\bd^2)$,
and taking an off-diagonal component---e.g., the $T^{12}$ component---to
eliminate $\bar{c}_U(\bd^2)$.
$D_U(\bd^2)$ will be the only form factor that survives.
We find:
\begin{multline}
  \label{eqn:DU}
  D_U(\bd^2)
  =
  \sum_{N=p,n}
  \int_0^\infty \d r \,
  \Bigg\{
    \frac{
      4
      A_N(\bd^2)
    }{
      \dl^2
    }
    j_2\left(\frac{\dl r}{2}\right)
    \Bigg[
      \frac{
        2 u^2(r)
        +
        8 w^2(r)
      }{r^2}
      -
      \frac{
        u(r) u'(r) + w(r) w'(r)
      }{r}
      +
      u(r) u''(r)
      \\
      +
      w(r) w''(r)
      -
      \big(u'(r)\big)^2
      -
      \big(w'(r)\big)^2
      \Bigg]
    -
    \frac{ 12 J_N(\bd^2) }{\dl}
    j_1\left(\frac{\dl r}{2}\right)
    \frac{ w^2(r) }{r}
    +
    2 D_N(\bd^2)
    j_0\left(\frac{\dl r}{2}\right)
    \Big[ u^2(r) + w^2(r) \Big]
    \Bigg\}
  \,.
\end{multline}
The forward limit can be found using some integration tricks:
\begin{align}
  \label{eqn:DU:0}
  D_U(0)
  =
  \sum_{N=p,n}
  \left\{
    2
    D_N(0)
    -
    2 \mathcal{P}_D
    J_N(0)
    +
    \left(
    \frac{1}{10}
    \mathcal{P}_S
    +
    \frac{1}{2}
    \mathcal{P}_D
    +
    \frac{2}{15}
    \int_0^\infty \d r \, r^2 \big[ u(r) u''(r) + w(r) w''(r) \big]
    \right)
    A_N(0)
    \right\}
  \,.
\end{align}
We see that the deuteron $D$-term is determined both by that of either nucleon
and contributions originating from the deuteron dynamics multiplied
with nucleon spin and momentum properties.
Note that in the last term the second derivatives of the radial S- and D-waves are
related to the nucleon-nucleon potential through Schr\"odinger's equation;
see Eq.~(\ref{eqn:uw:diff}).

The form factor $\bar{c}_U(\bd^2)$ can also be found by looking at
unpolarized ensembles via Eq.~(\ref{eqn:mff:U})
and contracting it with $\dl_i $
to eliminate contributions from $A_U(\bd^2)$ and
$D_U(\bd^2)$.
We thus obtain the formula:
\begin{multline}
  \label{eqn:cbarU}
  \bar{c}_U(\bd^2)
  =
  \sum_{N=p,n}
  \int_0^\infty \d r \,
  \Bigg\{
    \frac{A_N(\bd^2) }{2 m_N^2 \dl}
    j_1\left(\frac{\dl r}{2}\right)
    \Bigg[
      u'(r) u''(r)
      +
      w'(r) w''(r)
      \\
      -
      u(r)u'''(r)
      -
      w(r)w'''(r)
      -
      \frac{12 w^2(r)}{r^3}
      \Bigg]
    +
    \frac{ \bar{c}_N(\bd^2) }{2}
    j_0\left(\frac{\dl r}{2}\right)
    \Big[
      u^2(r) + w^2(r)
      \Big]
    \Bigg\}
  \,.
\end{multline}
The result does not identically vanish as required by local momentum conservation,
but this occurs because we have not accounted for stresses in the interaction itself.
It is not a flaw, but \emph{expected} that $\bar{c}_U(\bd^2)$ be non-zero
for partial contributions to the stress tensor.
In fact, it encodes forces acting on the subsystems---in this case,
the nucleons---through the Cauchy momentum equation.
The relationship to forces is more apparent if we
use the Schr\"odinger equation~(\ref{eqn:uw:diff})
to rewrite $\bar{c}_U(\bd^2)$ as:
\begin{multline}
  \label{eqn:cbar:potential}
  \bar{c}_U(\bd^2)
  =
  \sum_{N=p,n}
  \int_0^\infty \d r \,
  \Bigg\{
    -
    \frac{A_N(\bd^2) }{2 m_N \dl}
    j_1\left(\frac{\dl r}{2}\right)
    \Bigg[
      V_c'(r)
      u^2(r)
      +
      V_w'(r)
      w^2(r)
      +
      4 \sqrt{2}
      V_t'(r)
      u(r) w(r)
      \Bigg]
    \\
    +
    \frac{ \bar{c}_N(\bd^2) }{2}
    j_0\left(\frac{\dl r}{2}\right)
    \Big[
      u^2(r) + w^2(r)
      \Big]
    \Bigg\}
  \,,
\end{multline}
where
next to a contribution from the nucleon $\bar{c}_N$
weighted with deuteron structure
(similar in form to the unpolarized $A_U$ of Eq.~(\ref{eqn:AU})),
there is also a contribution proportional to $A_N$
where the integrand depends on derivatives of the potential.
We will make the relationship of this form factor to forces more concrete
and explore it further in Sec.~\ref{sec:force}.
The forward limit of $\bar{c}_U(\bd^2)$---which
will later be useful in calculating mechanical radii---can be written:
\begin{align}
  \label{eqn:cbarU:0}
  \bar{c}_U(0)
  =
  \sum_{N=p,n}
  \left\{
    \frac{1}{6 m_N^2}
    \int_0^\infty \d r \,
    \left( u(r) u''(r) + w(r) w''(r) - \frac{6 w^2(r)}{r^2} \right)
    A_N(0)
    +
    \frac{1}{2}
    \bar{c}_N(0)
    \right\}
\end{align}

There are two tensor-polarized D-like form factors,
and isolating them is a bit more involved.
These are both found by considering tensor-polarized ensembles via Eq.~(\ref{eqn:mff:T}).
The form factor $D_{T1}(\bd^2)$ can be isolated by evaluating the $T^{12}$ component of
Eq.~(\ref{eqn:mff:T}), dividing it by $\sin^2(\theta_\dl)\cos(\phi_\dl)\sin(\phi_\dl)$,
and integrating it with the spherical harmonic $Y_2^0(\Hat{\dl})$.
All structures besides the $D_{T1}(\bd^2)$ structure are eliminated by this projection.
The result is:
\begin{multline}
  \label{eqn:DT1}
  D_{T1}(\bd^2)
  =
  \sum_{N=p,n}
  \int_0^\infty \d r \,
  \Bigg\{
    \frac{ 48 m_N^2 A_N(\bd^2) }{\dl^4}
    j_4\left(\frac{\dl r}{2}\right)
    \Bigg[
      \sqrt{2}
      \Big( u(r) w''(r) + w(r) u''(r) - 2 u'(r) w'(r) \Big)
      \\
      -
      w(r) w''(r)
      +
      \big(w'(r)\big)^2
      +
      \frac{
        \sqrt{2}
        \Big(
        3 w(r) u'(r) - 5 u(r) w'(r)
        \Big)
        +
        w(r) w'(r)
      }{r}
      +
      \frac{
        6
        \Big( 2\sqrt{2} u(r) w(r) - w^2(r) \Big)
      }{r^2}
      \Bigg]
    \\
    +
    \frac{ 96 m_N^2 J_N(\bd^2)}{\dl^3}
    j_3\left(\frac{\dl r}{2}\right)
    \Bigg[
      \sqrt{2} \Big( u(r) w'(r) - w(r) u'(r) \Big)
      -
      \frac{ 2 \sqrt{2} u(r) w(r) - w^2(r) }{r}
      \Bigg]
    \\
    +
    \frac{
      24 m_N^2
      D_N(\bd^2)
    }{ \dl^2 }
    j_2\left(\frac{\dl r}{2}\right)
    \Big[
      2\sqrt{2} u(r) w(r) - w^2(r)
      \Big]
    \Bigg\}
  \,.
\end{multline}
This form factor is finite in the forward limit, and equal to:
\begin{multline}
  D_{T1}(0)
  =
  (2m_N)^2
  \sum_{N=p,n}
  \Bigg\{
    2 D_N(0) \,
    Q_d
    +
    \left(
    -
    \frac{4}{7} Q_d
    +
    \frac{\sqrt{2}}{35}
    \int_0^\infty \d r \,
    r^3
    \Big( u(r) w'(r) - w(r) u'(r) \Big)
    \right)
    J_N(0)
    \\
    +
    \frac{1}{42}
    \left(
    Q_d
    +
    \frac{1}{30}
    \int_0^\infty \d r \,
    r^4 \Big( 3 \sqrt{2} u(r) w''(r) + \sqrt{2} w(r) u''(r) - 2 w(r) w''(r) \Big)
    \right)
    A_N(0)
    \Bigg\}
  \,.
\end{multline}
The forward limit of $D_{T1}(0)$ is dominated by
the first term in this expression,
and accordingly is proportional to the deuteron's quadrupole moment.
Obtaining $D_{T1}(0) / D_U(0) \approx (2m_N)^2Q_d = A_T(0)/A_U(0)$
was the primary motivation for normalizing the form factor $D_{T1}(\bd^2)$
as we did.

The form factor $D_{T2}(\bd^2)$ can be isolated in a similar way.
We can take the $T^{12}$ component of Eq.~(\ref{eqn:mff:T}),
divide it by $\sin^2(\theta_\dl)\cos(\phi_\dl)\sin(\phi_\dl)$,
and integrate with the spherical harmonic $Y_0^0(\hat{\dl})$.
The result is:
\begin{multline}
  \label{eqn:DT2}
  D_{T2}(\bd^2)
  =
  \sum_{N=p,n}
  \int_0^\infty \d r \,
  \Bigg\{
    \frac{ 12 A_N(\bd^2) }{\dl^3 }
    j_3\left(\frac{\dl r}{2}\right)
    \Bigg[
      \frac{
        \Big( 2\sqrt{2} u''(r) - w''(r)\Big) w(r)
        -
        \Big( 2\sqrt{2} u'(r) - w'(r) \Big) w'(r)
      }{r}
      \\
      +
      \frac{
        \Big( 2\sqrt{2} u(r) + 5 w(r) \Big)
        w'(r)
      }{r^2}
      -
      \frac{18 w^2(r)}{r^3}
      \Bigg]
    +
    \frac{ 6 J_N(\bd^2) }{\dl^2}
    j_2\left(\frac{\dl r}{2}\right)
    \Bigg[
      \sqrt{2} \Big( u(r) w''(r) - w(r) u''(r) \Big)
      \\
      -
      \frac{ 4 \Big( \sqrt{2} u'(r) + w'(r) \Big) w(r) }{r}
      -
      \frac{\Big( 2\sqrt{2} u(r) w(r) - w^2(r) \Big)}{r^2}
      +
      \frac{3 w^2(r)}{r^2}
      \Bigg]
    \Bigg\}
  \,.
\end{multline}
It is worth remarking that, in contrast to $D_U(\bd^2)$ and $D_{T1}(\bd^2)$,
the nucleon $D_N(\bd^2)$ does not contribute to $D_{T2}(\bd^2)$.
The forward limit of this form factor is also finite.

There are also several non-conserved form factors present
in tensor-polarized ensembles.
The two $\overline{c}_T$ form factors can be isolated by
contracting of Eq.~(\ref{eqn:mff:T}) with $\dl_i \hat{z}_j + \dl_j \hat{z}_i$.
Dividing the result by $\cos(\theta_\dl)$ and integrating the spherical harmonic $Y_2^0(\hat{\dl})$
will isolate $\overline{c}_{T1}(\bd^2)$, giving:
\begin{multline}
  \label{eqn:cT1}
  \bar{c}_{T1}(\bd^2)
  =
  \sum_{N=p,n}
  \int_0^\infty \d r \,
  \Bigg\{
    \frac{ 6 A_N(\bd^2) }{\dl^3}
    j_3\left(\frac{\dl r}{2}\right)
    \Bigg[
      \sqrt{2}
      \Big(
      u'(r) w''(r)
      +
      w'(r) u''(r)
      -
      u(r) w'''(r)
      -
      w(r) u'''(r)
      \Big)
      +
      w(r) w'''(r)
      \\
      -
      w'(r) w''(r)
      +
      \frac{
        2\sqrt{2}
        \Big( u(r) w''(r) - w(r) u''(r) \Big)
      }{r}
      +
      \frac{
        6 \sqrt{2}
        \Big( u(r) w'(r) - w(r) u'(r) \Big)
      }{r^2}
      -
      \frac{
        12\Big( 2\sqrt{2} u(r) w(r) - w^2(r) \Big)
      }{r^3}
      \Bigg]
    \\
    +
    \frac{6\sqrt{2} J_N(\bd^2) }{\dl^2}
    j_2\left(\frac{\dl r}{2}\right)
    \Bigg[
      w(r) u''(r)
      -
      u(r) w''(r)
      +
      \frac{6 u(r) w(r)}{r^2}
      \Bigg]
    +
    \frac{
      6 m_N^2
      \bar{c}_N(\bd^2)
    }{\dl^2}
    j_2\left(\frac{\dl r}{2}\right)
    \Big[
      2\sqrt{2} u(r) w(r) - w^2(r)
      \Big]
    \Bigg\}
  \,.
\end{multline}
Doing the same as above, but with the spherical harmonic $Y_0^0(\hat{\dl})$,
isolates $\bar{c}_{T2}(\bd^2)$, giving:
\begin{multline}
  \label{eqn:cT2}
  \bar{c}_{T2}(\bd^2)
  =
  \sum_{N=p,n}
  \int_0^\infty \d r \,
  \Bigg\{
    \frac{ 3 A_N(\bd^2) }{m_N^2 \dl^2}
    j_2\left(\frac{\dl r}{2}\right)
    \Bigg[
      \frac{
        2\sqrt{2}
        \Big(
        u'(r) w''(r)
        -
        w(r) u'''(r)
        \Big)
        +
        w(r) w'''(r)
        -
        w'(r) w''(r)
      }{r}
      \\
      +
      \frac{
        2\sqrt{2}
        \Big( w(r) u''(r) - u(r) w''(r) \Big)
      }{r^2}
      -
      \frac{
        12\sqrt{2} w(r) u'(r)
      }{r^3}
      +
      \frac{
        12\Big( \sqrt{2} u(r) w(r) + w^2(r) \Big)
      }{r^4}
      \Bigg]
      \\
      +
      \frac{3\sqrt{2} J_N(\bd^2)}{4 m_N^2}
      j_2\left(\frac{\dl r}{2}\right)
      \Bigg[
        w(r) u''(r)
        -
        u(r) w''(r)
        +
        \frac{6 u(r) w(r)}{r^2}
        \Bigg]
      \Bigg\}
  \,.
\end{multline}

The final form factor, $\bar{s}(\bd^2)$, can be isolated
by taking the difference between components $T^{23}-T^{32}$
of Eq.~(\ref{eqn:mff:T}).
We find:
\begin{align}
  \label{eqn:sbar}
  \bar{s}(\bd^2)
  =
  \frac{6\sqrt{2} S_N(\bd^2)}{\dl^2}
  \int_0^\infty \d r \,
  j_2\left(\frac{\dl r}{2} \right)
  \left\{
    w(r) u''(r)
    - u(r) w''(r)
    +
    \frac{6 u(r) w(r)}{r^2}
    \right\}
  \,.
\end{align}
This form factor is related to tensor forces and spin-orbit coupling.
With the aid of the Schr\"odinger equation~(\ref{eqn:uw:diff}),
our result can be rewritten:
\begin{align}
  \label{eqn:sbar:2}
  \bar{s}(\bd^2)
  =
  \frac{6\sqrt{2} m_N S_N(\bd^2)}{\dl^2}
  \int_0^\infty \d r \,
  j_2\left(\frac{\dl r}{2} \right)
  \Big(
  2 V_t(r) - 6 V_{l2}(r) + 3 V_{ls}(r) - 9 V_{ls2}(r)
  \Big)
  u(r) w(r)
  \,,
\end{align}
where the functions appearing here are defined in Eq.~(\ref{eqn:potential}).
In the absence of spin- or angular momentum-dependent forces,
$\bar{s}(\bd^2)$ would vanish.

As a non-conserved form factor, $\bar{s}(\bd^2)$ must vanish when
summed over all constituents.
In this respect, it is no different from the three $\bar{c}(\bd^2)$ form factors.
However, as it arises from the antisymmetric part of the EMT,
$\bar{s}(\bd^2)$ receives contributions only from quarks~\cite{Cosyn:2019aio}.
In order for $\bar{s}(\bd^2)$ to vanish when exchange currents are introduced,
the exchange currents must include quarks.
Put another way, a non-zero $\bar{s}(\bd^2)$ in the one-body contributions
to the stress tensor cannot be induced by pure gluon exchange,
but can be induced by pion exchange (since pions consist of quarks).


\subsection{Numerical results for the EMT form factors}

In this section, we present numerical results for the EMT form factors
of the deuteron,
along with comparisons to the prior works of
Freese and Cosyn (FC)~\cite{Freese:2022yur},
He and Zahed (HZ)~\cite{He:2023ogg,He:2024vzz},
and Panteleeva \textsl{et al.} (PEGG)~\cite{Panteleeva:2024abz}.

\begin{figure}
  \includegraphics[width=\textwidth]{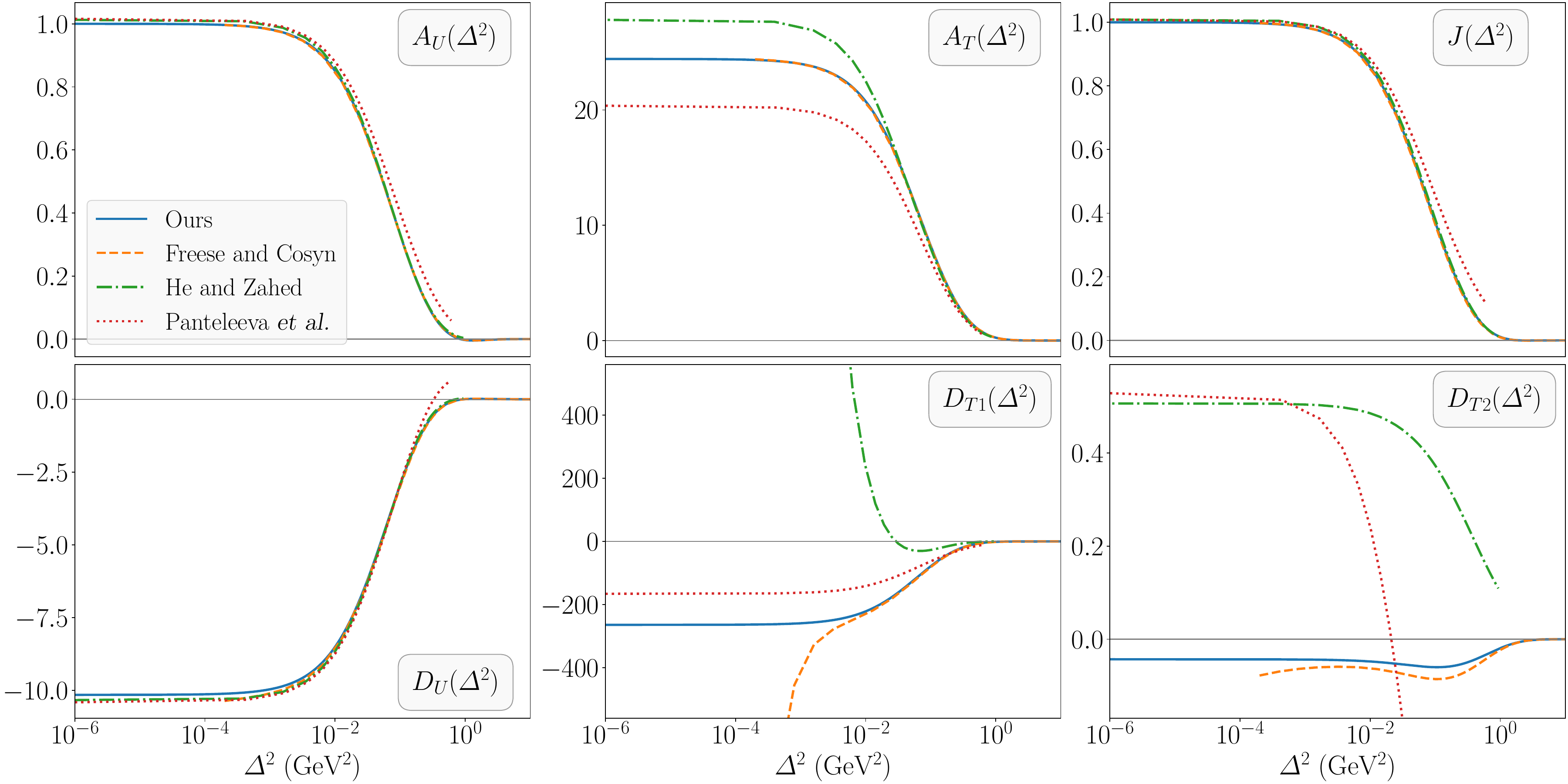}
  \caption{
    Numerical results for the six symmetric and conserved EMT-FFs of the deuteron.
    For our result,
    we used the AV18 deuteron wave function~\cite{Wiringa:1994wb}
    and the same nucleon EMT form factors as
    He and Zahed~\cite{He:2023ogg}.
    The results we compare to are from
    Refs.~\cite{Freese:2022yur,He:2024vzz,Panteleeva:2024abz},
    though the results of Ref.~\cite{Freese:2022yur}
    have been modified to use He and Zahed's nucleon form factors.
  }
  \label{fig:groups}
\end{figure}

We begin with a numerical calculation of the six conserved symmetric form
factors, since these were considered already in prior
works~\cite{Freese:2022yur,He:2023ogg,He:2024vzz,Panteleeva:2024abz}.
In order to make as direct a comparison as possible,
we utilize the same nucleon form factors as HZ,
which consist of the MIT lattice group's dipole fits of
$A_N(\bd^2)$ and $J_N(\bd^2)$~\cite{Hackett:2023rif}:
\begin{align}
  \begin{split}
    F_N(\bd^2)
    &=
    F_q(\bd^2)
    +
    F_g(\bd^2)
    \\
    F_f(\bd^2)
    &=
    \frac{\alpha_f}{(1 + \bd^2/\varLambda_f^2)^2}
    \,,
  \end{split}
\end{align}
where $F \in \{A,J\}$ and $f\in\{q,g\}$,
and where the particular values of $\alpha_f$ and $\varLambda_f$ are given in
Table III of the supplemental material of Ref.~\cite{Hackett:2023rif};
and a holographic QCD calculation of $D_N(\bd^2)$,
which uses an identical dipole form with
$D_q(0) = -1.30$,
$D_g(0) = -1.275$,
$\varLambda_q = 0.81$~GeV and
$\varLambda_g = 0.963$~GeV.
Notably, the calculation of PEGG was tuned to agree with HZ
through the choice of the low-energy constant $c_8$.
Additionally, we modified the code used in FC to use HZ's nucleon form factors.
Unfortunately, the code to generate the FC results is numerically unstable
at $\dl^2 \leq 10^{-4}$~GeV$^2$,
so we have truncated these results below this point.
This calculation and comparison are shown in Fig.~\ref{fig:groups}.

There is mixed agreement between our results and prior calculations.
All four calculations agree on $A_U(\bd^2)$
[top-left panel of Fig.~\ref{fig:groups}],
$J(\bd^2)$
[top-right panel of Fig.~\ref{fig:groups}]
and
$D_U(\bd^2)$
[bottom-left panel of Fig.~\ref{fig:groups}].
There appears to be some disagreement on $A_T(\bd^2)$
[top-middle panel of Fig.~\ref{fig:groups}],
but the disagreement can be entirely attributed
to the use of different wave functions,
which entail different quadrupole moments.
As we showed in Eq.~(\ref{eqn:forward:A}),
$A_T(0)$ is proportional to the deuteron quadrupole moment.
The AV18 wave function used here and by FC
gives a quadrupole moment of $0.2697$~fm$^2$,
while the Reid soft core potential used by HZ
gives a quadrupole moment of $0.3081$~fm$^2$.
These are respectively smaller and larger than the empirical quadrupole
moment, $0.2858$~fm$^2$~\cite{PhysRevA.81.042526}.
Similarly, the chiral EFT wave function of PEGG
gives an even smaller quadrupole moment: $0.2248$~fm$^2$.
The four $A_T(\bd^2)$ curves are nearly identical aside from scaling
by the quadrupole moment of each calculation.

On the other hand,
there is disagreement between the four calculations
for $D_{T1}(\bd^2)$ [bottom-middle panel of Fig.~\ref{fig:groups}]
and $D_{T2}(\bd^2)$ [bottom-right panel of Fig.~\ref{fig:groups}].
Our result and PEGG's result for $D_{T1}(\bd^2)$
agree up to the ratio of the deuteron quadrupole moments assumed in each work,
so these two calculations are consistent with each other.
Down to $\dl^2 \sim 10^{-2}$~GeV$^2$, our result also appears to agree with FC,
though the latter suddenly diverges at smaller $\dl^2$.
It is unclear to what extent this divergence is due to contamination by
non-conserved form factors
(which were not accounted for by FC, despite it being an impulse approximation calculation)
and to what extent it arises from numerical instability in the numerics.

The divergence in $D_{T1}(\bd^2)$ [bottom-middle panel of Fig.~\ref{fig:groups}]
for the HZ result
arises because they perform a substitution:
\begin{align*}
  P_N^\mu
  \rightarrow
  P_N^\mu - \frac{(\dl\cdot P_N)}{\dl^2} \dl^\mu
  \,.
\end{align*}
This substitution was used to impose local EMT conservation within their impulse approximation
and was also retained in the follow-up work that included exchange currents~\cite{He:2024vzz}.
The presence of a factor $\frac{1}{\dl^2}$ is responsible for the divergence
in $D_{T1}(\bd^2)$.

None of the models fully agree on $D_{T2}(\bd^2)$
[bottom-right panel of Fig.~\ref{fig:groups}].
In a sense, $D_{T2}(\bd^2)$ is the leftovers in the available tensor structures
from which the conserved part of the EMT can be built,
and accordingly may be especially sensitive to detailed differences
between the models---in contrast to the other form factors,
whose behavior is significantly constrained by known deuteron and nucleon properties.
Curiously, our result and the FC result,
which both use the AV18 deuteron wave function,
are fairly close, though they do not exactly coincide.
Also curiously, the HZ and PEGG results approach nearly the same forward limit,
which at the same time differs starkly from our and FC's forward limit.
Most remarkably, $D_{T2}(\bd^2)$ is extremely sensitive to the contribution
of exchange currents and higher-order corrections in chiral EFT;
see Refs.~\cite{He:2024vzz,Panteleeva:2024abz} for further details on this.

Our results for the deuteron EMT-FFs depend on two main inputs:
the deuteron wave function and the nucleon EMT-FF parameterizations.
The form factors $A_U(\bd^2)$ and $J(\bd^2)$ are
not very sensitive to the particular wave function or nucleon form factors,
while variance in $A_T(\bd^2)$ between wave functions is determined primarily
by the size of the deuteron's quadrupole moment.
The remaining form factors exhibit greater dependence on the choice of wave function
or nucleon EMT-FFs, so we consider these in more depth.

To illustrate wave function dependence,
we choose the AV18~\cite{Wiringa:1994wb}
and CD-Bonn~\cite{Machleidt:2000ge}
wave functions as representative examples of hard and soft wave functions.
We note that similarly hard wave functions (such as AV18 and Paris~\cite{Lacombe:1980dr})
produce nearly identical results, so these comparisons are not shown.
To illustrate dependence on the nucleon EMT-FFs,
we compare results using dipole forms for the nucleon form factors to pointlike nucleons.
For the former,
we use the same $A_N(\bd^2)$, $J_N(\bd^2)$ and $D_N(\bd^2)$ as above,
along with a dipole form for $S_N(\bd^2)$
with $S_N(0) = 0.204$ (from the JAM22 analysis~\cite{Cocuzza:2022jye})
and $\varLambda = 1.23$~GeV (from the PDG value of the $a_1$ mass~\cite{ParticleDataGroup:2024cfk}).
We also set $\bar{c}_N(\bd^2) = 0$ since we are implicitly summing over all parton flavors.
For a pointlike nucleon, we use $A_N(\bd^2) = 1$, $J_N(\bd^2) = S_N(\bd^2)= \frac{1}{2}$,
and $D_N(\bd^2) = \bar{c}_N(\bd^2) = 0$~\cite{Hudson:2017oul}.

\begin{figure}[htb]
  \centering
  \includegraphics[width=\linewidth]{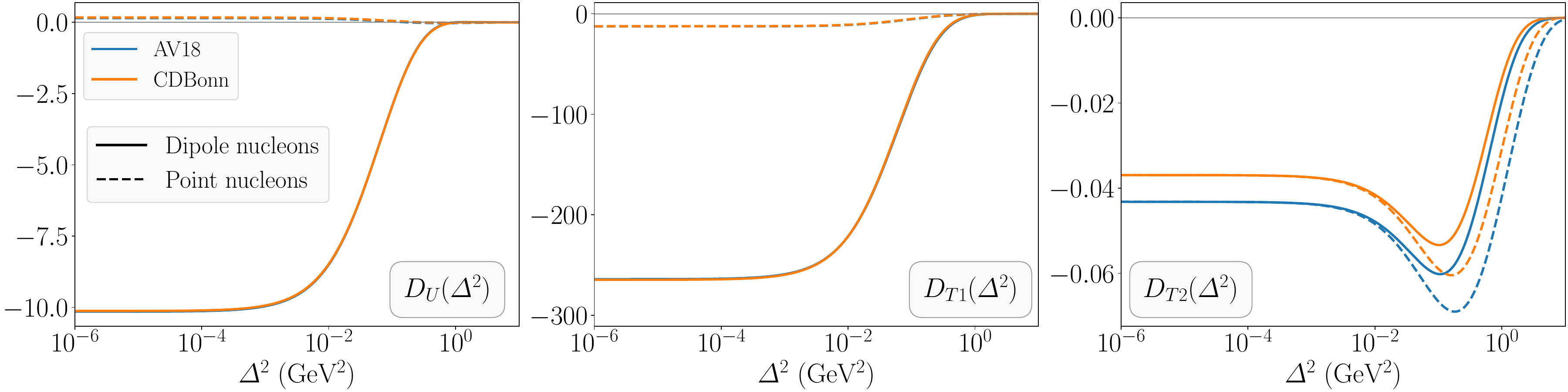}
  \caption{
    Comparisons between the deuteron $D$-like form factors
    for different deuteron wave functions
    and different nucleon EMT-FFs.
    Blue curves use the AV18 wave function~\cite{Wiringa:1994wb}
    while orange curves use CD-Bonn~\cite{Machleidt:2000ge}.
    Solid curves use dipole forms for the nucleon EMT-FFs,
    while dashed curves assume pointlike nucleons.
  }
  \label{fig:D4}
\end{figure}

We first show the wave function and nucleon EMT-FF dependence of the
$D$-like form factors in Fig.~\ref{fig:D4}.
The form factors $D_U(\bd^2)$ and $D_{T1}(\bd^2)$ show little dependence
on the deuteron wave function,
but significant dependence on the nucleon form factors.
These two form factors are dominated by the nucleon form factor $D_N(\bd^2)$,
and accordingly are extremely sensitive to it.
For pointlike nucleons in particular, this dominating contribution is absent.
In fact, for pointlike nucleons, we find $D_U(\bd^2) > 0$,
in apparent contradiction to the popular conjecture that $D(0) < 0$
is necessary for mechanical stability.
To be sure, nucleons in nature have a finite spatial extent and the
nucleon-nucleon force contains contributions from finite size effects---the
latter of which are technically incorporated into the wave function,
even in the pointlike nucleon calculation of Fig.~\ref{fig:D4}.
It is thus unclear whether a deuteron made up of pointlike nucleons---with these
finite-size effects removed---would actually be mechanically stable.
Thus, while the pointlike nucleon result in Fig.~\ref{fig:D4} appears to be
a counterexample to the $D(0) < 0$ stability criterion,
it lacks the forcefulness of other counterexamples---with the
hydrogen atom ground state being the preeminent
counterexample~\cite{Ji:2022exr,Czarnecki:2023yqd,Freese:2024rkr}.

The last $D$-like form factor, $D_{T2}(\bd^2)$,
does not depend on $D_N(\bd^2)$---see Eq.~(\ref{eqn:DT2})---%
but does depend on details of the deuteron wave function.
We accordingly see in the right panel of Fig.~\ref{fig:D4} that
a soft wave function (here CD-Bonn)
produces a smaller $D_{T2}(\bd^2)$ than
a hard wave function (here AV18).
On the other hand, the large-$\dl^2$ behavior is strongly affected by the nucleon EMT-FFs.
This further corroborates the discussion and observations in Fig.~\ref{fig:groups}
that $D_{T2}(\bd^2)$ is extremely sensitive to dynamics.

\begin{figure}[htb]
  \centering
  \includegraphics[width=\linewidth]{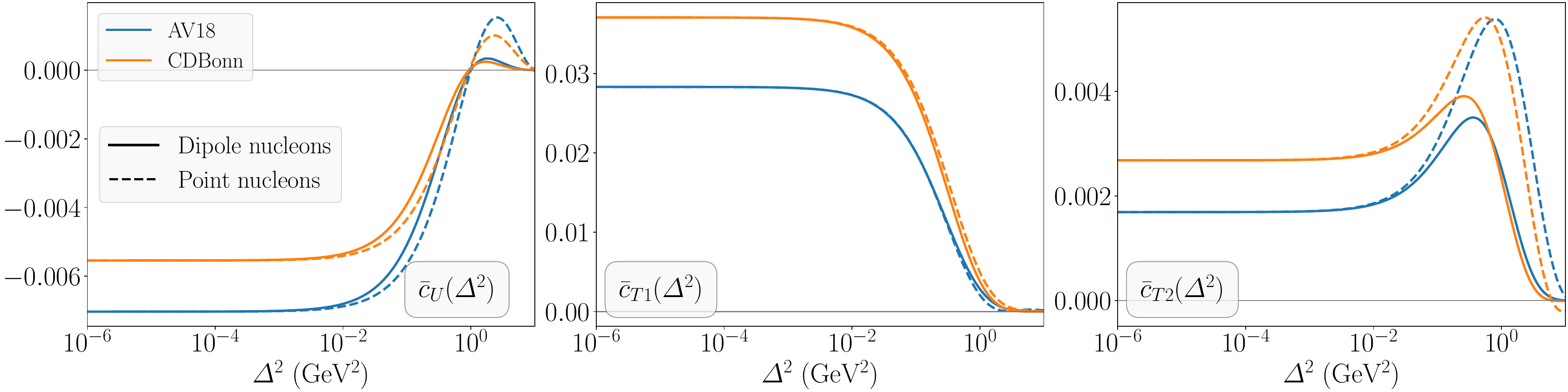}
  \caption{
    Comparisons between the deuteron $\bar{c}$-like form factors
    for different deuteron wave functions
    and different nucleon EMT-FFs.
    Blue curves use the AV18 wave function~\cite{Wiringa:1994wb}
    while orange curves use CD-Bonn~\cite{Machleidt:2000ge}.
    Solid curves use dipole forms for the nucleon EMT-FFs,
    while dashed curves assume pointlike nucleons.
  }
  \label{fig:cbar4}
\end{figure}

We next consider the non-conserved, $\bar{c}$-like form factors appearing in
the symmetric part of the stress tensor.
These results---which are new to the present work---are shown in Fig.~\ref{fig:cbar4}.
Since these quantify forces felt by subcomponents through
the Cauchy momentum equation~\cite{Polyakov:2018exb,Won:2023cyd,Freese:2024rkr,Kim:2025iis},
they are sensitive to the particular nucleon-nucleon force used.
For the same reason, they are sensitive to the nucleon EMT-FFs;
for a pointlike nucleon, the internucleon force will be felt only at the nucleon's exact location,
whereas for a finite-size nucleon,
the force will be distributed over its constituent quarks and gluons.

\begin{figure}[htb]
  \centering
  \includegraphics[width=0.7\linewidth]{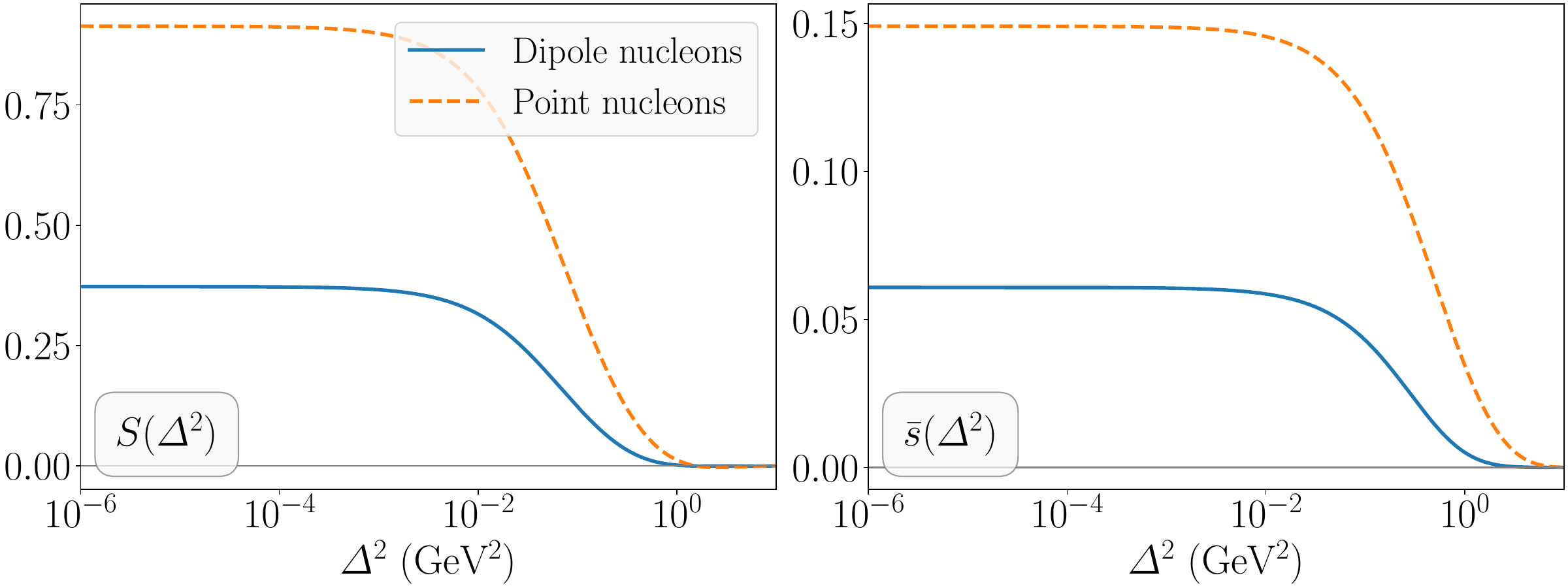}
  \caption{
    Comparisons between the EMT-FFs appearing in the antisymmetric
    part of the stress tensor,
    using either a dipole form for $S_N(\bd^2)$ (solid blue curve)
    or assuming pointlike nucleons (dashed orange curve).
  }
  \label{fig:antisymmetric}
\end{figure}

Lastly, we consider the two form factors appearing in the antisymmetric
part of the stress tensor;
these are shown in Fig.~\ref{fig:antisymmetric}.
These form factors are not particularly sensitive to the deuteron wave function,
so we only show dependence on the nucleon form factors,
which effectively amounts to the overall factor $S_N(\bd^2)$
in Eqs.~(\ref{eqn:S}) and (\ref{eqn:sbar}).
The forward limit $S(0)$ in particular gives the amount of
the deuteron's angular momentum
that is carried by intrinsic quark spin, which we find to be
$S(0) \approx 0.37$,
and hence $37\%$ of the deuteron's total spin.
This is slightly less than the $\sim 40\%$ of the proton's total spin carried
by quark intrinsic spin if we use the central JAM22 value
$S_N(0) = 0.204$~\cite{Cocuzza:2022jye}.
This occurs because of the deuteron depolarization factor,
as in Eq.~(\ref{eqn:JS:0}).


\section{Spatial densities of stresses}
\label{sec:density}

With the EMT form factors in hand,
we move on to considering the associated densities.
We will follow the trail cleared by
Refs.~\cite{Li:2022ldb,Freese:2022fat,Freese:2024rkr,Freese:2025tqd}
by formulating the quantum expectation value
$\langle \Psi | \hat{T}^{ij}(\bm{x},t) | \Psi \rangle$
of the stress tensor as a convolution relation,
in which internal densities are
boosted by the barycentric velocity and
smeared by the barycentric probability density.
This will allow identification of internal densities---including mass,
mass flux, momentum and stress distributions---as Fourier transforms
of the EMT form factors.

This formulation relies on the Galilei symmetry group of non-relativistic
quantum mechanics,
and the deuteron is accordingly treated as a non-relativistic system.
This raises the question of whether incorporating the quark and gluon
substructure of the nucleons is justified.
As a composite system, the nucleon is robustly relativistic,
which has led to considerable controversy about the appropriate way to
describe its internal densities;
see Refs.~\cite{Fleming:1974af,Burkardt:2000za,Miller:2018ybm,Lorce:2020onh,Jaffe:2020ebz,Freese:2021czn,Panteleeva:2022uii,Li:2024vgv}
for a variety of perspectives.
One helpful point---illustrated in the numerical examples of
Ref.~\cite{Freese:2022fat}---is that relativistic corrections to the
density formulas are washed out
in the quantum expectation value
if the nucleon's wave packet is
broader than its Compton wavelength,
about $1.3$~fm.
Since the deuteron's radius is about $2$~fm---and so its diameter
about $4$~fm---the nucleons in a deuteron are sufficiently
smeared so that relativistic corrections to their densities
can be neglected.

To help build intuition with concrete illustrations,
we will present numerical results for the densities along with
the exposition of formal analytic results.
These numerical results, like the EMT-FFs we found above,
will only contain one-body contributions,
and accordingly would receive corrections from exchange currents.
Just as above, we use the AV18 deuteron wave function~\cite{Wiringa:1994wb}.
For the nucleon form factors
$A_N(\bd^2)$, $J_N(\bd^2)$ and $D_N(\bd^2)$,
we use the meson dominance fit
of Broniowski and Ruiz Arriola~\cite{Broniowski:2025ctl}
to the MIT lattice form factor data~\cite{Hackett:2023rif}.
These form factors differ from those used in Sec.~\ref{sec:mff},
but do not lead to significant qualitative changes in the results.
The parametric form used by Broniowski and Ruiz Arriola has a strong
basis in the analytic properties expected of the energy-momentum tensor,
due to both
the large $-t$ asymptotics predicted from perturbative QCD
and the presence of specific
(isoscalar, spin-two and spin-zero)
meson poles in the positive $t$ region.
Additionally, the lattice data that the model's parameters are fit to
are state of the art,
and the resulting $D_N(\bd^2)$ in particular
is a better fit to the lattice data than the dipole fit
provided in Ref.~\cite{Hackett:2023rif}.
For $S_N(\bd^2)$,
we again use a dipole form
with $S_N(0) = 0.204$ (from the JAM22 analysis~\cite{Cocuzza:2022jye})
and $\varLambda = 1.23$~GeV (from the PDG value of the $a_1$ mass~\cite{ParticleDataGroup:2024cfk}).
As we are summing over parton flavors,
we set $\bar{c}_N(\bd^2) = 0$.


\subsection{Identification of internal densities}

For a deuteron prepared in a physical state $|\Psi\rangle$,
the expectation value of the stress tensor can be written---with
the aid of completeness relations---as:
\begin{align}
  \label{eqn:exp:mom}
  \langle \Psi | \hat{T}^{ij}(\bm{x},t) | \Psi \rangle
  =
  \sum_{s,s'}
  \int \frac{\d^3p }{(2\pi)^3}
  \int \frac{\d^3p'}{(2\pi)^3}
  \langle \Psi | p', s' \rangle
  \langle p', s' | \hat{T}^{ij}(0) | p, s \rangle
  \langle p, s | \Psi \rangle
  \e^{i(E_{\bm{p}'}-E_{\bm{p}})t}
  \e^{-i(\bm{p}'-\bm{p})\cdot\bm{x}}
  \,.
\end{align}
Here, and throughout this section,
we drop the subscript $d$ from momenta and spins,
with an implicit understanding that $\bm{p}$ and $s$
signify deuteron momentum and spin.
The Fourier transform of
$ \langle p, s | \Psi \rangle $
gives the wave function describing the location of the deuteron's barycenter,
which we denote as $\Phi_{s}(\bm{R},t)$ to distinguish it from the
wave function $\psi_d(\bm{r})$ describing its internal structure.
Through the Fourier transform:
\begin{align}
  \Phi_{s}(\bm{R},t)
  =
  \int \frac{\d^3p }{(2\pi)^3}
  \langle p, s | \Psi \rangle
  \e^{-iE_{\bm{p}}t}
  \e^{i\bm{p}\cdot\bm{R}}
  \,,
\end{align}
Eq.~(\ref{eqn:exp:mom}) can be rewritten:
\begin{align}
  \label{eqn:exp:coord}
  \langle \Psi | \hat{T}^{ij}(\bm{x},t) | \Psi \rangle
  =
  \sum_{s,s'}
  \int \d^3 R \,
  \int \frac{\d^3\dl}{(2\pi)^3}
  \Phi^*_{s'}(\bm{R},t)
  \langle p', s' | \hat{T}^{ij}(0) | p, s \rangle
  \Phi_{s}(\bm{R},t)
  \e^{-i\bd\cdot(\bm{x}-\bm{R})}
  \bigg|_{2i \bm{P} \rightarrow \blrn}
  \,,
\end{align}
where the average total momentum $\bm{P}$ is replaced by a two-sided
derivative acting on the wave packet;
see Refs.~\cite{Li:2022ldb,Freese:2022fat}
for step-by-step derivations of similar expressions.
The resulting expression can be broken down into terms that contain
two, one or zero two-sided derivatives,
since the form factor breakdown (\ref{eqn:mff}) contains terms
with two, one or zero factors of $\bm{P}$.
We can thus rewrite
Eq.~(\ref{eqn:exp:coord}) as a convolution relation between several
wave-packet-dependent smearing functions
and internal densities:
\begin{multline}
  \label{eqn:smear}
  \langle \Psi | \hat{T}^{ij}(\bm{x},t) | \Psi \rangle
  =
  \sum_{s,s'}
  \int \d^3 R \,
  \Bigg\{
    \left(
    -
    \Phi^*_{s'}(\bm{R},t)
    \frac{\lrn^i\lrn^j}{4M_d^2}
    \Phi_{s}(\bm{R},t)
    \right)
    \mathfrak{a}_{s's}(\bm{x}-\bm{R})
    -
    \frac{i \Phi^*_{s'}(\bm{R},t) \lrn^i \Phi_{s}(\bm{R},t) }{2M_d}
    \mathfrak{p}^j_{s's}(\bm{x}-\bm{R})
    \\
    -
    \frac{i \Phi^*_{s'}(\bm{R},t) \lrn^j \Phi_{s}(\bm{R},t) }{2M_d}
    \mathfrak{f}^i_{s's}(\bm{x}-\bm{R})
    +
    \Phi^*_{s'}(\bm{R},t)
    \Phi_{s}(\bm{R},t)
    \mathfrak{t}^{ij}_{s's}(\bm{x}-\bm{R})
    \Bigg\}
  \,,
\end{multline}
where the internal densities are:
\begin{align}
  \label{eqn:density:mass}
  \mathfrak{a}_{s's}(\bm{b})
  &=
  \epi_a
  \epf_b
  M_d
  \int \frac{\d^3\dl}{(2\pi)^3}
  \left[
    \delta^{ab}
    A_U(\bd^2)
    +
    Y_2^{ab}(\hat{\dl})\frac{\bd^2}{2M_d^2}
    A_T(\bd^2)
    \right]
  \e^{-i\bd\cdot\bm{b}}
  \\
  \label{eqn:density:momentum}
  \mathfrak{p}^i_{s's}(\bm{b})
  &=
  \epi_a
  \epf_b
  \int \frac{\d^3\dl}{(2\pi)^3}
  \left(\frac{ \delta^{ai} \dl^b - \delta^{bi} \dl^a }{2}\right)
  \Big( J(\bd^2) - S(\bd^2) \Big)
  \e^{-i\bd\cdot\bm{b}}
  \\
  \label{eqn:density:flux}
  \mathfrak{f}^i_{s's}(\bm{b})
  &=
  \epi_a
  \epf_b
  \int \frac{\d^3\dl}{(2\pi)^3}
  \left(\frac{ \delta^{ai} \dl^b - \delta^{bi} \dl^a }{2}\right)
  \Big( J(\bd^2) + S(\bd^2) \Big)
  \e^{-i\bd\cdot\bm{b}}
  \\
  \label{eqn:density:stress}
  \mathfrak{t}^{ij}_{s's}(\bm{b})
  &=
  \epi_a
  \epf_b
  \int \frac{\d^3\dl}{(2\pi)^3}
  \Bigg\{
    \frac{\dl^i \dl^j - \delta^{ij} \bd^2}{4M_d}
    \left[
      \delta^{ab}
      D_U(\bd^2)
      +
      Y_2^{ab}(\hat{\dl})
      \frac{\bd^2}{2M_d^2}
      D_{T1}(\bd^2)
      \right]
    \notag \\ & \qquad \qquad
    +
    \frac{\bd^2}{2M_d}
    \left[
      Q^{jlab} Y_2^{il}(\hat{\dl})
      +
      Q^{liab} Y_2^{lj}(\hat{\dl})
      -
      Q^{klab}
      Y_2^{kl}(\hat{\dl})
      \delta^{ij}
      -
      \frac{1}{3}
      Q^{ijab}
      \right]
    D_{T2}(\bd^2)
    \notag \\ & \qquad \qquad
    -
    M_d
    \delta^{ij}
    \left[
      \delta^{ab}
      \bar{c}_U(\bd^2)
      +
      Y_2^{ab}(\hat{\dl})
      \frac{\bd^2}{2M_d^2}
      \bar{c}_{T1}(\bd^2)
      \right]
    -
    M_d
    Q^{ijab}
    \bar{c}_{T2}(\bd^2)
    \notag \\ & \qquad \qquad
    +
    \frac{\bd^2}{4M_d^2}
    \Big(
    \delta^{a[i}_{\phantom{2}} Y_2^{j]b}(\hat{\dl})
    +
    \delta^{b[i}_{\phantom{2}} Y_2^{j]a}(\hat{\dl})
    \Big)
    \bar{s}(\bd^2)
    \Bigg\}
  \e^{-i\bd\cdot\bm{b}}
  \,,
\end{align}
with $\bm{b} = \bm{x} - \bm{R}$ being the three-dimensional
vector displacement from the deuteron's barycenter,
and where the tensors $Y_2^{ij}(\hat{\dl})$ and $Q^{ijab}$
were defined in Eqs.~(\ref{eqn:harmonic}) and (\ref{eqn:quadrupole}).
Strictly speaking,
these quantities can only be interpreted as actual densities
when traced with a physical spin density matrix;
off-diagonal components (for which $s'\neq s$) on their own
do not have a direct physical meaning,
but are important computational tools that must be accounted for
when the spin quantization axis and deuteron polarization are in different
directions, or when using arbitrary density matrices.

The densities listed in Eqs.~(\ref{eqn:density:mass}), (\ref{eqn:density:momentum}),
(\ref{eqn:density:flux}) and (\ref{eqn:density:stress})
can be respectively identified as
an internal mass density      $\mathfrak{a}     (\bm{b})$,
an internal momentum density  $\mathfrak{p}^i   (\bm{b})$,
an internal mass flux density $\mathfrak{f}^i   (\bm{b})$,
and an internal stress tensor $\mathfrak{t}^{ij}(\bm{b})$.
These particular identifications are motivated by
the Galilei boost formula for the stress tensor~\cite{Freese:2025tqd}\footnote{
  The use of $+$ as an index is motivated by the Galilei group being a subgroup
  of a $(4+1)$-dimensional Poincar\'{e} group.
  Non-relativistic time and mass correspond to the plus component of five-vectors
  in the $(4+1)$-dimensional spacetime acted on by the larger group.
  See Refs.~\cite{pinsky:gal,omote:gal,Santos:2004pq,Freese:2025tqd}
  for expositions on the five-vector formalism.
}:
\begin{align}
  \label{eqn:boost:stress}
  T^{ij}
  =
  T^{ij}_{\mathrm{rest}}
  +
  v^i T^{+j}_{\mathrm{rest}}
  +
  v^j T^{i+}_{\mathrm{rest}}
  +
  v^i v^j T^{++}_{\mathrm{rest}}
  \,,
\end{align}
where $\bm{v}$ is the velocity of the system,
and---in the system's rest frame---$T^{+j}_{\mathrm{rest}}$ is the momentum density,
$T^{i+}_{\mathrm{rest}}$ is the mass flux density,
and $T^{++}_{\mathrm{rest}}$ is the mass density~\cite{Freese:2025tqd}.
In this context,
\begin{align}
  \label{eqn:velocity}
  \bm{v}_{s's}
  =
  -
  \frac{i}{2M_d}
  \frac{
    \Phi^*_{s'}(\bm{R},t) \lrn^i \Phi_{s}(\bm{R},t)
  }{
    \Phi^*_{s'}(\bm{R},t) \Phi_{s}(\bm{R},t)
  }
\end{align}
is identified as the effective velocity by which the system is boosted.
This velocity formula also appears in the de Broglie-Bohm pilot wave interpretation
of quantum mechanics~\cite{db:pilot,Bohm:1951xw,Bohm:2006und}\footnote{
  See also Ref.~\cite{Barandes:2026hwq}
  for a recent argument that pilot wave theory is best understood
  as a hidden Markov model,
  in which the pilot wave is a latent variable
  representing the memory of the system.
  Similar reasoning about classical pilot wave hydrodynamics
  can be found in Ref.~\cite{Frumkin:2022jup}.
}.
This velocity is accordingly often referred to as the Bohmian velocity.

In terms of the Bohmian velocity,
the convolution formula (\ref{eqn:smear}) can be rewritten:
\begin{multline}
  \langle \Psi | \hat{T}^{ij}(\bm{x},t) | \Psi \rangle
  =
  \sum_{s,s'}
  \int \d^3 R \,
  \Bigg\{
    \Phi^*_{s'}(\bm{R},t)
    \Phi_{s}(\bm{R},t)
    \bigg(
    v_{s's}^i
    v_{s's}^j
    \mathfrak{a}_{s's}(\bm{x}-\bm{R})
    +
    v_{s's}^i
    \mathfrak{p}^j_{s's}(\bm{x}-\bm{R})
    \\
    +
    v_{s's}^j
    \mathfrak{f}^i_{s's}(\bm{x}-\bm{R})
    +
    \mathfrak{t}^{ij}_{s's}(\bm{x}-\bm{R})
    \bigg)
    +
    \mathcal{Q}_{s's}^{ij}(\bm{R},t)
    \mathfrak{a}_{s's}(\bm{x}-\bm{R})
    \Bigg\}
  \,.
\end{multline}
This almost has the form of a Galilei-boosted stress
tensor---as in Eq.~(\ref{eqn:boost:stress})---smeared
out by the probability density
$ \Phi^*_{s'}(\bm{R},t) \Phi_{s}(\bm{R},t)$,
aside from the addition of a quantum stress
tensor~\cite{Takabayasi:1952,Freese:2024rkr,Freese:2025tqd}:
\begin{multline}
  \label{eqn:stress:quantum}
  \mathcal{Q}_{s's}^{ij}(\bm{R},t)
  =
  \frac{1}{4M_d}
  \bigg\{
    (\nabla_i \mathscr{R}_{s})
    (\nabla_j \mathscr{R}_{s'})
    +
    (\nabla_i \mathscr{R}_{s'})
    (\nabla_j \mathscr{R}_{s})
    -
    \mathscr{R}_{s}
    (\nabla_i \nabla_j \mathscr{R}_{s'})
    -
    \mathscr{R}_{s'}
    (\nabla_i \nabla_j \mathscr{R}_{s})
    \\
    -
    i \mathscr{R}_{s} \mathscr{R}_{s'}
    \big(\nabla_i \nabla_j [\mathscr{S}_{s} - \mathscr{S}_{s'}]\big)
    \bigg\}
  \,,
\end{multline}
where $\mathscr{R}_{s}(\bm{R},t)$ and $\mathscr{S}_{s}(\bm{R},t)$ are
real-valued functions defined through a polar decomposition of the wave function:
\begin{align}
  \Phi_{s}(\bm{R},t)
  =
  \mathscr{R}_{s}(\bm{R},t)
  \e^{i\mathscr{S}_{s}(\bm{R},t)}
  \,.
\end{align}
This quantum stress tensor (\ref{eqn:stress:quantum})
is a generalization of the expression found in
Refs.~\cite{Takabayasi:1952,Freese:2024rkr,Freese:2025tqd},
the latter of which lacks spin indices.
Its meaning depends on the interpretation of quantum mechanics one adopts.
In the pilot wave interpretation
it describes quantum forces exerted on the deuteron by its wave packet,
while in Nelson's stochastic dynamics~\cite{Nelson:1966sp}
it also contains disordered stochastic motion
in the form of directionally symmetric osmotic velocities.
The exact interpretation of this quantum stress tensor does not affect
the isolation of internal densities in non-relativistic quantum mechanics,
and the results we obtain in this work thus do not depend on
any particular interpretation of quantum mechanics.

It is worth noting that off-diagonal components ($s'\neq s$)
of both the quantum stress tensor (\ref{eqn:stress:quantum})
and the Bohmian velocity (\ref{eqn:velocity})---which can also be written:
\begin{align}
  \bm{v}_{s's}
  =
  \frac{
    \bm{\nabla} \mathscr{S}_{s}
    +
    \bm{\nabla} \mathscr{S}_{s'}
  }{
    2M_d
  }
  +
  \frac{i}{2M_d}
  \bm{\nabla}
  \log\left(
  \frac{\mathscr{R}_{s'}}{\mathscr{R}_{s}}
  \right)
\end{align}
---will generally have imaginary parts.
Similar to the internal densities,
the Bohmian velocity and quantum stress tensor only have
a direct interpretation as a velocity or stress tensor, respectively,
when traced with a physical spin density matrix.
The off-diagonal components thus do not have a direct physical meaning,
but nonetheless are an important computational tool
that must be accounted for when the spin quantization axis
and the polarization direction are not the same.
On the other hand, the diagonal components ($s'=s$)
reduce to the standard expressions for the Bohmian velocity~\cite{Bohm:1951xw,Bohm:2006und}
and the quantum stress tensor~\cite{Takabayasi:1952,Freese:2024rkr,Freese:2025tqd}.


\subsection{Mass density, radius and quadrupole moment}
\label{sec:mass}

\begin{figure}
  \includegraphics[width=\textwidth]{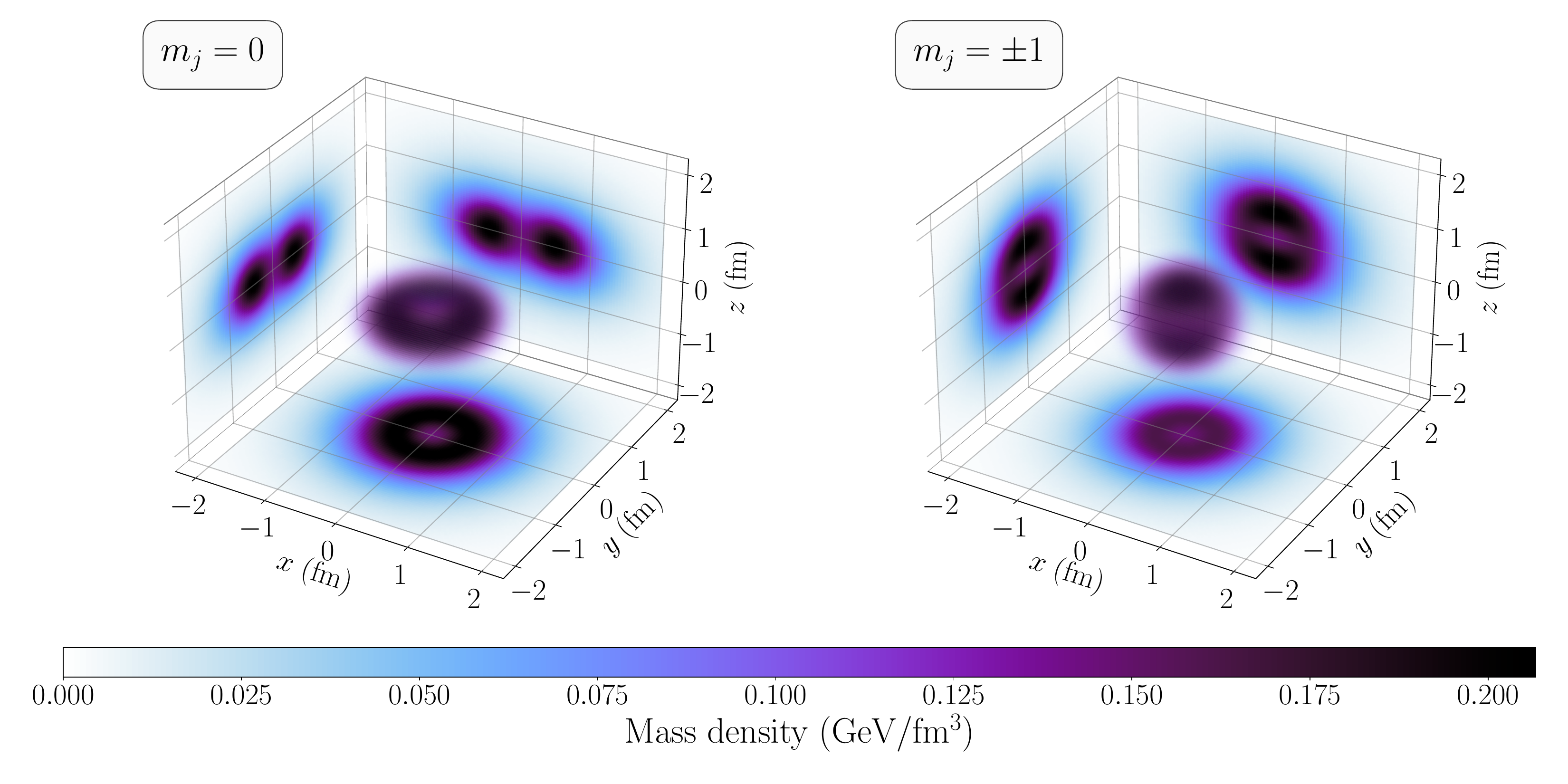}
  \caption{
    Mass density of a deuteron in an $m_j=0$ state (left panel)
    and an $m_j=\pm1$ state (right panel).
    The images on the walls of the plots are slices of the mass density at
    $x=0$ (left wall), $y=0$ (back wall) and $z=0$ (floor).
    This calculation uses the AV18 deuteron wave function~\cite{Wiringa:1994wb}
    and the meson dominance nucleon form factors of
    Broniowski and Ruiz Arriola~\cite{Broniowski:2025ctl}.
  }
  \label{fig:mass}
\end{figure}

We first obtain the mass, mass flux and momentum densities,
since these are simpler than the stress tensor.
We begin with the mass density.
In terms of the harmonic tensors%
---as defined in Eq.~(\ref{eqn:harmonic})---%
the internal mass density of the deuteron can be written:
\begin{align}
\label{eqn:mass_density}
  \mathfrak{a}_{s's}(\bm{b})
  &=
  \epi_a
  \epf_b
  \left(
  \delta^{ab}
  \mathfrak{a}_{U}(b)
  -
  \frac{1}{2}
  Y_2^{ab}(\hat{b})
  \mathfrak{a}_{T}(b)
  \right)
  \,,
\end{align}
with the unpolarized and tensor-polarized contributions
being given by Bessel transforms:
\begin{align}
  \begin{split}
    \mathfrak{a}_U(\bm{b})
    &=
    \frac{M_d}{2\pi^2}
    \int_0^\infty \d \dl \,
    \dl^2
    A_U(\bd^2)
    j_0(b\dl)
    \\
    \mathfrak{a}_T(\bm{b})
    &=
    \frac{M_d}{2\pi^2}
    \int_0^\infty \d \dl \,
    \frac{\dl^4}{M_d^2}
    A_T(\bd^2)
    j_2(b\dl)
    \,.
  \end{split}
\end{align}
The mass densities of pure states
(with spin projection along the $z$-axis for definiteness)
can be written as linear combinations
of these two contributions:
\begin{align}\label{eq.massDensitiesPure}
  \begin{split}
    \mathfrak{a}_{m_j=\pm1}(\bm{b})
    &
    =
    \mathfrak{a}_{U}(b)
    +
    \frac{1}{6}
    \left(\frac{3}{2}\cos^2(\theta_b)-\frac{1}{2}\right)
    \mathfrak{a}_{T}(b)
    \\
    \mathfrak{a}_{m_j=0}(\bm{b})
    &
    =
    \mathfrak{a}_{U}(b)
    -
    \frac{1}{3}
    \left(\frac{3}{2}\cos^2(\theta_b)-\frac{1}{2}\right)
    \mathfrak{a}_{T}(b)
    \,.
  \end{split}
\end{align}
Numerical results for these are shown in Fig.~\ref{fig:mass}.
The mass densities follow the well-known donut and dumbbell shapes of the deuteron's
$m_j=0$ and $m_j=\pm1$ states~\cite{Forest:1996kp,Mantysaari:2024xmy}, respectively,
but with blurring due to the finite size of the nucleons.

\begin{figure}
  \includegraphics[width=\textwidth]{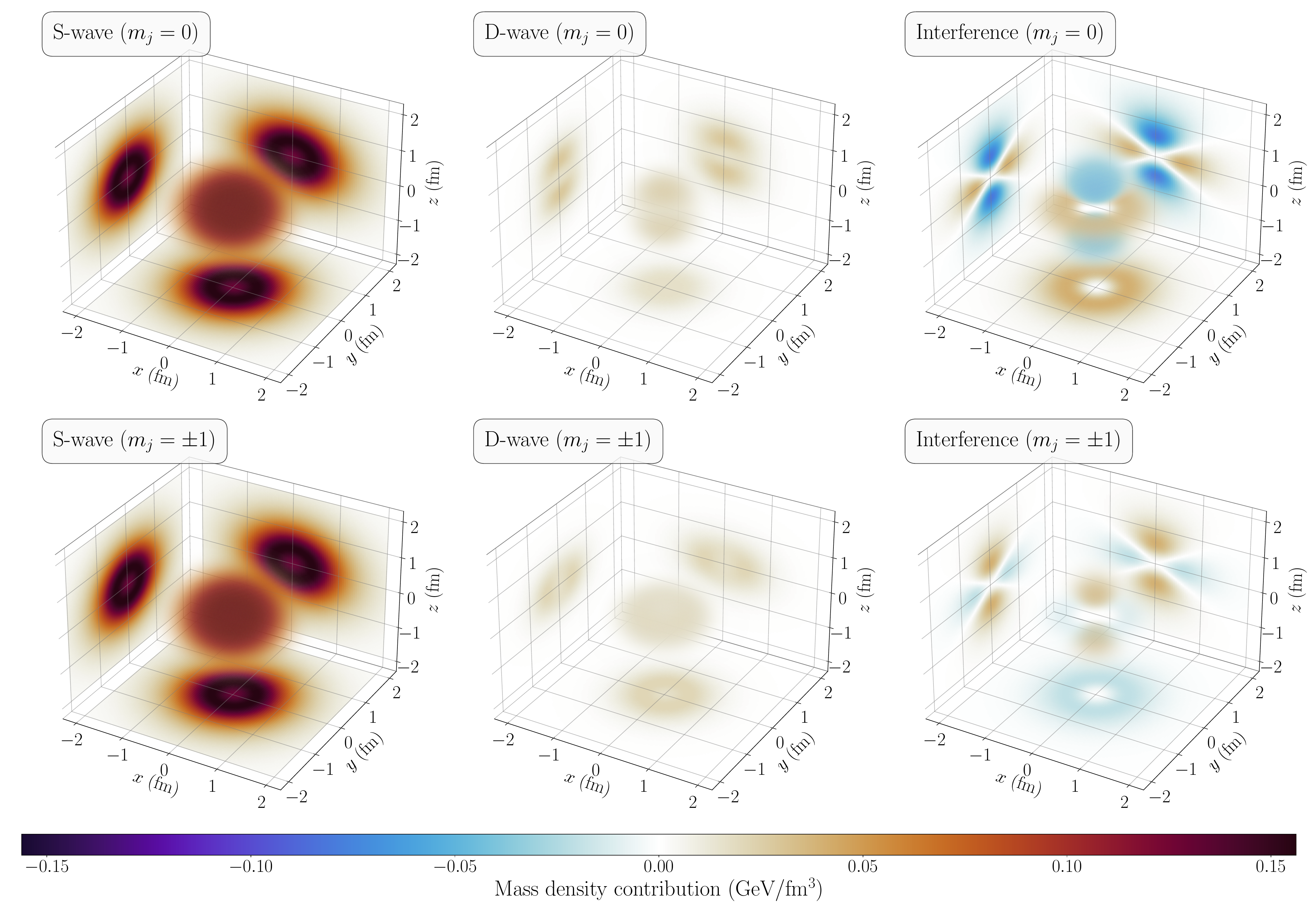}
  \caption{
    Separate S-wave (left column), D-wave (middle column) and interference (right column)
    contributions to the mass density of a deuteron
    an $m_j=0$ state (top row) and an $m_j=\pm1$ state (bottom row).
    The images on the walls of the plots are slices of the mass densities at
    $x=0$ (left wall), $y=0$ (back wall) and $z=0$ (floor).
    This calculation uses the AV18 deuteron wave function~\cite{Wiringa:1994wb}
    and the meson dominance nucleon form factors of
    Broniowski and Ruiz Arriola~\cite{Broniowski:2025ctl}.
  }
  \label{fig:interference}
\end{figure}

The mass radius is identified as the root mean squared radius
of the mass density in Eq.~(\ref{eqn:mass_density}).
Since $b^2$ is a scalar,
only the unpolarized part of the mass density $\mathfrak{a}_U$ contributes to the integrals,
and the mass radius becomes independent of the deuteron spin ensemble\footnote{
  This holds in general for any higher spin particle in the non-relativistic picture
  as all new structures can be identified with higher-order multipoles
  which integrate to zero against a scalar quantity.
}.
We can write
\begin{equation}
  \label{eqn:massRadius}
  \langle r^2 \rangle_{\mathrm{Mass}}
  =
  \frac{
    \int \d^3b \; b^2 \mathfrak{a}_U(\bm{b})
  }{
    \int \d^3b \; \mathfrak{a}_U(\bm{b})
  }
  =
  -6
  \frac{\d A_U(\bd^2)}{\d \bd^2}\Bigg|_{\bd=0}
  \,.
\end{equation}
In the impulse approximation,
we can use the result for $A_U$ of Eq.~(\ref{eqn:AU}), which leads to the relation:
\begin{equation}
  \label{eqn:massradiusIA}
  \langle r^2 \rangle_{\mathrm{Mass}}
  =
  \langle r_N^2 \rangle_{\mathrm{Mass}}
  +
  \langle r^2\rangle_{\mathrm{Matter}}
  \,.
\end{equation}
Additional contributions from two-body currents would also be added in quadrature.
Here,
$\langle r_N^2 \rangle_{\mathrm{Mass}} = -6 A_N'(0)$
signifies the square mass radius of the nucleon and
$\langle r^2\rangle_{\mathrm{Matter}}$
would be the mass radius assuming pointlike nucleons,
and is often called the matter radius in older literature~\cite{Sprung:1990zz,Wong:1991ks,Babenko:2008zz,Zhaba:2017syr}.
It is given explicitly by\footnote{
  The factor $\frac{1}{4}$ difference between Eq.~(\ref{eqn:radius:matter})
  and the numerator of Eq.~(\ref{eqn:massRadius}) can be understood from
  the different character of $\bm r$ (relative coordinate between the two nucleons)
  and $\bm b$ (coordinate of the nucleon relative to the barycenter).
  In a frame with the barycenter $\bm R=0$, we have $\bm b = \pm \frac{1}{2}\bm{r}$ for either nucleon.
}:
\begin{equation}
  \label{eqn:radius:matter}
  \langle r^2 \rangle_{\mathrm{Matter}}
  \equiv
  \bigg\langle \frac{m_p\bm \hat r_p^2+m_n\bm \hat r_n^2}{M_d}\bigg\rangle
  =
  \frac{1}{4} \int_0^\infty \d r \; r^2 [u^2(r)+w^2(r)]
  \,.
\end{equation}
Using the meson dominance form factors of Ref.~\cite{Broniowski:2025ctl},
$\sqrt{\langle r^2 \rangle}_{\mathrm{Mass}}= 2.04~\text{fm}$
for the one-body contribution to the deuteron's mean squared mass radius.

The mass quadrupole moment tells a more straightforward story.
As usual~\cite{Blatt:1952ije},
the quadrupole moment is defined as the mean value of
$(3 z^2 - b^2)$ for an $m_j=j=1$ state:
\begin{align}
  \label{eqn:quad:mass}
  Q_{\mathrm{Mass}}
  =
  \frac{
    \int \d^3 b \,
    (3 z^2 - b^2)
    \mathfrak{a}_{m_j=1}(\bm{b})
  }{
    \int \d^3 b \,
    \mathfrak{a}_{m_j=1}(\bm{b})
  }
  =
  \frac{3}{M_d}
  \int \d^3 b \,
  b^2
  Y_2^{zz}(\hat{b})
  \mathfrak{a}_{m_j=1}(\bm{b})
  \,,
\end{align}
with the mass density used as the density in question.
The result is the same whether the nucleons are pointlike
or have finite extent, and we find:
\begin{align}
  Q_{\mathrm{Mass}}
  =
  \frac{A_T(0)}{(2m_N)^2}
  =
  Q_d
  \,.
\end{align}
This is identical to the usual electric quadrupole moment,
as defined in Eq.~(\ref{eqn:wf:quad}).
The result not depending on nucleon structure likely relates
to the nucleons themselves not having quadrupole moments.

Before proceeding to the other densities,
it is worth dissecting the mass density further
and examining its separate S-wave, D-wave and interference contributions.
This examination will actually clarify several of our subsequent results.
The breakdown is shown in Fig.~\ref{fig:interference}.
The S-wave contribution is spherically symmetric,
though its density dips near the center,
making it more like a caramel truffle than a solid ball.
The D-wave contribution depends on the polarization state,
with a dumbbell-like shape for $m_j=0$ and a donut shape for $m_j=\pm1$---exactly
the opposite as for the full mass density.
The shapes of the full densities are determined predominantly
by the S-wave and the interference term.
The interference contribution in effect removes part of the truffle's shell
and reinforces the rest,
giving either a donut shape (for $m_j=0$)
or a dumbbell shape (for $m_j=\pm1$).


\subsection{Momentum and mass flux densities}

\begin{figure}
  \includegraphics[width=\textwidth]{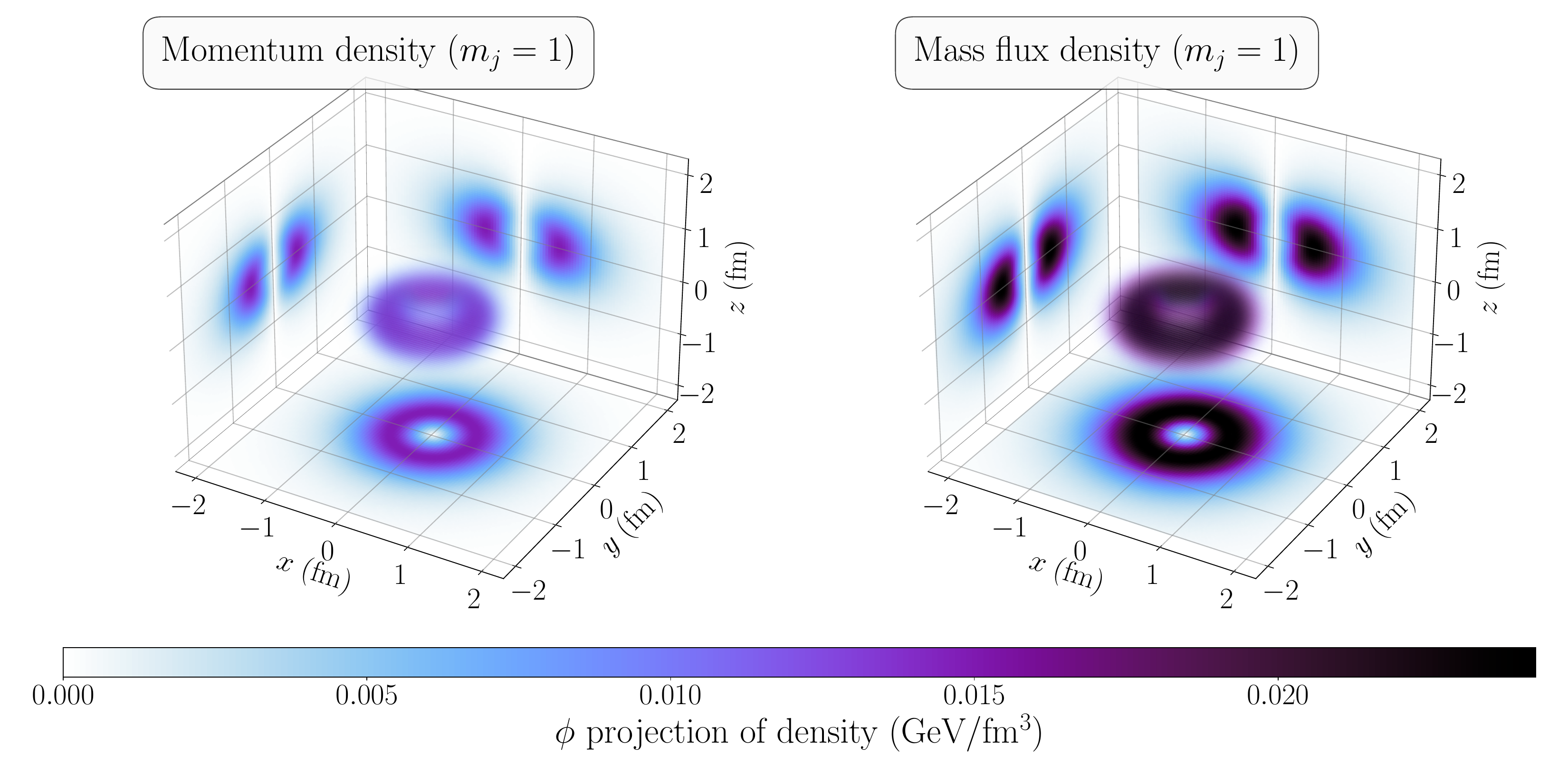}
  \caption{
    $\phi$-direction projection of the momentum density (left panel)
    and mass flux density (right panel)
    for an $m_j=1$ state.
    The images on the walls of the plots are slices of these densities at
    $x=0$ (left wall), $y=0$ (back wall) and $z=0$ (floor).
    This calculation uses the AV18 deuteron wave function~\cite{Wiringa:1994wb}
    and the meson dominance nucleon form factors of
    Broniowski and Ruiz Arriola~\cite{Broniowski:2025ctl}.
  }
  \label{fig:momentum}
\end{figure}

Next, for an $m_j=\pm1$ state, the momentum and mass flux densities are:
\begin{align}
  \label{eqn:momentum:density}
  \begin{split}
    \mathfrak{p}_{m_j=\pm1}(\bm{b})
    &=
    \pm
    \frac{
      \hat{z} \times \hat{b}
    }{4\pi^2}
    \int_0^\infty \d \dl \,
    \dl^3
    \big( J(\bd^2) - S(\bd^2) \big)
    j_1(b\dl)
    \\
    \mathfrak{f}_{m_j=\pm1}(\bm{b})
    &=
    \pm
    \frac{
      \hat{z} \times \hat{b}
    }{4\pi^2}
    \int_0^\infty \d \dl \,
    \dl^3
    \big( J(\bd^2) + S(\bd^2) \big)
    j_1(b\dl)
    \,.
  \end{split}
\end{align}
For $m_j=0$, these densities identically vanish,
as these are pure states without vector polarization.
Vector polarization is the only part of the density matrix that couples to $J$ and $S$;
see Eq.~(\ref{eqn:mff:V}).
Numerical results for both densities are shown for the $m_j=1$ state in
Fig.~\ref{fig:momentum}.

At first glance, it might seem quite peculiar that the momentum and mass flux densities
presented here
have only a $\phi$ component---since $\hat{z} \times \hat{b} = \sin\theta \hat{\phi}$.
There is thus apparently no momentum flow in the radial or polar directions.
This merits stressing that the densities obtained here---and, indeed,
all of the hadronic densities obtained through Fourier transforms of form factors---are
\emph{expectation values}.
In principle, nucleons in the deuteron may be moving radially,
but if---at any point in space---the nucleon is just as likely to move inwards as outwards,
the expectation value for the radial momentum will be zero.
Since the momentum and mass flux densities in
Eq.~(\ref{eqn:momentum:density}) and Fig.~\ref{fig:momentum}
are expectation values, only coherent ordered motion survives averaging.
Thus, the densities show only orbital motion around the spin quantization axis.

Whether there is symmetric, unordered motion in the radial and polar directions
depends on the interpretation of quantum mechanics one adopts.
In the pilot wave interpretation,
the momentum and mass flux densities of
Eq.~(\ref{eqn:momentum:density}) and Fig.~\ref{fig:momentum}
are taken at face value as the actual densities.
As we discussed above, however, other interpretations---such as
stochastic dynamics~\cite{Nelson:1966sp}---contain
additional unordered, symmetric motion that has been averaged out in
these densities.

Two other features of Fig.~\ref{fig:momentum} are worth pointing out.
Firstly, the mass flux density is much larger than the momentum density.
This occurs because we have used the asymmetric energy-momentum tensor.
As can be seen in Eq.~(\ref{eqn:momentum:density}),
the difference between the momentum and mass flux densities depends
on the form factor $S(\bd^2)$,
which comes from the antisymmetric part of the EMT;
see Eq.~(\ref{eqn:mff}).
In effect, quark spin does not contribute to the momentum density when using
the asymmetric EMT, but it does contribute to the mass flux density.

The other peculiar feature is the donut shape of both the momentum
and mass flux densities for the $m_j=1$ state.
This seems to be at odds with the mass distribution for this state
having a dumbbell shape.
However, the shapes of these distributions need not be the same
since nucleons at different locations may carry different momenta.
Owing in part to the factor $\hat{z}\times\hat{b}$,
and in part to the presence of $\dl j_1(b\dl)$ in the integrand,
the S-wave, D-wave and S-D interference contributions
to the mass flux and momentum densities
are all donut-shaped and of similar magnitude.


\subsection{Internal stress tensor}

Let us finally consider the internal stress tensor.
We follow (with minor modifications)
the notation of Refs.~\cite{Polyakov:2018rew,Polyakov:2019lbq},
and decompose the internal stress tensor as follows:
\begin{multline}
  \label{eqn:stress:ps}
  \mathfrak{t}^{ij}_{s's}(\bm{b})
  =
  \epi_a
  \epf_b
  \bigg\{
    p_U(b)
    \delta^{ab}
    \delta^{ij}
    +
    s_U(b)
    \delta^{ab}
    Y_2^{ij}(\hat{b})
    +
    \frac{1}{2M_d^2}
    Q^{klab}
    \partial_k \partial_l
    \Big[
      p_{T1}(b) \delta^{ij}
      +
      s_{T1}(b) Y_2^{ij}(\hat{b})
      \Big]
    +
    p_{T2}(b)
    Q^{ijab}
    \\
    +
    2
    s_{T2}(r)
    \Big(
    Q^{ilab} Y_2^{lj}(\hat{b})
    +
    Q^{jlab} Y_2^{li}(\hat{b})
    -
    \delta^{ij} Q^{lkab} Y_2^{lk}(\hat{b})
    \Big)
    +
    s_A(b)
    \Big(
    \delta^{a[i}_{\phantom{2}} Y_2^{j]b}(\hat{b})
    +
    \delta^{b[i}_{\phantom{2}} Y_2^{j]a}(\hat{b})
    \Big)
    \bigg\}
  \,.
\end{multline}
The irreducible harmonic tensor and quadrupole tensor
were defined further above in Eqs.~(\ref{eqn:harmonic}) and (\ref{eqn:quadrupole}).
Our notation differs from
Ref.~\cite{Polyakov:2019lbq}
in some factors multiplying the $p$ and $s$ functions,
and in the inclusion of a new antisymmetric structure
associated with the $\bar{s}(\bd^2)$ form factor.
The maps between their nomenclature and ours are:
\begin{align}
  \begin{split}
    p_{U}
    &=
    p_0
    \\
    s_{U}
    &=
    s_0
    \\
    p_{T1}
    &=
    -2
    p_3
    \\
    s_{T1}
    &=
    -2
    s_3
    \\
    p_{T2}
    &=
    p_2
    \\
    s_{T2}
    &=
    s_2
    \,.
  \end{split}
\end{align}
In effect, the only conventional difference in our choice of normalization
is the factor $-2$ in the $T1$ structures,
which we have chosen so that
$D_{T1}(0) / D_U(0) \approx (2m_N)^2 Q_d = A_{T}(0)/A_U(0)$.
The $p$ and $s$ functions depend only on the magnitude of $\bm{b}$,
and are all given by Bessel transforms of the EMT form factors:
\begin{align}
  \label{eqn:stress:bessel}
  \begin{split}
    p_U(b)
    &=
    -
    \frac{1}{2\pi^2}
    \int_0^\infty \d \dl \, \dl^2
    \left\{
      \frac{\dl^2}{6 M_d}
      D_U(\dl^2)
      +
      M_d
      \bar{c}_U(\dl^2)
      \right\}
    j_0(\dl b)
    \\
    p_{T1}(b)
    &=
    -
    \frac{1}{2\pi^2}
    \int_0^\infty \d \dl \, \dl^2
    \left\{
      \frac{\dl^2}{6 M_d}
      D_{T1}(\dl^2)
      +
      M_d
      \bar{c}_{T1}(\dl^2)
      \right\}
    j_0(\dl b)
    \\
    p_{T2}(b)
    &=
    -
    \frac{1}{2\pi^2}
    \int_0^\infty \d \dl \, \dl^2
    \left\{
      \frac{\dl^2}{6 M_d}
      D_{T2}(\dl^2)
      +
      M_d
      \bar{c}_{T2}(\dl^2)
      \right\}
    j_0(\dl b)
    \\
    s_U(b)
    &=
    -
    \frac{1}{2\pi^2}
    \int_0^\infty \d \dl \, \dl^2
    \frac{\dl^2}{4M_d}
    D_{U}(\dl^2)
    j_2(\dl b)
    \\
    s_{T1}(b)
    &=
    -
    \frac{1}{2\pi^2}
    \int_0^\infty \d \dl \, \dl^2
    \frac{\dl^2}{4M_d}
    D_{T1}(\dl^2)
    j_2(\dl b)
    \\
    s_{T2}(b)
    &=
    -
    \frac{1}{2\pi^2}
    \int_0^\infty \d \dl \, \dl^2
    \frac{\dl^2}{4M_d}
    D_{T2}(\dl^2)
    j_2(\dl b)
    \\
    s_A(b)
    &
    =
    -
    \frac{1}{2\pi^2}
    \int_0^\infty \d \dl \, \dl^2
    \frac{\dl^2}{4M_d}
    \bar{s}(\dl^2)
    j_2(\dl b)
    \,.
  \end{split}
\end{align}
The similarity in these formulas was part of the motivation
for the conventions we adopted in our EMT-FF breakdown~(\ref{eqn:mff}).

In Eq.~(\ref{eqn:stress:ps}), the functions $p_U$ and $s_U$
appear with the same tensor structures available for building
the intrinsic stress tensor of spin-zero and spin-half systems.
These functions are often respectively called the pressure
and shear in the hadron physics
literature~\cite{Polyakov:2018zvc,Burkert:2018bqq,Lorce:2018egm,Burkert:2021ith,Burkert:2023wzr,Lorce:2025oot},
though it is more precise to call them the isotropic pressure
and pressure anisotropy instead.
After all, $p_U$ is obtained
(for lower-spin systems or unpolarized ensembles)
by taking the isotropic average of the stress in all directions,
and $s_U$ gives the remaining traceless contributions to the stress tensor.
The other structures contribute for ensembles carrying tensor polarization.
The functions $p_{T1}$ and $s_{T1}$ closely mimic the $p_U,s_U$ behavior,
with the former appearing only in the isotropic average and the
latter providing the anisotropic leftovers.

The tensor structures multiplying $p_{T2}$ and $s_{T2}$,
on the other hand,
do not tell the same tale.
The tensor structure multiplying $p_{T2}$ is traceless,
and $p_{T2}$ accordingly contributes only to the pressure anisotropy.
The structure multiplying $s_{T2}$ is more complicated,
and contributes to both the isotropic pressure and the anisotropy.
The labels of $p$ and $s$ for these functions are adopted
for aesthetic reasons---namely, that they obey similar formulas
to the other $(p,s)$ pairs;
for example, see Eq.~(\ref{eqn:stress:bessel}) here
or Eq.~(37) of Polyakov and Sun~\cite{Polyakov:2019lbq}.

One difficulty associated with the $T1$ contributions to the stress tensor
is the presence of derivatives in Eq.~(\ref{eqn:stress:ps}).
For practical numerical calculations,
it is preferable to only perform Bessel transforms and avoid numerical derivatives.
To this end, we suggest an alternate means of writing the $T1$ contributions:
\begin{multline}
  \label{eqn:stress:alt}
  \mathfrak{t}^{ij}_{s's}(\bm{b})
  =
  \epi_a
  \epf_b
  \bigg\{
    p_U(b)
    \delta^{ab}
    \delta^{ij}
    +
    s_U(b)
    \delta^{ab}
    Y_2^{ij}(\hat{b})
    +
    \widetilde{p}_{T1}(b)
    Q^{klab}
    Y_2^{kl}(\hat{b})
    \delta^{ij}
    +
    \sum_{\mathclap{n=0,2,4}}
    \widetilde{s}^{(n)}_{T1}(b)
    X^{ijab}_n(\hat{b})
    \\
    +
    p_{T2}(b)
    Q^{ijab}
    +
    2
    s_{T2}(r)
    \Big(
    Q^{ilab} Y_2^{lj}(\hat{b})
    +
    Q^{jlab} Y_2^{li}(\hat{b})
    -
    \delta^{ij} Q^{lkab} Y_2^{lk}(\hat{b})
    \Big)
    +
    s_A(b)
    \Big(
    \delta^{a[i}_{\phantom{2}} Y_2^{j]b}
    +
    \delta^{b[i}_{\phantom{2}} Y_2^{j]a}
    \Big)
    \bigg\}
  \,.
\end{multline}
The new rank-four tensor structures appearing here are:
\begin{align}
  \begin{split}
    X_0^{ijab}(\hat{r})
    &=
    \frac{1}{15}
    \big( \delta^{ij} \delta^{ab} + \text{permutations} \big)
    -
    \frac{1}{9} \delta^{ij} \delta^{ab}
    \\
    X_2^{ijab}(\hat{r})
    &=
    \frac{1}{7}
    \big( Y_2^{ij} \delta^{ab} + \text{permutations} \big)
    -
    \frac{1}{3}
    \big( Y_2^{ij} \delta^{ab} + Y_2^{ab} \delta^{ij} \big)
    \\
    X_4^{ijab}(\hat{r})
    &=
    Y_4^{ijab}(\hat{r})
    \,,
  \end{split}
\end{align}
and effectively arise from taking the outer product
of two rank-two harmonic tensors:
\begin{align}
  Y_2^{ij}(\hat{r})
  Y_2^{ab}(\hat{r})
  &=
  \sum_{\mathclap{n=0,2,4}}
  X_n^{ijab}(\hat{r})
  \,.
\end{align}
The functions $\widetilde{p}_{T1}$ and $\widetilde{s}_{T1}^{(n)}$
are given by the following Bessel transforms:
\begin{align}
  \begin{split}
    \widetilde{p}_{T1}(b)
    &=
    -
    \frac{1}{2\pi^2}
    \int_0^\infty d\dl \; \dl^2
    \frac{\dl^2}{2M_d^2}
    \left\{
      \frac{\dl^2}{6 M_d}
      D_{T1}(\dl^2)
      +
      M_d
      \bar{c}_{T1}(\dl^2)
      \right\}
    j_2(\dl b)
    \\
    \widetilde{s}^{(n)}_{T1}(b)
    &=
    \frac{i^n}{2\pi^2}
    \int_0^\infty d\dl \;\dl^2
    \frac{\dl^4}{8M_d^3}
    D_{T1}(\dl^2)
    j_n(\dl b)
    \,.
  \end{split}
\end{align}
The alternate breakdown (\ref{eqn:stress:alt}) is more complicated than
the Polyakov-Sun breakdown (\ref{eqn:stress:ps}),
having a greater number of apparently more complicated tensor structures
and $b$-dependent functions.
However, the lack of derivatives makes the alternate breakdown more amenable
to numerical calculations.
We will accordingly use the alternate breakdown (\ref{eqn:stress:alt})
for the remainder of this work.


\subsection{Stress tensor in spherical coordinates}

We will now consider components of the internal stress tensor more explicitly
in spherical coordinates.
To aid with this, we will break the stress tensor into three pieces:
(1) an unpolarized piece that is the same for all polarization states;
(2) the symmetric part of the tensor-polarized contribution; and
(3) the antisymmetric part of the tensor-polarized contribution.
In spherical coordinates,
these pieces respectively have the following non-zero components:
\begin{align}
  \label{eqn:stress:spherical}
  \begin{split}
    \mathfrak{t}^{ij}_{U}
    &=
    \begin{bmatrix}
      p_r^{(U)} & 0 & 0 \\
      0 & p_\theta^{(U)} & 0 \\
      0 & 0 & p_\phi^{(U)}
    \end{bmatrix}
    \\
    \mathfrak{t}^{ij}_{TS}
    &=
    \begin{bmatrix}
      p_r^{(T)} & s^{(T)} & 0 \\
      s^{(T)} & p_\theta^{(T)} & 0 \\
      0 & 0 & p_\phi^{(T)}
    \end{bmatrix}
    \\
    \mathfrak{t}^{ij}_{TA}
    &=
    \begin{bmatrix}
      0 & -\tau^{(T)} & 0 \\
      \tau^{(T)} & 0 & 0 \\
      0 & 0 & 0
    \end{bmatrix}
    \,.
  \end{split}
\end{align}
The stress tensor of a pure state can be reconstructed through:
\begin{align}
  \begin{split}
    \mathfrak{t}^{ij}_{m_j=0}
    &=
    \mathfrak{t}^{ij}_U
    -
    \frac{1}{3}
    \Big(
    \mathfrak{t}^{ij}_{TS}
    +
    \mathfrak{t}^{ij}_{TA}
    \Big)
    \\
    \mathfrak{t}^{ij}_{m_j=\pm 1}
    &=
    \mathfrak{t}^{ij}_U
    +
    \frac{1}{6}
    \Big(
    \mathfrak{t}^{ij}_{TS}
    +
    \mathfrak{t}^{ij}_{TA}
    \Big)
    \,.
  \end{split}
\end{align}
In terms of the functions appearing in the breakdown (\ref{eqn:stress:alt}),
we find these components to be:
\begin{align}
  \begin{split}
    p_r^{(U)}
    &=
    p_U + \frac{2}{3} s_U
    \\
    p_\theta^{(U)}
    =
    p_\phi^{(U)}
    &=
    p_U - \frac{1}{3} s_U
    \\
    p_r^{(T)}
    &=
    \left( \frac{3}{2} \cos^2\theta - \frac{1}{2} \right)
    \left\{
      2 \widetilde{p}_{T1}
      -
      \frac{ 4}{15} \widetilde{s}_{T1}^{(0)}
      -
      \frac{ 8}{21} \widetilde{s}_{T1}^{(2)}
      -
      \frac{24}{35} \widetilde{s}_{T1}^{(4)}
      +
      2 p_{T2}
      +
      \frac{4}{3} s_{T2}
      \right\}
    \\
    p_\theta^{(T)}
    &=
    \left( \frac{3}{2} \cos^2\theta - \frac{1}{2} \right)
    \left\{
      2 \widetilde{p}_{T1}
      +
      \frac{ 4}{15} \widetilde{s}_{T1}^{(0)}
      +
      \frac{ 2}{ 5} \widetilde{s}_{T1}^{(4)}
      -
      2 p_{T2}
      -
      \frac{4}{3} s_{T2}
      \right\}
    +
    \left\{
      -
      \frac{ 2}{15} \widetilde{s}_{T1}^{(0)}
      +
      \frac{ 4}{21} \widetilde{s}_{T1}^{(2)}
      -
      \frac{ 2}{35} \widetilde{s}_{T1}^{(4)}
      +
      p_{T2}
      -
      \frac{4}{3} s_{T2}
      \right\}
    \\
    p_\phi^{(T)}
    &=
    \left( \frac{3}{2} \cos^2\theta - \frac{1}{2} \right)
    \left\{
      2 \widetilde{p}_{T1}
      +
      \frac{ 8}{21} \widetilde{s}_{T1}^{(2)}
      +
      \frac{ 2}{ 7} \widetilde{s}_{T1}^{(4)}
      -
      4 s_{T2}
      \right\}
    +
    \left\{
      \frac{ 2}{15} \widetilde{s}_{T1}^{(0)}
      -
      \frac{ 4}{21} \widetilde{s}_{T1}^{(2)}
      +
      \frac{ 2}{35} \widetilde{s}_{T1}^{(4)}
      -
      p_{T2}
      +
      \frac{4}{3} s_{T2}
      \right\}
    \\
    s^{(T)}
    &=
    \sin\theta \cos\theta
    \left\{
      \frac{ 2}{ 5} \widetilde{s}_{T1}^{(0)}
      +
      \frac{ 2}{ 7} \widetilde{s}_{T1}^{(2)}
      -
      \frac{24}{35} \widetilde{s}_{T1}^{(4)}
      -
      3 p_{T2}
      -
      2 s_{T2}
      \right\}
    \\
    \tau^{(T)}
    &=
    6
    \sin\theta \cos\theta
    s_A
    \,.
  \end{split}
\end{align}
In contrast to spin-zero and spin-half systems,
there are off-diagonal shear stresses in the deuteron
when using spherical coordinates.
This largely reflects the non-spherical shape of the deuteron.
A cubic material element of the deuteron, with
$(\hat{r}, \hat{\theta}, \hat{\phi})$
as its principal axes,
will be sheared in the $(r,\theta)$ plane.
If the symmetric stress tensor is used,
this shear can be eliminated through a local change of frame,
which we will consider next.


\subsection{Principal stresses of the symmetric stress tensor}

Any real symmetric square matrix can be diagonalized by real orthogonal transformations.
For the symmetric stress tensor, this corresponds to a local change of frame.
The axes defining the local frame are commonly called principal axes of
the symmetric stress tensor,
and their eigenvalues similarly are called
principal stresses~\cite{irgens2008continuum,jog2015continuum}.

\begin{figure}
  \begin{tabular}{ccc}
    \begin{tikzpicture}
      \coordinate (TL) at (-1, 1);
      \coordinate (TR) at ( 1, 1);
      \coordinate (BL) at (-1,-1);
      \coordinate (BR) at ( 1,-1);
      \coordinate (T) at ( 0, 1);
      \coordinate (B) at ( 0,-1);
      \coordinate (L) at (-1, 0);
      \coordinate (R) at ( 1, 0);
      \coordinate (FT) at ( 0-0.5, 1+1.0);
      \coordinate (FB) at ( 0+0.5,-1-1.0);
      \draw[-, color=ForestGreen, fill=ForestGreen!17!white] (TL) -- (TR) -- (BR) -- (BL) -- (TL);
      \draw[-Stealth, very thick, color=Purple] (FT) -- (T);
      \draw[-Stealth, very thick, color=Purple] (FB) -- (B);
      \draw[-Stealth, very thick, color=Purple] (T) -- +( 0.5, 0.0) node[above=0.1 cm] {$s$};
      \draw[-Stealth, very thick, color=Purple] (B) -- +(-0.5, 0.0);
      \draw[-Stealth, very thick, color=RoyalBlue] (T) -- +( 0.0,-1.0) node[right=0.1 cm] {$p$};
      \draw[-Stealth, very thick, color=RoyalBlue] (B) -- +( 0.0, 1.0);
    \end{tikzpicture}
    &
    % ...
    \begin{tikzpicture}
      \coordinate (TLp) at (-1+0.3, 1);
      \coordinate (TRp) at ( 1+0.3, 1);
      \coordinate (BLp) at (-1,-1);
      \coordinate (BRp) at ( 1,-1);
      \coordinate (Tp) at ( 0+0.3, 1);
      \coordinate (Bp) at ( 0,-1);
      \coordinate (Lp) at (-1+0.3/2, 0);
      \coordinate (Rp) at ( 1+0.3/2, 0);
      \coordinate (FTp) at ( 0-0.5+0.3, 1+1.0);
      \coordinate (FBp) at ( 0+0.5,-1-1.0);
      \draw[-, color=ForestGreen, fill=ForestGreen!17!white] (TLp) -- (TRp) -- (BRp) -- (BLp) -- (TLp);
      \draw[-Stealth, very thick, color=Purple] (FTp) -- (Tp);
      \draw[-Stealth, very thick, color=Purple] (FBp) -- (Bp);
      \draw[-Stealth, very thick, color=Purple] (Tp) -- +( 0.5, 0.0);
      \draw[-Stealth, very thick, color=Purple] (Bp) -- +(-0.5, 0.0);
      \draw[-Stealth, very thick, color=RoyalBlue] (Tp) -- +( 0.0,-1.0);
      \draw[-Stealth, very thick, color=RoyalBlue] (Bp) -- +( 0.0, 1.0);
    \end{tikzpicture}
    &
    % ...
    \begin{tikzpicture}
      \coordinate (TL) at (-0.866-0.5, 0.866-0.5);
      \coordinate (TR) at ( 0.866-0.5, 0.866+0.5);
      \coordinate (BL) at (-0.866+0.5,-0.866-0.5);
      \coordinate (BR) at ( 0.866+0.5,-0.866+0.5);
      \coordinate (T) at ( -0.5, 0.866);
      \coordinate (B) at (  0.5,-0.866);
      \coordinate (L) at (-0.866, -0.5);
      \coordinate (R) at ( 0.866,  0.5);
      \coordinate (FT) at ( -0.5-0.5, 0.866+1.0);
      \coordinate (FB) at ( +0.5+0.5,-0.866-1.0);
      \draw[-, color=ForestGreen, fill=ForestGreen!17!white] (TL) -- (TR) -- (BR) -- (BL) -- (TL);
      \draw[-Stealth, very thick, color=Purple] (FT) -- (T);
      \draw[-Stealth, very thick, color=Purple] (FB) -- (B);
      \draw[-Stealth, very thick, color=RoyalBlue] (T) -- +( 0.5,-0.866) node[right=0.1 cm] {$p'$};
      \draw[-Stealth, very thick, color=RoyalBlue] (B) -- +(-0.5, 0.866);
    \end{tikzpicture}
    \\
    ~~~~~~Unperturbed material element~~~~~~
    &
    ~~~~~~Deformed in non-principal frame~~~~~~
    &
    ~~~~~~Rotated to principal frame~~~~~~
  \end{tabular}
  \caption{
    Cartoon depiction of forces on a square-shaped material element
    from principal stresses.
    In the left and middle panels,
    the square is misaligned relative to the principal frame,
    so the principal stresses are oblique to the material's surfaces.
    The left panel shows the material before deformation,
    and the middle panel during deformation.
    The right panel shows a material element rotated to align
    with the principal axes of the stress tensor.
  }
  \label{fig:shear:cartoon}
\end{figure}
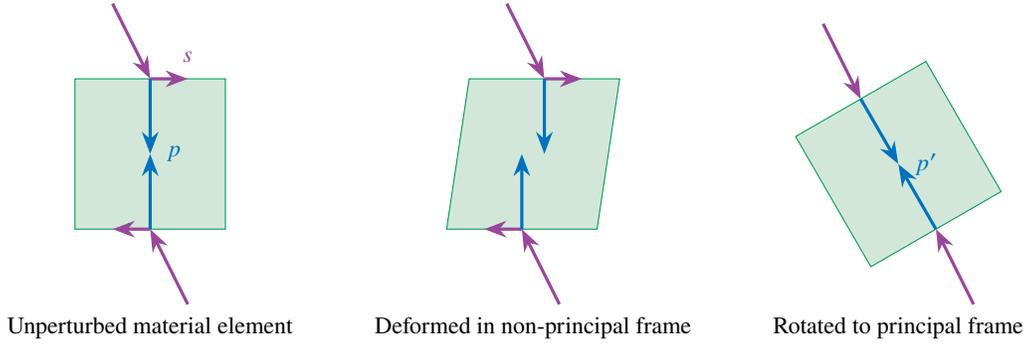

In frames other than the principal frame,
symmetric shear stresses arise precisely because
the principal stresses are oblique to normal surfaces aligned with the frame.
This is depicted in Fig.~\ref{fig:shear:cartoon}.
If a cube of material is aligned with a non-principal frame,
then it is acted on by forces oblique to its surfaces (left panel),
which causes the material to shear (middle panel).
The shearing can be avoided, however,
by rotating the material element to align with the principal axes (right panel).
By diagonalizing the symmetric part of the stress tensor,
we in effect are eliminating shear stresses by
reorienting hypothetical material elements in this manner.
To be sure, this only works for symmetric shear stresses;
we will discuss antisymmetric shear stresses in Sec.~\ref{sec:torsion}.

Since the only non-zero off-diagonal components of the symmetric stress tensor
are $\mathfrak{t}^{r\theta}=\mathfrak{t}^{\theta r}$,
the azimuthal pressure $p_\phi$ is already a principal stress.
To get the other two principal stresses,
we need only find the eigenvalues of the $2\times 2$ matrix:
\begin{align}
  \begin{bmatrix}
    p_r & s \\
    s & p_\theta
  \end{bmatrix}
  \,.
\end{align}
The principal stresses are thus:
\begin{align}
  \label{eqn:eigenpressure}
  p_\pm
  =
  \frac{1}{2}
  \left(
  p_r + p _\theta
  \pm
  \sqrt{ (p_r - p_\theta)^2 + 4s^2 }
  \right)
  \,.
\end{align}
As previously discussed in Ref.~\cite{Freese:2022yur}
in the context of the light front stress tensor,
there is an ambiguity in how the pair of principal stresses is chosen
as functions of space.
One could instead choose
\begin{align*}
  \bar{p}_\pm
  =
  \frac{1}{2}
  \left(
  p_r + p _\theta
  \pm
  (p_r - p_\theta)
  \sqrt{ 1 +\left(\frac{2s}{p_r-p_\theta}\right)^2 }
  \right)
  \,,
\end{align*}
which would have the benefit of reducing to the radial and polar pressures
$p_r$ and $p_\theta$ when $s=0$.
However, the pressures $\bar{p}_\pm$ would be discontinuous when
$p_r - p_\theta = 0$,
whereas $p_\pm$ as defined in Eq.~(\ref{eqn:eigenpressure}) are continuous everywhere.
In any case, $p_+$ and $p_-$ are respectively most like a radial and polar pressure,
so we shall call these the isoradial ($p_+$) and isopolar ($p_-$) pressures.
For the remainder of the text,
we will show results for $p_{\pm}$
and disregard the alternate pair $\tilde{p}_{\pm}$.

Let us call the local principal axes of the symmetric stress tensor
$\hat{e}_\pm$ for the respective principal stresses $p_\pm$.
These are unit vectors, i.e., $\hat{e}_\pm^2=1$.
These can be found by solving:
\begin{align}
  \begin{bmatrix}
    p_r & s \\
    s & p_\theta
  \end{bmatrix}
  \begin{bmatrix}
    \hat{e}_\pm\cdot\hat{r} \\ \hat{e}_\pm\cdot\hat{\theta}
  \end{bmatrix}
  =
  p_\pm
  \begin{bmatrix}
    \hat{e}_\pm\cdot\hat{r} \\ \hat{e}_\pm\cdot\hat{\theta}
  \end{bmatrix}
  \,.
\end{align}
The magnitudes of the components are:
\begin{align}
  \begin{split}
    (\hat{e}_\pm\cdot\hat{r})^2
    &=
    \frac{1}{2} \left(
    1 \pm \frac{p_r - p_\theta}{\sqrt{(p_r-p_\theta)^2 + 4s^2}}
    \right)
    \\
    (\hat{e}_\pm\cdot\hat{\theta})^2
    &=
    \frac{1}{2} \left(
    1 \pm \frac{p_\theta - p_r}{\sqrt{(p_r-p_\theta)^2 + 4s^2}}
    \right)
  \end{split}
\end{align}
and their relative sign is:
\begin{align}
  \mathrm{sgn}(\hat{e}_\pm\cdot\hat{\theta})
  =
  \pm \mathrm{sgn}(\hat{e}_\pm\cdot\hat{r})
  \mathrm{sgn}(s)
  \,.
\end{align}

\begin{figure}
  \includegraphics[width=\textwidth]{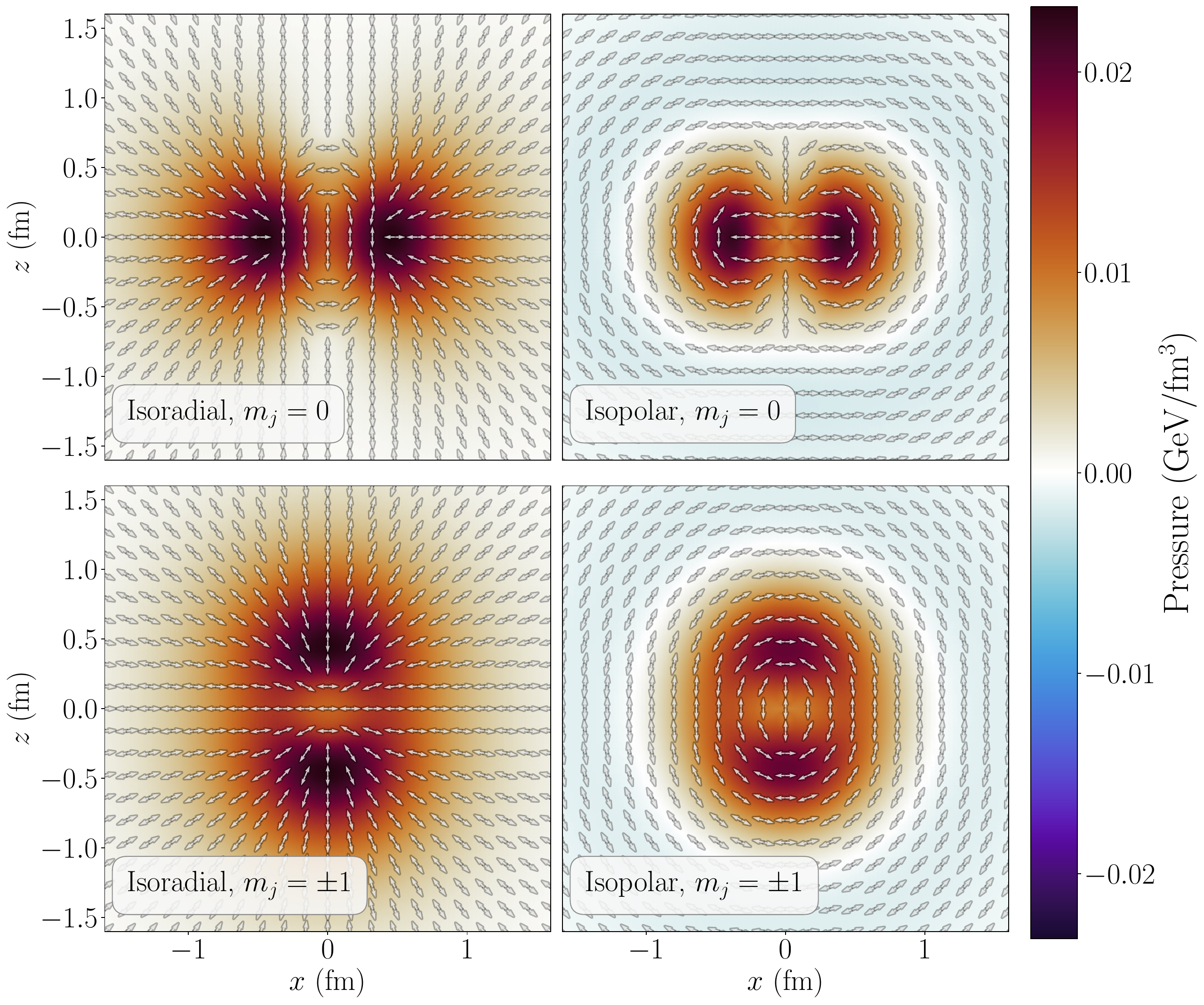}
  \caption{
    The principal stresses and principal axes
    in the deuteron at $y=0$.
    The top row shows the $m_j=0$ state,
    and the bottom row $m_j=\pm1$.
    The left column shows the isoradial pressure
    ($p_+$, with principal axis $\hat{e}_+$)
    and the right column the isopolar pressure
    ($p_-$, with principal axis $\hat{e}_-$).
    This calculation uses the AV18 deuteron wave function~\cite{Wiringa:1994wb}
    and the meson dominance nucleon form factors of
    Broniowski and Ruiz Arriola~\cite{Broniowski:2025ctl}.
  }
  \label{fig:eigenvectors}
\end{figure}

Numerical results for the principal stresses and principal axes are
shown in Fig.~\ref{fig:eigenvectors}.
A two-dimensional slice at $y=0$ was chosen to aid visualization
of the directions,
and two-sided double-arrows are used to emphasize that
pressure corresponds to balancing forces from opposite directions,
rather than a net force.
(Discussion of a net force from the divergence of the stress tensor
is deferred to Sec.~\ref{sec:force}.)
The principal axes are distorted from the radial and polar directions
to instead follow the non-spherical shape of the deuteron.

One disadvantage of the choice $p_\pm$ for the principal stresses
is that while the principal stresses themselves are continuous,
the principal axes can be discontinuous when
$s=0$---in particular, exhibiting sudden 90 degree turns
along the spin quantization axis and in the equatorial plane.
The shear stress becoming zero in place places occurs
by virtue of the factor $\sin\theta \cos\theta$.
Discontinuities in the principal axes occur in the $m_j=0$ state specifically,
as is apparent in the top row of Fig.~\ref{fig:eigenvectors}.

\begin{figure}
  \includegraphics[width=\textwidth]{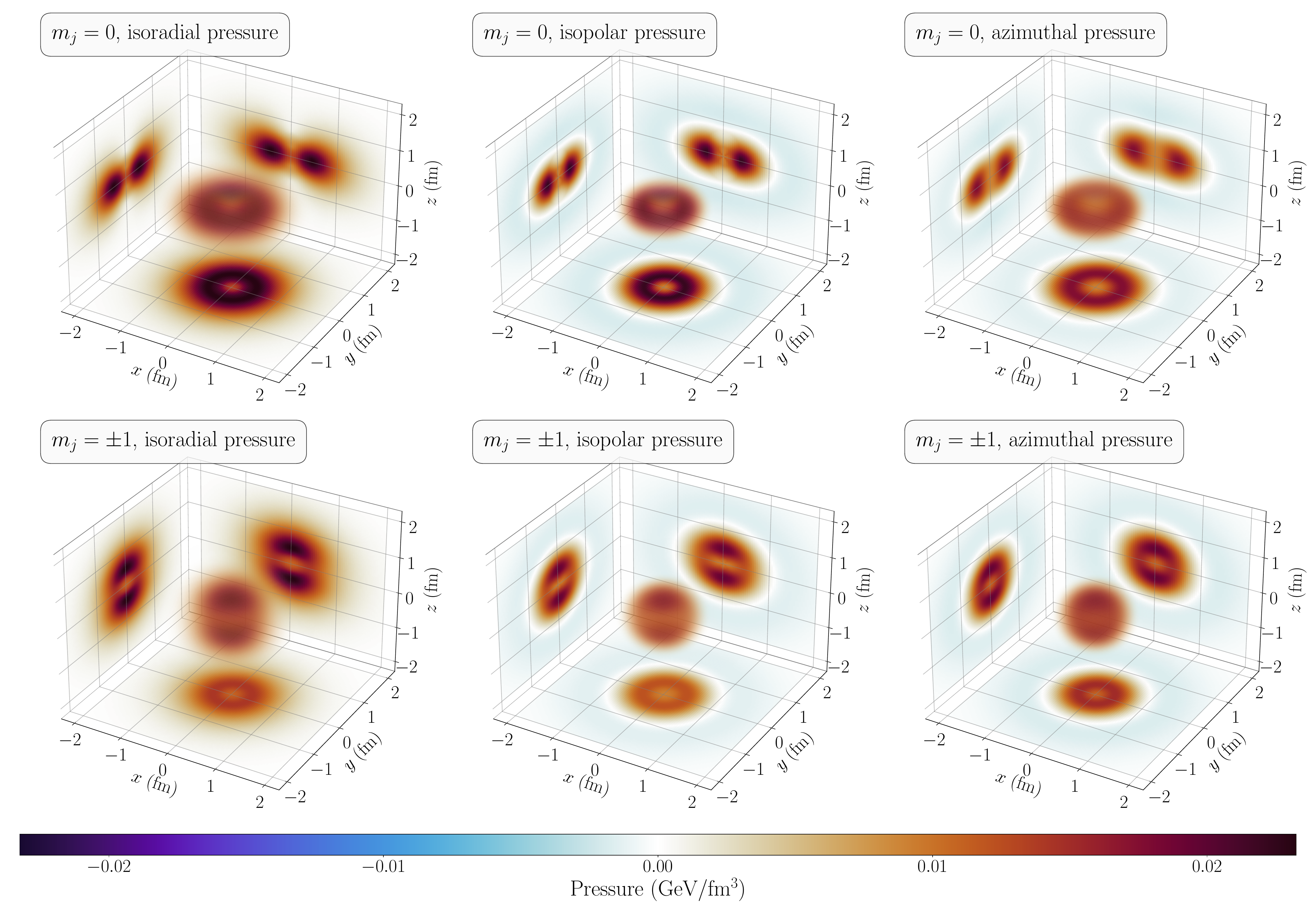}
  \caption{
    Three-dimensional principal stress distributions in the deuteron.
    The top row is for the $m_j=0$ state,
    and the bottom row for $m_j=\pm1$.
    The columns from left to right depict the isoradial,
    isopolar and azimuthal pressures;
    see Fig.~\ref{fig:eigenvectors} for a visualization
    of the isoradial and isopolar directions.
    The images on the walls of the plots are slices of the pressure distributions at
    $x=0$ (left wall), $y=0$ (back wall) and $z=0$ (floor).
    This calculation uses the AV18 deuteron wave function~\cite{Wiringa:1994wb}
    and the meson dominance nucleon form factors of
    Broniowski and Ruiz Arriola~\cite{Broniowski:2025ctl}.
  }
  \label{fig:pressure}
\end{figure}

Finally, with the meaning and direction of the principal stresses made clear,
we present numerical results for the three principal stresses in the deuteron
in Fig.~\ref{fig:pressure}.
The shapes of the pressure distributions largely follow the shapes of the
mass densities for each polarization state; see Fig.~\ref{fig:mass}.
This can be understood as a consequence of the $D_N(\bd^2)$ form factor
dominating the $D_U(\bd^2)$ and $D_{T1}(\bd^2)$ form factors---both
of which are larger than $D_{T2}(\bd^2)$---%
appearing in the pressures; see Fig.~\ref{fig:D4}
and the discussion around it.
As one can see from the formulas for $D_U(\bd^2)$~(\ref{eqn:DU})
and $D_{T1}(\bd^2)$~(\ref{eqn:DT1}),
the dominating terms effectively smear out the nucleon's internal pressures
by the same probability density that describes the deuteron's mass distribution.

Quite similarly to the nucleon,
where the radial pressure is strictly positive
but the tangential pressures flip from positive to
negative~\cite{Polyakov:2018zvc,Lorce:2018egm},
the deuteron's isoradial pressure is strictly positive
while the isopolar and azimuthal pressures flip sign.
These patterns are also known to hold for large nuclei~\cite{Sauer:1976zzf}.
We remind the reader that a positive pressure corresponds to a tendency
of a system to expand,
which is counteracted in static scenarios by the application of compressive forces;
whereas a negative pressure corresponds to a tendency
of a system to contract,
which is counteracted by pulling
and is commonly otherwise known as tension.
The presence of tangential tension far from the deuteron's center is,
in some respects, qualitatively similar to a liquid drop,
which has a thin layer of tangential surface tension.
In contrast to a liquid drop, however---and much like a nucleon---the
tangential surface tension is extremely diffuse.
This reflects the deuteron having a fuzzy rather than a sharp boundary.


\subsection{Mechanical radius and quadrupole moment}

The mechanical radius of a composite system is typically defined
as the mean-squared radius of the radial pressure distribution~\cite{Polyakov:2018zvc}.
Recalling that the radial pressure is the normal stress in the radial direction,
the mechanical radius can also be interpreted as the radius of the
normal force per unit area over any sphere around the deuteron's center.
Similarly to the mass radius,
the fact that $b^2$ is scalar results in only
the unpolarized radial pressure
$p^{(U)}_r(b)$ contributing to the mechanical radius.
The mechanical radius is independent of deuteron spin ensemble
in the non-relativistic case considered here,
so the resulting formula coincides with the spin-zero and spin-half cases:
\begin{equation}
  \label{eqn:radius:mech}
  \langle r^2 \rangle_{\mathrm{Mech}}
  =
  \frac{
    \int \d^3b\; b^2 p_r^{(U)}(b)
  }{
    \int \d^3b \; p_r^{(U)}(b)
  }
  =
  \frac{
    \int \d^3b \; b^2 \left[\frac{2}{3}s_U(b)+p_U(b)\right]
  }{
    \int \d^3b \; \left[\frac{2}{3}s_U(b)+p_U(b)\right]
  }
  =
  \frac{
    6 D_U(0)
    -
    24M_d^2
    \frac{\d \bar{c}_U(\bd^2) }{ \d \bd^2 }\Big|_{\bd=0}
  }{
    \int_0^\infty \d\bd^2 D_U(\bd^2) + 4M_d^2 \bar{c}_U(0)
  }
\end{equation}
The $\bar{c}_U$ terms are present because our calculation accounts only for one-body contributions.
Using the meson dominance nucleon form factors of Ref.~\cite{Broniowski:2025ctl},
we find the mechanical radius
of the deuteron
to be $\sqrt{\langle r^2 \rangle}_{\mathrm{Mech}} = 1.61$~fm.

\begin{table}
  \renewcommand{\arraystretch}{2.0}
  \caption{
    Comparison of different deuteron radii and quadrupole moments.
    Definitions of the radii are found in Eqs.~(\ref{eqn:massRadius}),
    (\ref{eqn:radius:matter}) and (\ref{eqn:radius:mech}),
    while the quadrupole moments are defined in Eqs.~(\ref{eqn:quad:mass})
    and (\ref{eqn:quad:mech}).
    Note that the mass and matter quadrupole moments are equal
    because nucleons do not carry quadrupole moments.
    The meson dominance nucleon EMT-FFs of Ref.~\cite{Broniowski:2025ctl}
    and the AV18 deuteron wave function~\cite{Wiringa:1994wb}
    were used to obtain these values.
  }
  \begin{tabular}{cccc}
    \toprule
    ~~~~ Radius ~~~
    &
    ~~~ Value (fm) ~~~
    &
    ~~~~ Quadrupole moment ~~~
    &
    ~~~ Value (fm$^2$) ~~~
    \\
    \hline
    $\sqrt{\langle r^2 \rangle}_{\mathrm{Mass}}$   &
    2.04 &
    $Q_{\mathrm{Mass}}$   &
    0.269
    \\
    $\sqrt{\langle r^2 \rangle}_{\mathrm{Matter}}$  &
    1.97 &
    $Q_{\mathrm{Matter}}$   &
    0.269
    \\
    $\sqrt{\langle r^2 \rangle}_{\mathrm{Mech}}$   &
    1.61 &
    $Q_{\mathrm{Mech}}$   &
    0.344
    \\
    \bottomrule
  \end{tabular}
  \label{tab:deuteronRadii}
\end{table}

The mechanical radius obtained here is smaller than
the corresponding mass and matter radii of the deuteron.
This behavior is unexpected, given that for the nucleon the mechanical radius is known to exceed the mass radius.
We note, however, that a consistent inclusion of all interaction contributions
would modify the values of $D_U$ (to include force carrier contributions)
and $\bar{c}_U$ (to be zero).
We therefore expect that incorporating these interaction effects
(pion cloud etc.)
could lead to a larger mechanical radius than the values reported above.
At the same time, the deuteron has different dynamics at play than the proton,
and there is no a priori reason that mechanical radii should exceed mass radii in general.
It will therefore be interesting to see whether our finding that
$\langle r^2 \rangle_{\mathrm{Mech}} < \langle r^2 \rangle_{\mathrm{Mass}}$
persists when exchange current contributions are incorporated.

In addition to a mean squared radius,
a mechanical quadrupole moment can also be also defined.
Similarly to the mechanical radius,
the radial pressure distribution is used to define the mechanical quadrupole moment;
and similarly to the mass quadrupole moment,
we evaluate the mean value of $3z^2-b^2$ for an $m_j=1$ state:
\begin{align}
  \label{eqn:quad:mech}
  Q_{\mathrm{Mech}}
  =
  \frac{
    \int \d^3 b \; (3z^2-b^2) p_r^{(m_j=1)}(\bm{b})
  }{
    \int \d^3 b \; p_r^{(m_j=1)}(\bm{b})
  }
  =
  \frac{
    \frac{6}{5 M_d^2}
    \int_0^\infty \d\bd^2 \, D_{T1}(\bd^2)
    +
    4 \bar{c}_{T1}(0)
    +
    \frac{4}{5}
    \left(
    D_{T2}(0)
    -
    4 M_d^2 \frac{\d \bar{c}_{T2}(\bd^2) }{ \d \bd^2 }\Big|_{\bd=0}
    \right)
  }{
    \int_0^\infty \d\bd^2 \, D_U(\bd^2)
    +
    4 M_d^2 \bar{c}_U(0)
  }
  \,.
\end{align}
As when we examined the mechanical radius,
the denominator only receives contributions from the unpolarized part of $p_r(\bm{b})$.
By contrast, the numerator receives only contributions from the tensor-polarized part---and
at that, is numerically dominated by contributions from $D_{T1}(\bd^2)$ in particular.
Using the AV18 wave function
and the meson dominance nucleon form factors of Ref.~\cite{Broniowski:2025ctl},
we find the mechanical quadrupole moment to be $Q_{\mathrm{Mech}} = 0.344~\mathrm{fm}^2$.
This is a fair amount larger than the electric
and mass quadrupole moment for the AV18 wave function, $0.269~\mathrm{fm}^2$.
The quadrupole moments are summarized in Table~\ref{tab:deuteronRadii}
along with the corresponding radii.


\subsection{Justification for stress tensor interpretation}

While the spatial components $T^{ij}$ of the EMT are commonly interpreted as quantifying
mechanical stresses in hadrons~\cite{Polyakov:2002yz,Polyakov:2018zvc,Burkert:2018bqq,Shanahan:2018nnv,Lorce:2018egm,Kumericki:2019ddg,Freese:2021czn,Burkert:2021ith,Panteleeva:2021iip,Pefkou:2021fni,Lorce:2021xku,Duran:2022xag,Burkert:2023wzr,Li:2024vgv,Hu:2024edc,Lorce:2025oot,Broniowski:2025ctl}.
skepticism has recently been expressed about this interpretation~\cite{Ji:2021mfb,Ji:2025gsq,Ji:2025qax}.
It is thus prudent to take a moment to justify our interpretation of
the densities we have obtained as constituting genuine stresses within the deuteron.

The energy-momentum tensor of QCD can be derived in a variety of ways,
but especially pertinent to the current discussion is the derivation through
local spacetime translations using Noether's second theorem.
By a local spacetime translation, we mean a reconfiguration of the quark and gluon fields
in which the fields are translated by different amounts $\xi^\mu(x)$ at every spacetime point,
but in which spacetime itself is not transformed.
The asymmetric, gauge-invariant EMT of QCD was derived in Ref.~\cite{Freese:2025glz}
using local spacetime translations alone,
while Ref.~\cite{BarroseSa:2025uxe}
shows that supplementing local translations with an internal transformation
of the vierbein allows one to obtain a symmetric EMT.
In either case, the change of the QCD action under such a transformation is\footnote{
  Our sign differs from Ref.~\cite{Freese:2025glz}
  because we consider an active local translation,
  whereas Ref.~\cite{Freese:2025glz}
  considers a passive local translation.
}:
\begin{align}
  \delta_\xi S_{\mathrm{QCD}}
  =
  -
  \int \d^4x \,
  T^{\mu\nu}_{\mathrm{QCD}}(x)
  \partial_\mu \xi_\nu(x)
  \,,
\end{align}
where
$T^{\mu\nu}_{\mathrm{QCD}}(x)$
is asymmetric under local translations alone~\cite{Freese:2025glz}
and symmetric under a combined local translation and vierbein rotation~\cite{BarroseSa:2025uxe}.

Under a local, time-independent spatial translation
(i.e., a static spacetime transformation), the Lagrangian transforms as:
\begin{align}
  \delta_\xi L_{\mathrm{QCD}}
  =
  \int \d^3x \,
  T^{ij}_{\mathrm{QCD}}(x)
  \partial_i \xi^j(x)
  \,.
\end{align}
In general, the quantity $\partial_i \xi^j(x)$ can be decomposed into symmetric
and antisymmetric pieces under exchange of indices:
\begin{align}
  \begin{split}
    \partial_i \xi^j(x)
    &=
    E_{ij}(x)
    +
    R_{ij}(x)
    \\
    E_{ij}(x)
    &=
    \frac{1}{2}
    \big(
    \partial_i \xi^j(x)
    +
    \partial_j \xi^i(x)
    \big)
    \\
    R_{ij}(x)
    &=
    \frac{1}{2}
    \big(
    \partial_i \xi^j(x)
    -
    \partial_j \xi^i(x)
    \big)
    \,.
  \end{split}
\end{align}
Crucially, $E_{ij}(x)$ is, to linear order, the strain tensor of classical continuum
mechanics~\cite{fetter2003theoretical,chung2007general,irgens2008continuum,jog2015continuum},
while $R_{ij}(x)$ is a local rigid rotation~\cite{fetter2003theoretical,irgens2008continuum}.
If the symmetric EMT is used, the antisymmetric tensor $R_{ij}(x)$
contracts with the stress tensor to zero.
If the change in the Lagrangian is attributed to an effective
potential energy\footnote{
  The division of energy into kinetic and potential energy depends on the
  degrees of freedom under consideration.
  Random motion of microscopic constituents may furnish kinetic energy
  at a fundamental level,
  but could be considered (for instance) as heat or chemical potential energy
  at a coarse-grained level.
  The effective potential energy we define---and thus the QCD stress tensor---%
  generally includes quark and gluon motion,
  in the same way that air and water pressure includes random molecular motion.
},
we obtain:
\begin{align}
  \delta_\xi
  \mathcal{U}_{\mathrm{eff}}
  =
  -
  T^{ij}_{\mathrm{sym}}(x)
  E_{ij}(x)
  \,,
\end{align}
which is exactly the form of the strain energy density
of a classical system under small deformations~\cite{fetter2003theoretical,chung2007general,irgens2008continuum}.
This justifies interpreting the symmetric part of
$T^{ij}_{\mathrm{QCD}}(x)$ as a genuine stress tensor---%
it quantifies how the effective potential energy changes
in response to deformations of the quark and gluon fields,
and thus the work necessary to effect these deformations.

The antisymmetric part of $T^{ij}_{\mathrm{QCD}}(x)$, on the other hand,
does not have a strict classical analogy.
The classical stress tensor is provably symmetric
under the joint assumptions
of linear momentum and angular momentum conservation~\cite{chung2007general,jog2015continuum}.
However, intrinsic spin is a strictly quantum mechanical phenomenon
that is not accounted for in these proofs.
The classical definition of angular momentum, $\bm{L} = \bm{r}\times\bm{p}$,
only accounts for orbital angular momentum and is not generally conserved
when particles with intrinsic spin are present.
An antisymmetric stress could exist in quantum mechanical systems
by transforming orbital angular momentum into spin and vice-versa.
We shall consider this next, and show that the antisymmetric part
of the deuteron's stress tensor does precisely this.


\subsection{Torsion stress in the asymmetric stress tensor}
\label{sec:torsion}

The asymmetric stress tensor
cannot be diagonalized by a local change of frame.
Broadly speaking, this is a consequence of orthogonal transformations
not mixing symmetric and antisymmetric matrices.
In the case at hand, the shear stresses appear only in the $(r,\theta)$
components of the stress tensor;
see Eq.~(\ref{eqn:stress:spherical}).
Antisymmetric $2\times2$ matrices are invariant under
rotations in the plane,
so the antisymmetric part of the stress tensor is the same whether
using a spherical frame or the principal axes
of the symmetric part of the stress tensor.

\begin{figure}
  \begin{center}
    \begin{tikzpicture}[scale=0.76]
      \coordinate (O) at (0,0);
      \draw[-stealth](O) -- +(0,6) node[below left] {\large$z$};
      \draw[-stealth](O) -- +(6,0) node[below left] {\large$x$};
      \coordinate (c) at (3,3);
      \draw[very thick, color=RoyalBlue, fill=RoyalBlue!11!white](c) circle (1);
      \coordinate (fn) at (3+0,3+1);
      \coordinate (fe) at (3+1,3+0);
      \coordinate (fs) at (3+0,3-1);
      \coordinate (fw) at (3-1,3+0);
      \coordinate (fne) at (3+0.7,3+0.7);
      \coordinate (fnw) at (3-0.7,3+0.7);
      \coordinate (fse) at (3+0.7,3-0.7);
      \coordinate (fsw) at (3-0.7,3-0.7);
      \draw[-Stealth, thick, color=Orange] (fn) -- +(-2, 0);
      \draw[-Stealth, thick, color=Orange] (fe) -- +( 0, 2);
      \draw[-Stealth, thick, color=Orange] (fs) -- +( 2, 0);
      \draw[-Stealth, thick, color=Orange] (fw) -- +( 0,-2);
      \draw[-Stealth, thick, color=Orange] (fne) -- +(-1.4, 1.4);
      \draw[-Stealth, thick, color=Orange] (fnw) -- +(-1.4,-1.4);
      \draw[-Stealth, thick, color=Orange] (fse) -- +( 1.4, 1.4);
      \draw[-Stealth, thick, color=Orange] (fsw) -- +( 1.4,-1.4);
    \end{tikzpicture}
  \end{center}
  \caption{
    A depiction of torsion stresses arising from the
    antisymmetric part of the stress tensor.
    A hypothetical surface is depicted in blue,
    and momentum fluxes at points along the surface
    for $\tau > 0$ (right-handed twisting) are depicted in orange.
    The momentum flux associated with this contribution
    is always orthogonal to the surface normal,
    and is vortical in the sense that it circles
    around the surface of a material element.
  }
  \label{fig:torsion:cartoon}
\end{figure}
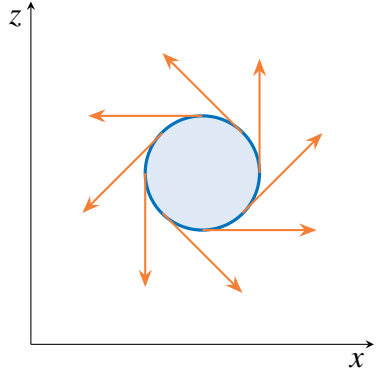

To make sense of the antisymmetric shear stress,
consider its interpretation as a momentum flux density.
Given a plane with normal $\hat{n}$,
the momentum flux through the plane
due to the antisymmetric part of the stress tensor
is:
\begin{align}
  f_p^j
  =
  \Big(
  (\hat{n}\cdot\hat{\theta}) \hat{r}^j
  -
  (\hat{n}\cdot\hat{r}) \hat{\theta}^j
  \Big)
  \tau
  =
  (\hat{\phi} \times \hat{n})^j
  \tau
  \,,
\end{align}
where $\tau_{m_j=0} = -\frac{1}{3} \tau^{(T)}$
and $\tau_{m_j=\pm1} = +\frac{1}{6} \tau^{(T)}$.
The momentum flux is always orthogonal to the surface normal,
and will tend to circle around any material element in a vortical fashion;
see Fig.~\ref{fig:torsion:cartoon} for a cartoon depiction of
such stresses on a circular material element.
A stress of this kind can be induced by twisting an object,
and will induce torsion strain in the object;
accordingly, the stress can be called torsion stress.

The sign of of the torsion stress $\tau$
corresponds to the direction of the twisting.
In spherical coordinates, $\tau > 0$ corresponds to twisting
as the fingers on one's right hand would curl
if the thumb were pointed in the $\hat{\phi}$ direction,
whereas $\tau < 0$ corresponds to twisting
as the fingers on the left hand would curl instead.
\begin{figure}
  \includegraphics[width=\textwidth]{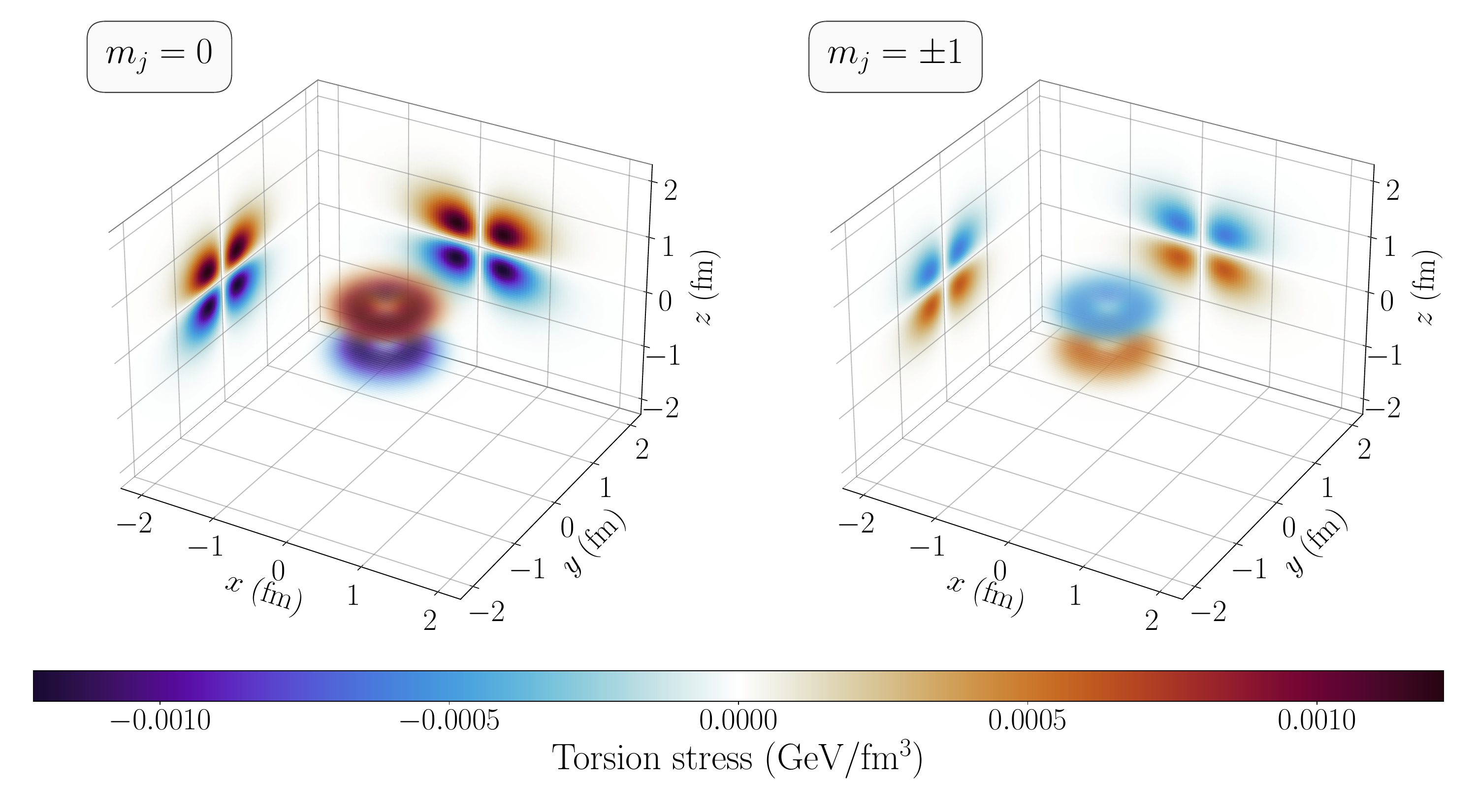}
  \caption{
    Torsion distribution in a deuteron in an $m_j=0$ state (left panel)
    and an $m_j=\pm1$ state (right panel).
    The images on the walls of the plots are slices of the torsion distribution at
    $x=0$ (left wall), $y=0$ (back wall) and $z=0$ (floor),
    although the last is identically zero.
    This calculation uses the AV18 deuteron wave function~\cite{Wiringa:1994wb}
    and a dipole parametrization for $S_N(\bd^2)$.
  }
  \label{fig:torsion}
\end{figure}

Numerical results for the torsion stress are shown in Fig.~\ref{fig:torsion}.
Up to a relative factor $-2$, the torsion is the same for both $m_j=0$
and $m_j=\pm1$ states,
and accordingly does not follow the shape of the mass distribution.
However---much as we saw with the momentum and mass flux densities---this
can occur because only nucleons in certain positions may be experiencing torsion.

In the deuteron, torsion is caused by spin-dependent forces that reorient
the nucleons' (and thus their quarks') spin directions.
We already saw in Eq.~(\ref{eqn:sbar:2}) that the $\bar{s}(\bd^2)$ form factor
arises from spin-dependent parts of the potential,
but the identification of torsion with spin reorientation can be made more
manifest by dissecting how exactly it arises from
the nucleon matrix element~(\ref{eqn:mff:half}).
The antisymmetric piece of the nucleon matrix element is:
\begin{align}
  -
  \frac{
    i
    (\bd\times\bm{\sigma}_{s's})^{[i}
    P_N^{j]}
  }{2m_N}
  S_N(\bd^2)
  =
  -
  \frac{
    i
    (\bd\times\bm{\sigma}_{s's})^{[i}
    P_d^{j]}
  }{4m_N}
  S_N(\bd^2)
  -
  \frac{
    i
    (\bd\times\bm{\sigma}_{s's})^{[i}
    k^{j]}
  }{2m_N}
  S_N(\bd^2)
  \,,
\end{align}
with the first term on the right-hand side contributing to
$S(\bd^2)$ and the second to $\bar{s}(\bd^2)$.
One can use the identities $(\bm{A}\times\bm{B})^i = \epsilon_{ijk} A^j B^k$
and $A^{[i} B^{j]} = \epsilon_{ijk} (\bm{A}\times\bm{B})^k$
to rewrite the term appearing in $\bar{s}(\bd^2)$ as:
\begin{align*}
  - \frac{i \epsilon_{ijk}}{2m_N}
  \Big(
  \sigma_{s's}^k (\bd\cdot\bm{k})
  -
  \dl^k (\bm{\sigma}_{s's}\cdot\bm{k})
  \Big)
  \,.
\end{align*}
The second term in this quantity is conserved,
as it is orthogonal to the momentum transfer $\bd$,
and accordingly cannot contribute to $\bar{s}(\bd^2)$.
When jumping from momentum space to coordinate space,
one makes the substitutions
$\bm{k} \rightarrow -\frac{i}{2} \blrn$
and
$\bd \rightarrow 2i \bm{\nabla}$,
with the former placed between $\psi_d^\dagger$ and $\psi_d$
and the latter outside the entire expression.
The contribution to $\bar{s}(\bd^2)$ thus takes the form:
\begin{align*}
  - \frac{i \epsilon_{ijk}}{2m_N}
  \nabla_l \Big[
    \psi_d^\dagger(\bm{r})
    \sigma^k
    \big(
    \nabla_l
    \psi_d(\bm{r})
    \big)
    -
    \big(
    \nabla_l
    \psi_d^\dagger(\bm{r})
    \big)
    \sigma^k
    \psi_d(\bm{r})
    \Big]
  =
  - \frac{i \epsilon_{ijk}}{2m_N}
  \Big(
    \psi_d^\dagger(\bm{r})
    \sigma^k
    \big(
    \nabla^2
    \psi_d(\bm{r})
    \big)
    -
    \big(
    \nabla^2
    \psi_d^\dagger(\bm{r})
    \big)
    \sigma^k
    \psi_d(\bm{r})
    \Big)
  \,.
\end{align*}
The Schr\"{o}dinger equation (\ref{eqn:schrodinger})
allows this to be rewritten:
\begin{align*}
  \frac{i \epsilon_{ijk}}{2}
  \psi_d^\dagger(\bm{r})
  [ \sigma^k, V(\bm{r},\bm{\sigma}_p,\bm{\sigma}_n) ]
  \psi_d(\bm{r})
  \,,
\end{align*}
and since the kinetic energy commutes with the Pauli matrices,
this can at last be written:
\begin{align*}
  \epsilon_{ijk}
  \psi_d^\dagger(\bm{r})
  i [\hat{H}, \varSigma^k_N ]
  \psi_d(\bm{r})
  \,,
\end{align*}
where $\varSigma^k = \frac{1}{2}\sigma^k_N$ is the spin operator of the
probed nucleon.
In the Heisenberg picture,
$i[\hat{H}, \varSigma^k_N] = \dot{\varSigma}^k_N$,
so this operator in effect quantifies a rate of change for the nucleon spin.
The $\bar{s}(\bd^2)$ form factor---and torsion stress in the
deuteron---thus indeed arises from reorientation of nucleon spin.
Moreover, this form factor can receive contributions only from terms in the
nucleon-nucleon potential that fail to commute with the nucleon spin
operator---that is, from spin-dependent forces,
such as the tensor force and spin-orbit coupling.

Since torsion stress arises from spin reorientation by spin-dependent forces,
we should expect torsion to be present only where said forces are in fact
inducing spin flips.
It thus makes sense that the torsion would be present at the boundary of
where the D-wave probability is large,
and where the interference between the S and D waves switches from
constructive to destructive;
compare to Fig.~\ref{fig:interference}.
Moreover, the spin state of a nucleon can be represented as a point on a sphere%
---the Bloch sphere~\cite{Nielsen:2012yss}---%
with the point corresponding to the nucleon spin direction.
The evolution of quantum states is continuous\footnote{
  State reduction during measurement may be the sole exception to this,
  but not all formulations of quantum mechanics have discontinuous state reduction;
  see the many worlds~\cite{Everett:1957hd,Saunders:2012zz}
  and pilot wave~\cite{db:pilot,Bohm:1951xw,Bohm:2006und}
  interpretations for instance.
  In any case, an isolated deuteron in its ground state is not being
  subjected to measurement.
},
so a nucleon can only transition
from pointing up along the $z$-axis
(as it might in the S-wave for an $m_j=+1$ state)
to pointing down
(as it would in the D-wave for the same $m_j=+1$ state)
through a continuous path on the Bloch sphere.
Our result for the torsion stress tells us that this spin reorientation
between the S- and D-waves
happens in the $(r,\theta)$ plane.


\subsection{Forces felt by subcomponents}
\label{sec:force}

For a continuum system, the net force density on any material element
is a sum of forces from adjacent elements---given by
$-\nabla_i \mathfrak{t}^{ij}$---and an external body force density $f^j$.
For a static system, the net force must be zero, entailing:
\begin{align}
  f^j(\bm{b})
  =
  \nabla_i
  \mathfrak{t}^{ij}(\bm{b})
  \,,
\end{align}
and thus providing a means of determining the external force density.
This is commonly known as the Cauchy momentum equation,
and sometimes as Cauchy's first law
of motion~\cite{chatterjee1999mathematical,irgens2008continuum}.
It has been explored in
Refs.~\cite{Polyakov:2018exb,Won:2023cyd,Kim:2025iis}
as a way of mapping out the average force felt by quarks in a hadron,
and was shown by Ref.~\cite{Freese:2024rkr} to successfully
reproduce the Coulomb force law when applied to the hydrogen atom.
The application of the Cauchy momentum equation to spin-one systems
was thoroughly explored by Kim and Kim~\cite{Kim:2025iis},
so we borrow their formalism to investigate force distributions in the deuteron.

One matter to bear in mind throughout this discussion is that
since nucleons have finite extent%
---as encoded by the use of realistic EMT-FFs---%
we will be obtaining force distributions
\emph{within} the nucleon owing to the nucleon-nucleon force.
At the same time, all parton flavors are being summed over,
so these are not internal forces, but an external force---again,
from the inter-nucleon interaction---distributed over quarks and gluons in the nucleon.
It is also crucial to recall that these forces are exerted by force carriers
(e.g., pions at long distances),
and that---by Newton's third law---the force exerted \emph{on} these force
carriers by the nucleons must be equal and opposite to the forces we will map out.
This is effectively equivalent to saying that the $\bar{c}$ and $\bar{s}$
form factors must identically vanish if we include contributions from
the force carriers,
and that the force distributions we will find are non-zero
because the nucleons are an open subsystem of the deuteron.

Bearing this in mind,
we use Eq.~(23) of Ref.~\cite{Kim:2025iis} as a starting point
to decompose the force density:
\begin{align}
  \label{eqn:force:breakdown}
  f^j_{s's}(\bm{b})
  =
  \epi_a
  \epf_b
  \left\{
    \delta^{ab}
    Y_1^j(\hat{b})
    f_0(b)
    +
    Q^{jlab}
    Y_1^l(\hat{b})
    f_2(b)
    +
    Q^{klab}
    Y_3^{jkl}(\hat{b})
    f_3(b)
    \right\}
  \,.
\end{align}
Taking the divergence of the internal stress tensor, as given in
Eq.~(\ref{eqn:density:stress}), results in:
\begin{multline}
  f^j_{s's}(\bm{b})
  =
  i
  M_d
  \epi_a
  \epf_b
  \int \frac{\d^3 \dl}{(2\pi)^3}
  |\bd|
  \Bigg\{
    \delta^{ab}
    Y_1^j(\hat{\dl})
    \bar{c}_U(\bd^2)
    -
    Q^{klab} Y_3^{jkl}(\hat{\dl})
    \frac{\dl^2}{2M_d^2}
    \Big( \bar{c}_{T1}(\bd^2) - \bar{s}(\bd^2) \Big)
    \\
    +
    Q^{jlab} Y_1^j(\hat{\dl})
    \left[
      \left(
      \bar{c}_{T2}(\bd^2)
      -
      \frac{\dl^2}{2M_d^2} \bar{s}(\bd^2)
      \right)
      -
      \frac{\dl^2}{5M_d^2}
      \Big( \bar{c}_{T1}(\bd^2) - \bar{s}(\bd^2) \Big)
      \right]
    \Bigg\}
  \e^{-i\bm{b}\cdot\bd}
  \,.
\end{multline}
All of the angular dependence in the integrand is contained in harmonic tensors,
making it straightforward to turn the Fourier transform into a collection
of Bessel transforms.
The functions in the force breakdown (\ref{eqn:force:breakdown}) can thus be written:
\begin{align}
  \label{eqn:f023}
  \begin{split}
    f_0(b)
    &=
    \frac{M_d}{2\pi^2}
    \int_0^\infty \d \dl \,
    \dl^3
    \bar{c}_U(\dl^2)
    j_1(b \dl)
    \\
    f_2(b)
    &=
    \frac{M_d}{2\pi^2}
    \int_0^\infty \d \dl \,
    \dl^3
    \left[
      \left(
      \bar{c}_{T2}(\bd^2)
      -
      \frac{\dl^2}{2M_d^2} \bar{s}(\bd^2)
      \right)
      -
      \frac{\dl^2}{5M_d^2}
      \Big( \bar{c}_{T1}(\bd^2) - \bar{s}(\bd^2) \Big)
      \right]
    j_1(b \dl)
    \\
    f_3(b)
    &=
    \frac{M_d}{2\pi^2}
    \int_0^\infty \d \dl \,
    \frac{\dl^5}{2 M_d^2}
    \Big( \bar{c}_{T1}(\bd^2) - \bar{s}(\bd^2) \Big)
    j_3(b \dl)
    \,.
  \end{split}
\end{align}
The peculiar linear combinations
$\bar{c}_{T1}(\bd^2) - \bar{s}(\bd^2)$
and
$\bar{c}_{T2}(\bd^2) - \frac{\bd^2}{2M_d^2} \bar{s}(\bd^2)$
in effect amount to replacing $J_N(\bd^2) \rightarrow J_N(\bd^2) - S_N(\bd^2)$
in the formulas for $\bar{c}_{T1}(\bd^2)$ [Eq.~(\ref{eqn:cT1})]
and $\bar{c}_{T2}(\bd^2)$ [Eq.~(\ref{eqn:cT2})].
This occurs because, when using the asymmetric EMT,
quark spin is not counted towards the momentum density.
Since force constitutes a rate of change of momentum, changes in quark spin
likewise do not count towards the force distribution
when using the asymmetric EMT.

From here, it is most instructive---as it was with the other densities we
considered---to look separately at unpolarized and tensor-polarized ensembles.
As before, the force density in pure states can be reconstructed through
appropriate linear combinations of these.

For an unpolarized ensemble, only $f_0(b)$ survives,
and the force density is given by:
\begin{align}
  f_U^j(\bm{b})
  =
  \hat{b}^j
  f_0(b)
  \,,
\end{align}
which is a central force.
An especially instructive case to consider is pointlike nucleons,
for which $A_N(\bd^2) = 1$ and $\bar{c}_N(\bd^2) = 0$.
If we use the expression (\ref{eqn:cbar:potential}) for $\bar{c}_U(\bd^2)$,
but without the sum over $N$
(in order to count the force felt by a single nucleon),
then with the aid of
Eq.~(\ref{eqn:f023})
and the identity
\begin{align}
  \int_0^\infty \d \dl \,
  \dl^2
  j_l(\dl b)
  j_l\left(\frac{\dl r}{2}\right)
  =
  \frac{\pi}{b^2}
  \delta(r-2b)
  \,,
\end{align}
we find:
\begin{align}
  f_U^j(\bm{b})
  =
  -
  \frac{1}{4\pi r^2}
  \Big(
  V_c'(r)
  u^2(r)
  +
  V_w'(r)
  w^2(r)
  +
  4 \sqrt{2}
  V_t'(r)
  u(r) w(r)
  \Big)
  \bigg|_{r=2b}
  \,,
\end{align}
where the potential functions are defined in Eqs.~(\ref{eqn:potential})
and (\ref{eqn:uw:diff}).
For a finite-size nucleon, this force density must be additionally smeared
through a convolution with the Fourier transform of $A_N(\bd^2)$.
Since
$\frac{1}{4\pi r^2} u^2(r)$
and
$\frac{1}{4\pi r^2} w^2(r)$
give the probability densities of the S- and D-waves,
it would seem that $-V_c'(r)$ and $-V_w'(r)$ give the central force experienced
by a nucleon in the S- and D-wave, respectively---at least,
when polarization states are averaged over.
There is an additional interference force proportional to $-V'_t(r)$,
which is manifestly quantum mechanical
and cannot be cleanly attributed to the nucleon being in
a particular orbital state.

For a tensor-polarized ensemble, the force distribution is:
\begin{align}
  f_T^j(\bm{b})
  =
  \left(
  3
  \cos^2\theta
  -
  1
  \right)
  \left[
    f_2(b)
    +
    \frac{3}{5}
    f_3(b)
    \right]
  \hat{b}^j
  -
  3
  \sin\theta \cos\theta
  \left[
    f_2(b)
    -
    \frac{2}{5}
    f_3(b)
    \right]
  \hat{\theta}^j
  \,.
\end{align}
The central part reproduces Eq.~(31a) of Ref.~\cite{Kim:2025iis},
and the polar part contains a term $-\frac{2}{5}f_3(b)$
missing from their Eq.~(31b)\footnote{
  The authors of Ref.~\cite{Kim:2025iis}
  have confirmed by private communication that
  the $-\frac{2}{5}f_3(b)$ should be present.
}.

\begin{figure}
  \includegraphics[width=\textwidth]{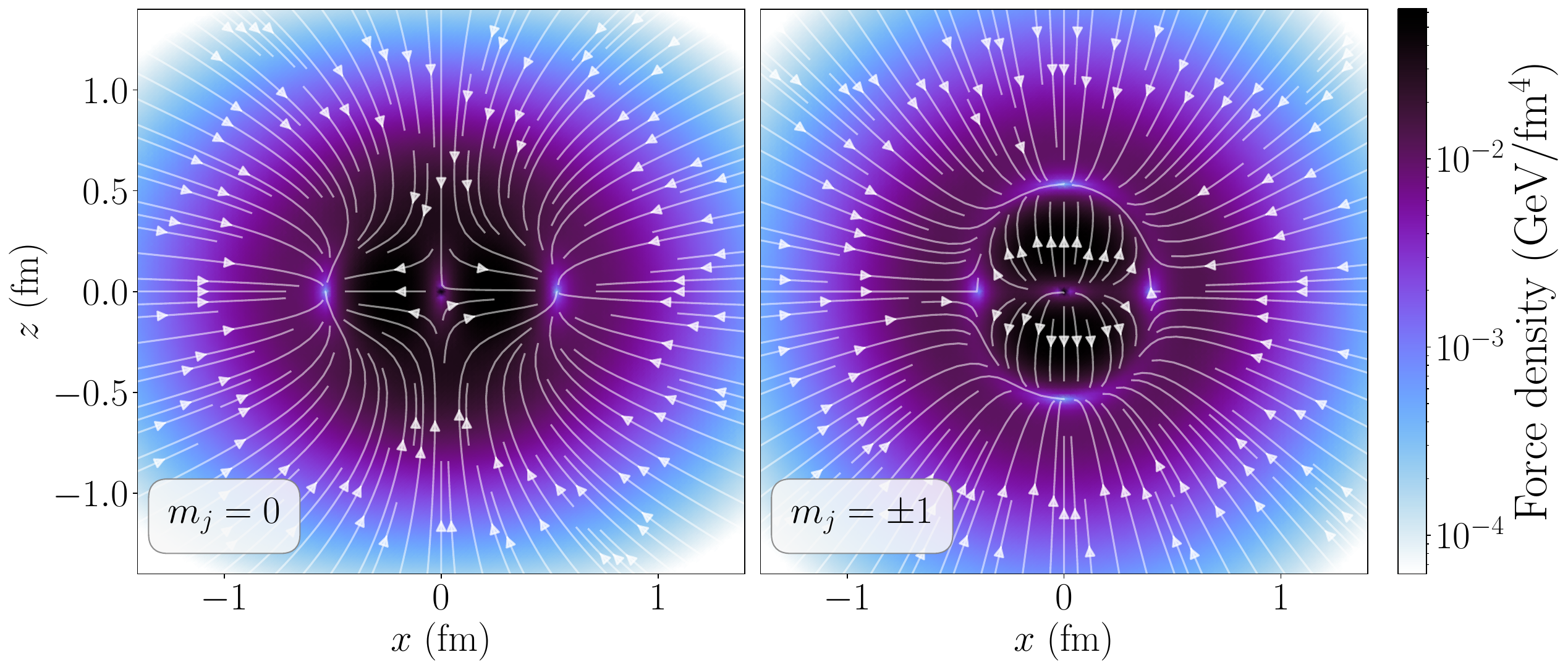}
  \caption{
    The force distribution within nucleons in the deuteron.
    The color indicates the magnitude of the force on a logarithmic scale,
    while white semitransparent streamlines signify the direction of the force.
    This calculation uses the AV18 deuteron wave function~\cite{Wiringa:1994wb},
    along with the meson dominance nucleon form factors of
    Broniowski and Ruiz Arriola~\cite{Broniowski:2025ctl}
    and a dipole parametrization for $S_N(\bd^2)$.}
  \label{fig:force}
\end{figure}

With the unpolarized and tensor-polarized force distributions in hand,
we can reconstruct force densities for pure states through:
\begin{align}
  \begin{split}
    f^j_{m_j=0}(\bm{b})
    &=
    f^j_U(\bm{b})
    -
    \frac{1}{3}
    f^j_T(\bm{b})
    \\
    f^j_{m_j=\pm1}(\bm{b})
    &=
    f^j_U(\bm{b})
    +
    \frac{1}{6}
    f^j_T(\bm{b})
    \,.
  \end{split}
\end{align}
Numerical results for these force distributions is shown in Fig.~\ref{fig:force}.
The qualitative story being told by these images is interesting.
As expected, there is a short-range repulsion close to the center of the deuteron,
and attraction further out.
At some intermediate distance, around $\sim0.5$~fm from the center,
the forces become predominantly polar, and are directed towards
either the equator in the $m_j=0$ case or the poles for $m_j=\pm1$.
These are of course the regions where the probability and the mass are
largely concentrated; compare Fig.~\ref{fig:mass}.


\section{Summary and outlook}
\label{sec:end}

In this work, we calculated the one-body contributions to
the eleven form factors of the deuteron's asymmetric energy-momentum tensor.
Since two-body contributions have been left out,
the four non-conserved form factors
($\bar{c}_U$, $\bar{c}_{T1}$, $\bar{c}_{T2}$ and $\bar{s}$)
are non-zero.
However, rather than constituting a flaw,
these form factors allow us to
map out the distribution of the inter-nucleon force within the nucleon
through the Cauchy momentum equation.

Six of the eleven form factors
($A_U$, $A_T$, $J$, $D_U$, $D_{T1}$ and $D_{T2}$)
have been calculated previously~\cite{Freese:2022yur,He:2023ogg,He:2024vzz,Panteleeva:2024abz}.
We have compared our results to prior works, finding mixed agreement.
Several discrepancies can be explained by
the use of different deuteron wave functions,
or by the use of ad hoc substitutions made in prior calculations.
However, all existing calculations of $D_{T2}$ differ,
likely owing to this form factor's sensitivity to fine details of dynamics.
The other five form factors
($\bar{c}_U$, $\bar{c}_{T1}$, $\bar{c}_{T2}$, $S$ and $\bar{s}$)
were calculated for the first time in this work.

From the form factors, we obtained spatial distributions of mechanical properties
following the methods developed in Refs.~\cite{Li:2022ldb,Freese:2022fat,Freese:2024rkr,Freese:2025tqd}.
These included mass, mass flux, momentum, stress and force distributions.
The stress distributions were parametrized using the framework of
Polyakov and Sun~\cite{Polyakov:2019lbq},
which was extended to include torsion stress
from the antisymmetric part of the stress tensor.
The symmetric part of the stress tensor was locally diagonalized to obtain principal stresses,
and the torsion stress of the antisymmetric part was shown to induce spin reorientation.
Finally, the framework of Kim and Kim~\cite{Kim:2025iis}
was used to obtain force distributions in the nucleons through the Cauchy momentum equation.

There are several directions that the work here can be extended and built upon.
One of these is the calculation of two-body currents,
which has been considered by He and Zahed~\cite{He:2024vzz}
and Panteleeva \textsl{et al.}~\cite{Panteleeva:2024abz}.
Ideally, the inclusion of these currents will restore local momentum conservation,
so that the non-conserved form factors
($\bar{c}_U$, $\bar{c}_{T1}$, $\bar{c}_{T2}$ and $\bar{s}$)
vanish when these contributions are included.
However, this vanishing requires a complete and self-consistent calculation.
For instance, merely including pion exchange contributions when the potential
includes phenomenological short-distance repulsion cannot be expected to restore
local momentum conservation;
\emph{all} of the interactions must be included in the two-body current for this to happen.
Future research in this direction will likely involve either a simplified model of
inter-nucleon interactions,
or investigation into an effective Lagrangian that can reproduce the AV18 potential.

Another potential avenue for future research is
the calculation of energy and energy flux densities of the deuteron
within a Galilei-covariant framework,
akin to the framework considered for non-relativistic fermions in Ref.~\cite{Freese:2025tqd}.
This would require the calculation of additional form factors,
due to Galilei symmetry being less constraining than Lorentz symmetry.
Since nature is Lorentz covariant rather than Galilei covariant,
these additional form factors are in a sense spurious,
and their size could be used as to estimate the error arising
from the use of non-relativistic quantum mechanics.

It may additionally be interesting to analyze the separate
quark and gluon contributions to the deuteron stress and force distributions.
This would of course require separating the nucleon form factors into
quark and gluon contributions,
and accordingly non-zero $\bar{c}_N^{(q,g)}$ would need to be considered.
This could lead to significant changes to the stress and force distributions,
and a comparative study between a quark-gluon breakdown for proton and deuteron stresses
could be especially enlightening.

One final avenue for future research is the application of the methods
developed here to calculating stress distributions in heavy spin-one quarkonia.
Many of the formulas we have derived can be used as-is,
with the Cornell potential in place of the AV18 potential.
In principal, one could also apply the framework to heavy spin-zero quarkonia
by dropping terms related to the D-wave and to spin.


\makeatletter{
\def\addcontentsline#1#2#3{}%
\def\tocsection#1#2#3{#3}%
\begin{acknowledgments}
  We warmly acknowledge helpful correspondence with
  Fangcheng He, Jun-Young Kim, Julia Panteleeva,
  Jian-Wei Qiu, Peter Schweitzer, Kemal Tezgin, Christian Weiss,
  Robert Wiringa and Ismail Zahed.
  We especially thank Fangcheng He and Julia Panteleeva for providing
  numerical results from their deuteron form factor calculations.
  AF was supported by
  the Center for Nuclear Femtography,
  operated by the Southeastern Universities Research Association
  in Washington, D.C.\ under an appropriation from the Commonwealth of Virginia;
  by the DOE contract No.~DE-AC05-06OR23177,
  under which Jefferson Science Associates, LLC operates Jefferson Lab;
  and by the Scientific Discovery through Advanced Computing (SciDAC) award
  \textsl{Femtoscale Imaging of Nuclei using Exascale Platforms}.
  AS and WC are supported by the National Science Foundation award PHY-2239274.
  Feyncalc~\cite{Shtabovenko:2023idz} was used in the derivation of results.
  This paper was written without the use of AI.
\end{acknowledgments}
}
\makeatother


\makeatletter{
\def\addcontentsline#1#2#3{}%
\def\tocsection#1#2#3{#3}%
\section*{Data availability}

The data and code used to produce the numerical results in this work are
publicly available in the open-source \textbf{deupack}
repository~\cite{Freese_deupack_2026}.
The deuteron EMT-FF results of past
works~\cite{Freese:2022yur,He:2024vzz,Panteleeva:2024abz}
are included in this repository with permission of the respective authors.
A Mathematica notebook that reproduces many of the analytic results of this work
is included in the ancillary material.
}
\makeatother

\appendix


\section{Comparison to past form factor breakdowns}
\label{sec:conventions}

\begin{table}
  \renewcommand{\arraystretch}{2.0}
  \caption{
    Dictionary for translating the EMT form factors
    defined in this work to conventions that have been utilized
    in prior calculations of deuteron EMT form factors.
  }
  \begin{tabular}{ccccc}
    \toprule
    ~This work ~&
    ~Cosyn \textsl{et al.}~\cite{Cosyn:2019aio} ~&
    ~Polyakov and Sun~\cite{Polyakov:2019lbq} ~&
    ~He and Zahed~\cite{He:2023ogg} ~&
    ~Panteleeva \textsl{et al.}~\cite{Polyakov:2019lbq} ~
    \\
    \hline
    $A_U$ &
    $\mathcal{G}_1 + \frac{\bd^2}{6 M_d^2} \mathcal{G}_2$ &
    $A_0 - \frac{\bd^2}{12 M_d^2} A_1$ &
    $A$ &
    $E_0$
    \\
    $A_T$ &
    $\mathcal{G}_2 $ &
    $-\frac{1}{2} A_1$ &
    $-Q$ &
    $-2E_2$
    \\
    $J$ &
    $\frac{1}{2} \mathcal{G}_5$ &
    $J$ &
    $J$ &
    $J$
    \\
    $D_U$ &
    $\mathcal{G}_3 + \frac{\bd^2}{6 M_d^2}\mathcal{G}_4 - \frac{2}{3}\mathcal{G}_6$ &
    $-D_0 - \frac{\bd^2}{12M_d^2} D_1 + \frac{4}{3} E$ &
    $D_0$ &
    $-4D_0$
    \\
    $D_{T1}$ &
    $\mathcal{G}_4$ &
    $-\frac{1}{2}D_1$ &
    $-\frac{2M_d^2}{\bd^2} D_{3}$ &
    $8D_3$
    \\
    $D_{T2}$ &
    $\frac{1}{2}\mathcal{G}_6$ &
    $-E$ &
    $D_2$ &
    $2D_2$
    \\
    $\bar{c}_U$ &
    $-\frac{1}{2}\mathcal{G}_7 - \frac{1}{2}\mathcal{G}_8 + \frac{\bd^2}{12M_d^2} \mathcal{G}_9$ &
    $\frac{3}{4}\bar{f} - \frac{1}{2}\bar{c}_0 - \frac{\bd^2}{24M_d^2} \bar{c}_1$ &
    --- &
    ---
    \\
    $\bar{c}_{T1}$ &
    $\frac{1}{2}\mathcal{G}_9$ &
    $-\frac{1}{2} \bar{c}_1$ &
    --- &
    ---
    \\
    $\bar{c}_{T2}$ &
    $\frac{1}{2}\mathcal{G}_7$ &
    $\bar{f}$ &
    --- &
    ---
    \\
    $S$ &
    $-\frac{1}{2} \mathcal{G}_{10}$ &
    --- &
    --- &
    ---
    \\
    $\bar{s}$ &
    $\mathcal{G}_{11}$ &
    --- &
    --- &
    ---
    \\
    \bottomrule
  \end{tabular}
  \label{tab:mff}
\end{table}

Table~\ref{tab:mff} presents a dictionary for translating our
EMT form factors into the nomenclature of several previous works.
The form factor breakdown of
Cosyn \textsl{et al.}~\cite{Cosyn:2019aio}
was based on prior work by Taneja \textsl{et al.}~\cite{Taneja:2011sy},
but was extended to incorporate antisymmetric structures in the asymmetric EMT.
This breakdown was used a starting point for
the deuteron helicity amplitudes used by Freese and Cosyn~\cite{Freese:2022yur}.
The form factor breakdown
of Polyakov and Sun~\cite{Polyakov:2019lbq}
was used as a starting point in works both by
He and Zahed~\cite{He:2023ogg,He:2024vzz}
and by
Panteleeva \textsl{et al.}~\cite{Panteleeva:2024abz}.
Both groups, however, primarily use auxiliary form factors that differ from
the Polyakov-Sun nomenclature.
For ease of comparison,
we also include their auxiliary form factors in Table~\ref{tab:mff}.

It is also helpful to be able to quickly convert the
light front helicity amplitudes
of Ref.~\cite{Freese:2022yur}
into EMT-FFs.
The helicity amplitudes are not form factors as such,
and there are multiple combinations of helicity amplitudes
that in principle give the same form factor.
The particular combinations we used to produce the ``Freese and Cosyn''
result in Fig.~\ref{fig:groups} are:
\begin{align}
  \begin{split}
    A_U
    &=
    \mathcal{A}_{++} + \frac{1}{3} \mathcal{A}_{-+}
    \\
    A_T
    &=
    - \frac{4 M_d^2}{\bd^2} \mathcal{A}_{-+}
    \\
    J
    &=
    \mathcal{J}_{++}
    \\
    D_U
    &=
    \frac{
      2 \mathcal{D}_{++} + \mathcal{D}_{00}
      - \frac{\bd^2}{4M_d^2}
      \Big( \mathcal{D}_{-+} + \mathcal{D}_{++} \Big)
    }{
      3\left(1-\frac{\bd^2}{4M_d^2}\right)
    }
    \\
    D_{T1}
    &=
    - \frac{4 M_d^2}{\bd^2} \mathcal{D}_{-+}
    \\
    D_{T2}
    &=
    \frac{
      \mathcal{D}_{00}
      -
      \left(
      1 - \frac{\bd^2}{2M_d^2}
      \right)
      \mathcal{D}_{++}
      -
      \mathcal{D}_{-+}
    }{
      2 \left( 1 - \frac{\bd^2}{4M_d^2} \right)
    }
    \,.
  \end{split}
\end{align}
These expressions can be derived using the explicit formulas
in Appendix B of Ref.~\cite{Freese:2022yur},
along with the dictionary in Table~\ref{tab:mff}.

Several prior conventions for EMT-FFs in the conserved part of the
symmetric EMT have been tabulated in Table I
of Polyakov and Sun~\cite{Polyakov:2019lbq},
who also provide a dictionary into their notation.
We do not include these prior conventions in Table~\ref{tab:mff} here
since they have not been used in recent deuteron EMT-FF calculations.


\section{Deuteron wave function}
\label{sec:wf}

This appendix reviews basic properties of the deuteron wave function.
Since the deuteron is a spin-triplet state,
it is helpful to recast the wave function
$\psi_d^{(s;s_p,s_n)}(\bm{r})$---which depends on two spin-half indices
$s_p$ and $s_n$---into
a $3\times3$ matrix
$\Psi_d(\bm{r})$
in the spin triplet representation
whose elements depend only on $s_p+s_n$.
Elements of this matrix are defined as:
\begin{align}
  \label{eqn:psid_msmj}
  \begin{split}
    \langle m_s | \Psi_d(\bm{r}) | m_j \rangle
    =
    \sum_{s_p, s_n}
    \left\langle 1, m_s \middle| \tfrac{1}{2}, s_p; \tfrac{1}{2}, s_n \right\rangle
    \psi_d^{(m_j;s_p,s_n)}(\bm{r})
    \,,
  \end{split}
\end{align}
where
$\left\langle 1, m_s \middle| \tfrac{1}{2}, s_p; \tfrac{1}{2}, s_n \right\rangle$
are Clebsch-Gordan coefficients.
Here the columns have fixed $m_j$
(deuteron magnetic number)
and the rows have fixed $m_s =s_p+s_n$
(two-nucleon spin magnetic number).
In this matrix form,
the radial and angular dependence of the wave function can be broken down as:
\begin{align}
  \label{eqn:wf}
  \Psi_d(\bm{r})
  =
  \mathcal{Y}_{101}(\hat{r})
  \frac{u(r)}{r}
  +
  \mathcal{Y}_{121}(\hat{r})
  \frac{w(r)}{r}
  \,,
\end{align}
where $u(r)$ and $w(r)$ are respectively the S- and D-wave radial wave functions,
and where
\begin{align}
  \langle m_s | \mathcal{Y}_{jls}(\hat{r}) | m_j \rangle
  =
  \sum_{m_l}
  \langle l, m_l; s, m_s | j, m_j \rangle
  Y_l^{m_l}(\hat{r})
\end{align}
are commonly called spinor spherical harmonics.
Here, $Y_l^m(\hat{r})$ are the spherical harmonics.

A few other elementary properties of the deuteron wave function
are worth reviewing.
Given the convention in Eq.~(\ref{eqn:wf}),
the S- and D-waves are normalized jointly as:
\begin{align}
  \label{eqn:wf:norm}
  \int_0^\infty \d r \,
  \Big(
  u^2(r) + w^2(r)
  \Big)
  =
  1
  \,.
\end{align}
There is no $r^2$ in the integration element,
because it's effectively been absorbed by the definition of $u(r)$ and $w(r)$.
Integrals of $u^2(r)$ or $w^2(r)$ by themselves give the S- and D-wave probabilities:
\begin{align}
  \label{eqn:sd:prob}
  \begin{split}
    \int_0^\infty \d r \,
    u^2(r)
    &=
    \mathcal{P}_S
    \\
    \int_0^\infty \d r \,
    w^2(r)
    &=
    \mathcal{P}_D
    \,.
  \end{split}
\end{align}

The quadrupole moment of the deuteron is defined
as the mean value of $\frac{1}{4} (3z^2 - r^2)$ in an $s_d=+1$ state,
and is given in terms of the S- and D-waves by:
\begin{align}
  \label{eqn:wf:quad}
  Q_d
  =
  \frac{1}{20}
  \int_0^\infty \d r \,
  r^2
  \Big[
    2\sqrt{2} u(r) w(r) - w^2(r)
    \Big]
  \,.
\end{align}


\subsection{Spinor spherical harmonics}

As matrices, the spinor spherical harmonics can be written:
\begin{align}
  \begin{split}
    \mathcal{Y}_{101}(\hat{r})
    &=
    Y_0^0(\hat{r})
    \mathbbm{1}_{3\times3}
    \\
    \mathcal{Y}_{121}(\hat{r})
    &=
    \begin{bmatrix}
      \sqrt{\tfrac{1}{10}} Y_2^0(\hat{r})
      & -\sqrt{\tfrac{3}{10}} Y_2^{1*}(\hat{r})
      & \sqrt{\tfrac{3}{5}} Y_2^{2*}(\hat{r})
      \\
      \sqrt{\tfrac{3}{10}} Y_2^{1}(\hat{r})
      & -\sqrt{\tfrac{2}{5}} Y_2^0(\hat{r})
      & \sqrt{\tfrac{3}{10}} Y_2^{1*}(\hat{r})
      \\
      \sqrt{\tfrac{3}{5}} Y_2^{2}(\hat{r})
      & \sqrt{\tfrac{3}{10}} Y_2^{1}(\hat{r})
      & \sqrt{\tfrac{1}{10}} Y_2^0(\hat{r})
    \end{bmatrix}
    \,.
  \end{split}
\end{align}
The spinor spherical harmonics---and thus the matrix form of the deuteron
wave function (\ref{eqn:wf})---are Hermitian.

Neglecting charge symmetry breaking terms,
the nucleon-nucleon potential
can generically be written in the form:
\begin{align}
  \label{eqn:potential}
  V(\bm{r}, \bm{\sigma}_p, \bm{\sigma}_n)
  =
  V_d(r)
  +
  (\bm{\sigma}_p\cdot\bm{\sigma}_n)
  V_\sigma(r)
  +
  \hat{S}_{pn}
  V_t(r)
  +
  \bm{L}^2
  V_{l2}(r)
  +
  (\bm{L}\cdot\bm{S})
  V_{ls}(r)
  +
  (\bm{L}\cdot\bm{S})^2
  V_{ls2}(r)
  \,,
\end{align}
where $\bm{S} = \bm{\sigma}_p + \bm{\sigma}_n$ and
\begin{align}
  \hat{S}_{pn}
  =
  3(\bm{\sigma}_p\cdot\hat{r})(\bm{\sigma}_n\cdot\hat{r})
  -
  \bm{\sigma}_p \cdot \bm{\sigma}_n
\end{align}
is the tensor operator.
For the spin-triplet channel,
$(\bm{\sigma}_p\cdot\bm{\sigma}_n)$ effectively equals $1$,
and the functions $V_d(r)$ and $V_\sigma(r)$ can be combined
into a central potential:
\begin{align}
  V_c(r) = V_d(r) + V_\sigma(r)
  \,.
\end{align}
The actions of the remaining operators in the potential (\ref{eqn:potential})
on the spinor spherical harmonics are:
\begin{align}
  \renewcommand{\arraystretch}{1.5}
  \begin{array}{ccc}
    \mathbbm{1}
    \mathcal{Y}_{101}
    =
    \mathcal{Y}_{101}
    &
    \qquad \qquad
    &
    \mathbbm{1}
    \mathcal{Y}_{121}
    =
    \mathcal{Y}_{121}
    \\
    \hat{S}_{pn}
    \mathcal{Y}_{101}
    =
    \sqrt{8}
    \mathcal{Y}_{121}
    &
    \qquad \qquad
    &
    \hat{S}_{pn}
    \mathcal{Y}_{121}
    =
    \sqrt{8}
    \mathcal{Y}_{101}
    -
    2
    \mathcal{Y}_{121}
    \\
    \bm{L}^2
    \mathcal{Y}_{101}
    =
    0
    &
    \qquad \qquad
    &
    \bm{L}^2
    \mathcal{Y}_{121}
    =
    6
    \mathcal{Y}_{121}
    \\
    (\bm{L}\cdot\bm{S})
    \mathcal{Y}_{101}
    =
    0
    &
    \qquad \qquad
    &
    (\bm{L}\cdot\bm{S})
    \mathcal{Y}_{121}
    =
    -3
    \mathcal{Y}_{121}
    \\
    (\bm{L}\cdot\bm{S})^2
    \mathcal{Y}_{101}
    =
    0
    &
    \qquad \qquad
    &
    (\bm{L}\cdot\bm{S})^2
    \mathcal{Y}_{121}
    =
    9
    \mathcal{Y}_{121}
    \,.
  \end{array}
\end{align}
Accordingly, the Schr\"odinger equation for the deuteron wave function
\begin{align}
  \label{eqn:schrodinger}
  -\frac{\bm{\nabla^2}}{m_N}
  \Psi_d(\bm{r})
  +
  V(\bm{r},\bm{\sigma}_p,\bm{\sigma_n})
  \Psi_d(\bm{r})
  =
  E_d
  \Psi_d(\bm{r})
  \,,
\end{align}
where $E_d < 0$ is the deuteron binding energy,
can be rewritten as coupled second-order differential equations
for $u(r)$ and $w(r)$:
\begin{align}
  \label{eqn:uw:diff}
  \begin{split}
    u''(r)
    &=
    m_N
    \Big(
    -E_d + V_c(r)
    \Big)
    u(r)
    +
    \sqrt{8} m_N V_t(r)
    w(r)
    \\
    w''(r)
    &=
    m_N
    \left(
    -E_d + V_w(r)
    +
    \frac{6}{m_N r^2}
    \right)
    w(r)
    +
    \sqrt{8} m_N V_t(r)
    u(r)
    \\
    V_w(r)
    & \equiv
    V_c(r) - 2 V_t(r) + 6 V_{l2}(r) - 3 V_{ls}(r) + 9 V_{ls2}(r)
    \,.
  \end{split}
\end{align}
These relations are helpful for numerical calculations
of the deuteron EMT-FFs,
since several formulas depend on the second derivatives
of $u(r)$ and $w(r)$.


\subsection{Polarization Vectors and Projectors}

The polarization vectors $\epi_a(s_d)$ for a non-relativistic deuteron
with spin-quantization axis along $\hat{\bm z}$ are:
\begin{align}
  \epi_a(\pm 1)
  =
  \frac{\hat{x}_a\pm i\hat{y}_a}{\sqrt{2}}
  \; ,\qquad
  \epi(0)
  =
  \hat{z}_a
  \,.
\end{align}
Note that, in contrast to the relativistic case,
these polarization vectors do not depend on the deuteron's momentum.

The unpolarized projector is:
\begin{equation}
    \frac{1}{3}\sum_{s_d}
  \epi_a(s_d)
  \epf_b(s_d)= \frac{1}{3} \delta_{ab}
\end{equation}
The vector-polarized projector is:
\begin{equation}
 \frac{1}{2}\left[
  \epi_a(1)
  \epf_b(1)-
  \epi_a(-1)
  \epf_b(-1)\right]= -\frac{1}{2}\epsilon^{abc}\hat{z}_c
\end{equation}
The tensor-polarized projector is:
\begin{equation}
  \epi_a(1) \epf_b(1)
  +
  \epi_a(-1)
  \epf_b(-1)
  -
  2 \epi_a(0) \epf_b(0)
  =
  \delta^{ab}
  -
  3
  \hat{z}^a\hat{z}^b
\end{equation}


\subsection{One-body operators in the spin-triplet representation}

Besides the deuteron wave function itself,
nucleon-spin-dependent operators also can be converted to
matrices in the triplet representation,
and sandwiched between $\Psi_d^\dagger(\bm{r})$ and $\Psi_d(\bm{r})$.
The trace of the result can be taken with a spin density matrix to get
e.g.\ the expectation value for unpolarized or tensor-polarized
ensembles of deuterons.
This provides an especially practical pathway for calculations of EMT-FFs,
by replacing spin sums by matrix operations.
In such a spin matrix representation,
Eq.~(\ref{eqn:1body:coordinate}) would be rewritten:
\begin{align}
  \label{eqn:1body:matrix}
  \langle \bm{p}'_d | \hat{T}^{ij}_p | \bm{p}_d \rangle
  =
  \int \d^3 r \,
  \e^{i\frac{\bd\cdot\bm{r}}{2}}
  \Psi_d^\dagger(\bm{r})
  \langle \bm{p}'_p | \hat{T}^{ij}_p | \bm{p}_p \rangle
  \Psi_d(\bm{r})
  \bigg|_{
    \bm{P}_p
    =
    \tfrac{1}{2} \big(
    \bm{P}_d - i\blrn
    \big)
  }
  \,,
\end{align}
where
$\langle \bm{p}'_p | \hat{T}^{ij}_p | \bm{p}_p \rangle$
is a spin-triplet version of the nucleon matrix element,
defined by introducing a Kronecker delta
for the spectator spin and coupling to the
appropriate Clebsch-Gordan coefficients:
\begin{align}\label{eqn:T_triplet}
  \langle p_p'; m_s' | \hat{T}^{ij}(0) | p_p; m_s \rangle
  \equiv
  \sum_{\substack{s_p, s'_p \\ s_n, s'_n}}
  \left\langle 1, m'_s \middle| \tfrac{1}{2}, s'_p; \tfrac{1}{2}, s'_n \right\rangle
  \langle p_p', s'_p | \hat{T}^{ij}(0) | p_p, s_p \rangle
  \left\langle \tfrac{1}{2}, s_p; \tfrac{1}{2}, s_n \middle| 1, m_s \right\rangle
  \delta_{s_n, s'_n}
  \,.
\end{align}
In Eq.~(\ref{eqn:T_triplet}), the coupling of the scalar neutron spin operator $\delta_{s_n,s'_n}$ with the scalar and vector proton spin operators in Eq.~(\ref{eqn:mff:half}) amounts to substituting the proton spin operators with their counterparts in the spin-1 representation:
\begin{align}
  \label{eqn:sigma:triplet}
  \begin{split}
    \delta_{s's}
    &\rightarrow
    \mathbbm{1}_{3\times3}
    \\
    (\sigma_x)_{s's}
    &\rightarrow
    \frac{1}{\sqrt{2}}
    \begin{bmatrix}
      0 & 1 & 0 \\
      1 & 0 & 1 \\
      0 & 1 & 0
    \end{bmatrix}
    \\
    (\sigma_y)_{s's}
    &\rightarrow
    \frac{1}{\sqrt{2}}
    \begin{bmatrix*}[r]
      0 & -i & 0 \\
      i & 0 & -i \\
      0 & i & 0
    \end{bmatrix*}
    \\
    (\sigma_z)_{s's}
    &\rightarrow
    \frac{1}{\sqrt{2}}
    \begin{bmatrix*}[r]
      1 & 0 & 0 \\
      0 & 0 & 0 \\
      0 & 0 &-1
    \end{bmatrix*}
    \,.
  \end{split}
\end{align}
A version of the breakdown (\ref{eqn:mff:half})
with these substitutions can be inserted directly into
the matrix equation (\ref{eqn:1body:matrix}).


\bibliography{references.bib}

\end{document}